\documentclass[11pt,fleqn]{report}
\usepackage{rsfs}
\usepackage[dvipdfm]{graphicx}
\usepackage{mediabb}
\usepackage{amsmath}\usepackage{longtable}
\usepackage{amssymb}
\usepackage[all]{xy}

\newcommand{\na}{\nabla}
\newcommand{\beq}{\begin{equation}}
\newcommand{\eeq}{\end{equation}}
\newcommand{\bit}{\begin{itemize}}
\newcommand{\eit}{\end{itemize}}
\newcommand{\ben}{\begin{enumerate}}
\newcommand{\een}{\end{enumerate}}
\newcommand{\la}{\langle}
\newtheorem{criterion}{Homogeneity Criterion}

\newcommand{\ra}{\rangle}

\newcommand{\ch}{{\cal{H}}}

\newcommand{\bc}{{\boldsymbol{C}}}
\newcommand{\cu}{{\cal{U}}}
\newcommand{\cm}{{\cal{M}}}

\newtheorem{Theorem}{Theorem}

\newcommand{\fh}{\mathfrak{H}}

\newcommand{\co}{{\cal{O}}}
\newcommand{\bs}{\boldsymbol}
\newcommand{\f}{\frac}

\newcommand{\mb}{\mbox}

\newcommand{\bz}{{\boldsymbol{Z}}}

\begin{document}
\begin{titlepage}

\begin{flushright}
\today
\end{flushright}

\vspace{1in}

\begin{center}

{\bf Time and a Temporally Statistical Quantum Geometrodynamics}

\vspace{1in}

\normalsize

{Eiji Konishi\footnote{E-mail address: konishi.eiji.27c@st.kyoto-u.ac.jp}}

\normalsize
\vspace{.5in}

{\em Faculty of Science, Kyoto University, Kyoto 606-8502, Japan}

\end{center}

\vspace{1in}

\baselineskip=24pt
\begin{abstract}
This paper is an exposition of the author's recent work 
%(arXiv:1001.3382 [hep-th], arXiv:1005.5430 [cond-mat.dis-nn], arXiv:1212.4956 [quant-ph])
 on modeling M-theory vacua and quantum mechanical observers in the framework of a temporally statistical description of quantum geometrodynamics, including measurement processes based on the canonical theory of quantum gravity. In this paper we deal with several fundamental issues of time: the time-less problem in canonical quantum gravity; the physical origin of state reductions; and time-reversal symmetry breaking. We first model the observers and consider the time-less problem by invoking the time reparametrization symmetry breaking in the quantum mechanical world as seen by the observers. We next construct the hidden time variable theory, using a model of the gauged and affinized S-duality symmetry in type IIB string theory, as the statistical theory of time and explain the physical origin of state reductions using it. Finally, by the extension of the time reparametrization symmetry to all of the temporal hidden variables, we treat the issue of time reversal symmetry breaking as the spontaneous breaking of this extended time reparametrization symmetry. The classification of unitary time-dependent processes and the geometrizations of unitary and non-unitary time evolutions using the language of the derived category are also investigated.
\end{abstract}

\vspace{.7in}
 
\end{titlepage}
\tableofcontents

\chapter{Introduction}

{\it{Geometrodynamics}} is a theoretical framework in physics to describe gravity, other fundamental interactions in nature and their sources in terms of a {{curved empty space}} and nothing more.\cite{MW,Wheeler0} In this paper, we review the author's previous attempts to construct a temporally statistical description of {{quantum}} geometrodynamics including measurement processes,\cite{Konishi2,Konishi,Konishi3} reveal its fundamental structures and attempt to resolve several fundamental issues of time.

In our quantum geometrodynamics, quantum gravity plays the central role, since it deals with the dynamical properties of time. As already seen in De Witt's celebrated three early papers on the quantum theory of gravity,\cite{WDW1,DW2,DW3} two kinds of theories of quantum gravity are recognized: the canonical theory and the background-dependent manifestly covariant theory. The former can be regarded as a non-perturbative theory and the latter is treated in De Witt's second and third papers\cite{DW2,DW3} as a perturbative theory. Before the second revolution,\cite{WP1,WP2} string theory\cite{GSW,GSW2} (except for string field theories\cite{HIKKO1,Witten1}) was an example of the latter type of theory, and loop quantum gravity theory,\cite{loop3,loop4,loop5,loop6,loop7,loop8} which is the Wilson loop representation of the Ashtekar reformulation of the general theory of relativity in terms of connection variables,\cite{loop1,loop2} was an example of the former type of theory. The remarkable feature of the canonical theory is the background independence, like that of the general theory of relativity. In the construction of our quantum geometrodynamics, which is of course background-independent, we invoke the canonical theory of quantum gravity using a model of S-duality symmetry\cite{Sen,jhs} in the gauged and affinized form\cite{Konishi2,K}, which was proposed and investigated by the author in the context of type IIB string theory.

The most central physical concept of the canonical theory of quantum gravity is the {\it{wave function of the Universe}}, which was invented by Wheeler and De Witt in the 1960s\cite{WDW1,WDW2,WDW3}. In the 1980s, inspired by the cosmological scenario of inflation \cite{Guth}, Vilenkin\cite{Vilenkin1,Vilenkin2,Vilenkin3} and Hartle and Hawking\cite{WDW4} applied this concept to the cosmology of the early Universe, where quantum gravity effects are dominant, with differing boundary conditions (tunneling from nothing and no-boundary, respectively). It is well-known that, in the fundamental equation in the canonical quantum theory of gravity (the {\it{Wheeler-De Witt equation}} of the wave function of the Universe), the Time Reparametrization Symmetry (TRpS) means that there is no concept of a Newtonian, external time and time is assumed to be generated by an internal clock, such as (in the most addressable example) the one provided by an expanding Universe. The wave function of the Universe naturally includes the behaviors of the wave functions of matter in quantum field theories. The internal clock works for these matter wave functions but it still requires the time reparametrization invariance and rejects the existence of a Newtonian, external time. This fundamental issue of the absence of a Newtonian, external time in the canonical theory of quantum gravity is the so-called `time-less problem'\cite{UW,Isham} and its details will be studied in this paper.

 The second most central physical concept of our theory is the {\it{observer}}. Indeed, any quantum theory including matter dynamics requires an observer. A wave function of matter has been interpreted as a stochastic de Broglie wave. The absolute square value of a coefficient of its eigenfunction expansion for an observable gives the measurement probability for the corresponding eigenvalue of the observable. Here, the measurement is made by an observer. The wave functions of matter develop unitarily until measurement from outside induces non-unitary development. The unitary time-dependent processes of the wave functions of quantum fields are studied in quantum field theory which searches for the fundamental gauge interactions and has provisionally resulted in the standard model of particle physics.\cite{Weinberg1,Weinberg2} On the other hand, for a long time, the non-unitary processes themselves were not taken up as an object of physical research. However, in the 1990s Roger Penrose proposed that the non-unitary processes seen in measurement processes are caused by quantum gravitational effects on time and the superposition of the wave functions is dynamically reduced.\cite{Penrose} In this thesis, the wave functions of matter are always physical entities, including times when they are subject to non-unitary processes.

Based on the Penrose thesis, we aim to resolve three fundamental problems of time in a quantum geometrodynamics including measurement processes:
\begin{enumerate}
\item {\bf{The time-less problem in canonical quantum gravity.}}

 As mentioned above, the Wheeler-De Witt equation of the wave function of the Universe has no time and possesses the TRpS. Consequently, the dynamics of the wave function of the Universe does not admit a description in terms of a Newtonian, external time. This time-less problem will be resolved in 
      {{Chapter 4}}
 %Section 4
  by invoking TRpS breaking in the quantum mechanical world as seen by the observer, which is assumed to be due to the classical and quantum mechanical self-identities of the observer. This TRpS breaking is the most important assumption of any quantum physical theory which presupposes the existence of observers.
\item {\bf{How many kinds of time-dependent processes exist? Classification and geometrizations of time-dependent processes.}}

Time-dependent processes have the following two types.
\begin{enumerate}
\item Unitary processes. We will construct a model in which the structure of the interactions in these processes is determined by the symmetry breaking of a fundamental discrete modular symmetry, which is a symmetry consequence in the quantized model of the gauged and affinized S-duality symmetry, due to the presence of a non-zero vacuum energy.\cite{Konishi2,K,KM1,KM2} (See 
Chapter 5.)
%Section 5.)
\item Non-unitary processes. To explain these processes in terms of the Penrose thesis, in 
{{Chapter 5}},
%Section 5,
we construct a statistical theory of time by using the hidden time variable and an infinite number of hidden parameters
as non-local hidden variables.
In 
Chapter 6,
%Section 6,
 the time-dependent processes in the quantum mechanical world, including the non-unitary processes, are formulated in the language of the derived category, where the objects of this derived category represent the 
 time-dependent processes of the matter systems.
\end{enumerate}
In the final step, both types of processes are geometrized using the derived category of the Homotopy Associative ($A_\infty$) category\cite{Fukaya,Seidel,FOOO}  in
      {{Chapter 6}}.
%Section 6.
 There we introduce the non-linear potential and reformulate the derived category model by giving the morphisms priority over the objects.
\item {\bf{How the Time Reversal Symmetry (TRS) is broken.}}

The laws of quantum physics are invariant under the combination of charge conjugation, space inversion and time reversal operations (CPT theorem).\cite{CPT} Although CP invariance is not maintained,\cite{CPviolate} the violation is very small so time reversal invariance is a good approximation. However, we know that in nature there is a very obvious distinction between the future and past directions of time. We will attempt to resolve this issue in 
{{Chapter 7}}
%Section 7
 by the proposal that the vacuum expectation values of newly introduced scalar particles $\Theta$ for the spontaneously broken strong TRpS of all of the temporal hidden variables (i.e., the hidden time variable and the temporal non-local hidden variables)
 give the temporal non-local hidden variables of the model
and lead to non-zero statistical variance of time increments determined by the values of the non-local hidden variables.
The inverse of the gradients of the time increment by the temporal non-local hidden variables are the Goldstone modes of these scalar particles $\Theta$. The vacuum expectation values of $\Theta$ produce the non-unitary time processes and break the TRS.
\end{enumerate}

Here, we note that, throughout this paper, we repeatedly use the following principal scheme of the Goldstone theorem to resolve issues.\cite{Goldstone,gsw}
\begin{equation}
{\mbox{Exact\ Symmetry}}\longrightarrow{\mbox{Spontaneous\ Breakdown}}\longrightarrow {\mbox{Goldstone\ Modes}}\;.
\end{equation}
In this scheme, each exact symmetry indicates a potential structure in the model of nature, and its spontaneous breakdown with Goldstone modes indicates the variability of such structures.

Next, we raise two miscellaneous concerns, which are related to the second problem in the above list:
\begin{enumerate}

\item {\bf{Non-linear potentials and the dynamics of morphisms.}}

We make a drastic change to the view of the non-perturbative transitions between type IIB string theory vacua in our model of the gauged and affinized S-duality symmetry: in the conclusive categorical formulation of the quantum mechanical world given in 
Chapter 6,
%Section 6,
 it is not objects themselves but the relationships between objects via the dynamical morphisms that control the non-perturbative transitions between vacua. The equation of the dynamical non-linear potentials for the morphisms
will be derived in 
Chapter 6.
%Section 6.
\item {\bf{Covariance and equivalence principles for the interactions.}}

The 
loop
part of the gauged and affinized S-duality symmetry is associated with the conventional Chan-Paton gauge symmetries\cite{WP2,CP} of the many body systems of Dirichlet-branes and fundamental open strings. There are two principles used to formulate the Chan-Paton gauge interactions among these systems geometrically.
First is the covariance principle, which states that in the D-brane field theory, the Chan-Paton gauge interactions are formulated as an $A_\infty$ category. Second is the equivalence principle, which states that the minimal model of this $A_\infty$ category,
 which is a gauge fixing of the gauge symmetries in the action or the equation of motion of string fields,
 gives the local elimination of the Chan-Paton gauge interactions, whose non-perturbative distortions are given by the non-linear potential, as an $A_\infty$ functor. The existence of such a minimal model is ensured by a mathematical theorem.\cite{Kad,Kajiura}
(See Chapter 6.)
%(See Section 6.)
\end{enumerate}

The organization of this paper is as follows.

In the next 
chapter, 
%section, 
we present minimal accounts of the basic notions: the Wheeler-De Witt equation and the Penrose thesis on state reductions, which will be revisited in 
Chapter 5
%Section 5
in the language of the hidden time variable theory.

In
Chapter 3,
%Section 3,
 we present a model of human brain-type quantum mechanical observers. These observers are modeled as neural-glial networks and have the ability to complete measurement.\footnote{Throughout this paper, we refer to the network of astrocytes, which are one of the glial cells in the human brain, as the {\it{glial network}}.}\cite{Konishi}

In 
Chapter 4,
%Section 4,
 we propose a scenario to resolve the time-less problem in canonical quantum gravity. In this scenario, we are forced to recognize clearly that the description of any quantum theory is based on an existence of an observer and thus TRpS is spontaneously broken in the quantum mechanical world as seen by the observer, although TRpS in the full quantum mechanical world is not broken. We reformulate the role of the Penrose thesis on the observer's conscious activities in terms of TRpS breaking.

Chapter 5
%Section 5
 is the core part of this exposition. It lays the foundations of the statistical theory of time, which explains how the state reductions are generated by the Penrose thesis, using a statistical model of time increments via an infinite number of hidden parameters and the hidden time variable for a linear combination of the Virasoro conserved charges accompanying the gauged and affinized S-duality symmetry in type IIB string theory.
  First, we explain the reason why the gauged and affinized S-duality symmetry is fundamental.\cite{Konishi2} Next, in the Becchi-Rouet-Stora-Tyutin (BRST) formalism of this gauge theory,\cite{BRST1,BRST2,BRST3,BRSTreview} by invoking the Hanada-Kawai-Kimura (HKK) result\cite{HKK}, we introduce all the ingredients of the model, that is, the spatial expanses, space-time dimensionality, the cosmic time increment as a function of the hidden time variable, time development of matter and the structure of the interactions between matter, by using the vacuum wave function that is interpreted as the wave function of the Universe.

In 
Chapter 6,
%Section 6,
 we formulate the sequences of non-unitary time processes in the language of derived categories. As explained above, we drastically change the dynamics of the category by putting morphism structure into the main position and introducing a non-linear potential. In the latter half of this 
chapter,
%section,
 we further investigate this derived category formulation in the language of the $A_\infty$ category.
  We also invoke the two principles of $A_\infty$ covariance and equivalence. The equivalence principle is for Chan-Paton quantized interactions and is based on the mathematical minimal model theorem of $A_\infty$ categories. Standing on this stage, we can view our constructions as a quantum geometrodynamics where the derived category of the $A_\infty$ category is curved empty space and nothing more.

In 
Chapter 7,
%Section 7,
 we summarize the results in the 
chapters 
%sections
so far. We introduce the strong version of TRpS for all of the temporal hidden variables  and propose a scenario in which this symmetry is spontaneously broken by the vacuum expectation values of newly introduced scalar particles $\Theta$, as explained above. This simultaneously breaks TRS. We survey all of the ingredients in our theory as a list. This list may be used as a guide to read the other 
chapters.
%sections.

Besides this main content, we need three appendices.

In Appendix A we present minimal accounts of the physiological elements of the neural-glial system in the human brain, which are required in 
Chapter 3.
%Section 3.

In Appendix B we calculate the transition rates of observer's quantum state based on the model of observers that is described in detail in 
Chapter 3.
%Section 3.

In Appendix C we give the definition of the renormalizations of vacua used in 
Chapter 5
%Section 5
by the Whitham deformation method.\cite{Wh1,Wh2,WKB}

Finally, although most of the ideas expressed in this paper are based on previously published work,\cite{Konishi2,Konishi,Konishi3} this paper does contain original material. In reviewing these ideas, the connections between them, which were not suggested in original papers, have been clarified by positioning them within the complete theory. Some new arguments and explanations are also given.

\subsection*{Acknowledgements}
I dedicate this paper to my mentor Yoshitaka Yamamoto and express my sincere gratitude to him for his continuous warm encouragement.
\chapter{Basic Notions}
Before we start to discuss our theory, we give here the minimum necessary accounts of the Wheeler-De Witt equation and the Penrose thesis on state reductions. Of course, readers who are familiar with these notions may skip this 
chapter.
%section.
\footnote{In this 
      chapter, the next two chapters
%section, the next two sections
and Appendix A, we denote time by $t$.}

\section{Wheeler De-Witt Equation}
In this 
section,
%subsection,
 we present a brief account of the basic ideas of the Wheeler-De Witt equation in four-dimensional space-time, using Planck units.\footnote{To simplify the explanation, here we consider the case in which the cosmological constant is absent.}\cite{Konishi3}
 (In the connection to the model in 
       Chapter 5,
% Section 5,
  this set up corresponds to the type IIB quantum string cosmology compactified on a Calabi-Yau threefold.\cite{KM2})
This equation is the fundamental equation of canonical quantum gravity and is the central object in this paper.
The {\it{Wheeler-De Witt equation}} of the wave functions of the Universe $\Psi$\cite{WDW1,WDW2,WDW3}, which is the temporal part of the 
diffeomorphism invariance (i.e., the time reparametrization invariance) 
condition on $\Psi$, takes the form
\begin{equation}\hat{{\cal{H}}}\Psi(h,{\mbox{Other\ Variables}})=0\;,\end{equation}
for the quantum Hamiltonian operator $\hat{{\cal{H}}}$ of general relativity. The quantum Hamiltonian operator is obtained by the canonical quantization of general relativity about the spatial metric $h$:
\begin{equation}
p_h\longrightarrow \hat{p}_h=-i\hbar\frac{\delta}{\delta h}\;,
\end{equation}
for the canonical conjugate momentum $p_h$ under the Arnowitt-Deser-Misner (ADM) decomposition of the space-time metric $g$\cite{ADM}, which treats the temporal development of space-time as foliations of three-dimensional hypersurface by introducing two kinds of Lagrange multipliers in the Einstein-Hilbert Lagrangian density
\begin{equation}
{\cal{L}}=\sqrt{-g}R_{g}\;,\end{equation}
for the four-dimensional space-time scalar curvature $R_{g}$, that is, a temporal lapse function and a spatial shift vector in the space-time metric.

To solve the Wheeler-De Witt equation, in the full infinite dimensional moduli space (called {\it{superspace}}) of all variables in the spatial metric is not possible. So, usually, we consider the mini-superspace of the scale factor variable $a$ in the spatial metric $h$ in the context of a homogeneous and isotropic Universe:
\begin{equation}
ds^2=-N^2(t)dt^2+a^2(t)d\Omega_3^2\;,
\end{equation}
where $d\Omega_3^2$ is the metric on the three-dimensional unit spherical surface and $N(t)$ is the temporal lapse function.
The gravitational part of the action of such a model is
\begin{equation}
S_{{{gra}}}=\frac{1}{2}\int dt \biggl(\frac{N}{a}\biggr)\Biggl(\biggl(\frac{a}{N}\frac{da}{dt}\biggr)^2-U(a)\Biggr)\;,\label{eq:Hartle-Hawking}
\end{equation}
where the ``potential'' part in the absence of a cosmological term is
\begin{equation}U(a)=-\sqrt{h(a)}R_{h}(a)\;,\end{equation}
for the three-dimensional spatial scalar curvature $R_{h}$.
When we write down the ``kinetic'' part of the Wheeler-De Witt equation corresponding to Eq.(\ref{eq:Hartle-Hawking}), there is an operator ordering issue for the scale factor $a$. By choosing the trivial operator ordering for the scale factor $a$, the Wheeler-De Witt equation is
\begin{equation}(-\hbar^2\partial_a^2+U(a)+\hat{{\cal{H}}}_q)\Psi(a,q)=0\;,\end{equation} where $q$ and ${\cal{H}}_q$ are the quantum mechanical variables of the matter systems being considered and their Hamiltonian, respectively.

In the semiclassical regime for the variable $a$, the scale factor of the Universe is treated as a clock. We now explain the derivation of the Schr${\ddot{{\rm{o}}}}$dinger equation\footnote{In the following, 
 we refer to the {{Schr${{\ddot{\mbox{o}}}}$dinger equations}} in the 
 coordinate-type representation.} for the matter wave function $\chi$, in this semiclassical regime for the scale factor only, following Vilenkin's paper.\cite{Vilenkin4}
We adopt the WKB form ansatz for the wave function of the Universe:
\begin{equation}\Psi(a,q)=\Psi_0(a)\chi(a,q)\;,\ \ \Psi_0(a)=A(a)e^{iS(a)/\hbar}\;.\label{eq:ansatz}\end{equation}
Here $S(a)$ is of order $\hbar^0$. For the $\Psi_0$ part of the ansatz in Eq.(\ref{eq:ansatz}), the Wheeler-De Witt equation gives
\begin{equation}(-\hbar^2 \partial_a^2+U(a))\Psi_0(a)=0\;.\end{equation}
The parts of order $\hbar^0$ and order $\hbar^1$ are
\begin{equation}-(\partial_a{S})^2+U(a)=0\;,\ \ iA\partial_a^2{S}+2i\partial_aA\partial_a S =0\;.\end{equation}
The matter part of the Wheeler-De Witt equation is
\begin{equation}2i\hbar\partial_aS\partial_a\chi-\hat{{\cal{H}}}_q\chi=0\;.\label{eq:Sch01}\end{equation}
By recognizing the expanding Universe as a clock, we introduce time $t$ by
\begin{equation}i\hbar\partial_t a=2N\partial_a S\;,\label{eq:Sch02}\end{equation}for lapse function $N=N(t)$ that is the Lagrange multiplier representing the arbitrariness of the time coordinate, and due to the time reparametrization invariance, $dt$ can appear only in the combination $N(t)dt$. When we consider $t$ as the cosmic time, $N=1$. Here we note that, in Eq.(\ref{eq:Sch02}), $\partial_aS$ is recognized as a canonically dual variable of the scale factor in the mini-superspace.
By substituting Eq.(\ref{eq:Sch02}) in Eq.(\ref{eq:Sch01}), we obtain
\begin{equation}i\hbar \partial_t\chi=N\hat{{\cal{H}}}_q\chi\;.\label{eq:Sch03}\end{equation}
As will be discussed in 
Chapter 4,
%Section 4,
 when the TRpS is broken and we choose $N=1$, Eq.(\ref{eq:Sch03}) is the Schr${\ddot{{\rm{o}}}}$dinger equation for the matter wave function $\chi$.
 \section{Penrose Thesis on State Reductions}
Sixteen years ago, Penrose and Hameroff proposed a scenario describing observer systems, which in this paper are human brains, to be a clue to the resolution of the unsatisfactory conceptual feature of the Copenhagen interpretation of quantum mechanics as seen in the infamous measurement problem.\cite{HP1,HP2} Among their ideas, the one relevant to our study can be summarized as follows.\cite{Konishi}
\ben
\item The conscious activities of the human brains contain a non-computable and non-algorithmic process.
\item As far as is known, a candidate for such a process is the collapse of a superposition of wave functions, and they adopt it.
\item Quantum gravity effects concerning the fluctuation of time increments cause the objective reduction of wave functions.
\een 
When we generalize their statement 1 to include the singular measurement property of the observers, their description of quantum mechanics does without the concept of an {\it{abstract ego}}. The measurement activity of the abstract ego, that is, the projection hypothesis in von Neumann's infinite regress of measurement,\cite{vNeumann1,vNeumann2,Wigner} is replaced by the quantum fluctuation of the time increment.
Then, the system of the quantum mechanical world plus quantum mechanical observers can be seen in total as an object of study in quantum physics.\cite{Konishi}

 In this 
       section,
 %subsection,
  we present a brief account for the Penrose thesis on the role of the effects of quantum gravity on the state reduction.\cite{Konishi}
In order to find the concrete form of the statement of the Penrose thesis, for a Hamiltonian $\hat{{\cal{H}}}$ and the wave function $\psi({\boldsymbol{x}},t)$, we rewrite the inverse of the derivative by time $t$ in the Schr${\ddot{{\mbox{o}}}}$dinger equation (where TRpS breaking is assumed)
\begin{equation}
i\hbar\frac{\partial \psi({\boldsymbol{x}},t)}{\partial t}=\hat{{\cal{H}}}\psi({\boldsymbol{x}},t)\;,
\end{equation} as an average over a normal stochastic variable $\delta t$:
\begin{equation}
\langle\psi({\boldsymbol{x}},t)\rangle=\exp\biggl(-\frac{it}{\hbar} \hat{{\cal{H}}} -\frac{\sigma t}{2\hbar^2}\hat{{\cal{H}}}^2\biggr)\langle\psi({\boldsymbol{x}},0)\rangle\;,\label{eq:Est}
\end{equation}
where the average is defined by the following recursion equation
\begin{equation}
\langle \psi({\boldsymbol{x}},\mu)\rangle=\int dt^\prime\exp\biggl(-\frac{i\delta t^\prime}{\hbar}{\hat{\cal{H}}}\biggr)f(\delta t^\prime)\langle\psi({\boldsymbol{x}},0)\rangle\;.\label{eq:Ave2}
\end{equation}
In Eq.(\ref{eq:Ave2}) we make an average over a normal stochastic variable $\delta t$ with mean $\mu$, quantum variance $\sigma\mu$ and distribution function $f(\delta t^\prime)$. In this paper, we often refer to the second exponential factor in Eq.(\ref{eq:Est}) as the {\it{quantum variance effect of the time increment}} $\delta t$. Owing to the facts that time increments can be accumulated and time has  the aspect of `now', the scale of the quantum variance of the time increment is 
determined
 by the scale of the vacuum expectation value of time in TRpS breaking (see 
Chapters 4 and 5).
%Sections 4 and 5).
 {\it{It is characteristic to existing observers and is considered as a new physical constant.}}

In quantum mechanics, a Hamiltonian $\hat{{\cal{H}}}$ is a Hermitian operator. Thus for the eigenvalues $\{\lambda\}$ of $\hat{{\cal{H}}}$, there exists a unique spectral family $\{d\hat{{\cal{H}}}(\lambda)\}$, and the spectral decomposition is\begin{equation}
\hat{{\cal{H}}}=\int \lambda d\hat{{\cal{H}}}(\lambda)\;.\label{eq:sch2}
\end{equation}
From the elementary property of the spectral components $\hat{{\cal{H}}}(\lambda)$ in Eq.(\ref{eq:sch2}),
\begin{equation}
\hat{{\cal{H}}}({\lambda_1})\hat{{\cal{H}}}({\lambda_2})=\delta_{\lambda_1\lambda_2}\hat{{\cal{H}}}({\lambda_1})\;,
\end{equation}
it follows that,
\begin{equation}
\hat{{\cal{H}}}^2=\int \lambda^2d\hat{{\cal{H}}}(\lambda)\;,
\end{equation}
and the time development in Eq.(\ref{eq:Est}) satisfies the properties of a contraction semigroup in the parameter $t$.

The degree of freedom of collapses of the superposition of wave functions \begin{equation}\psi=\sum_\lambda c_\lambda \psi^\lambda\;,\label{eq:Super3}\end{equation} is the spectral component $\hat{{\cal{H}}}(\lambda)$. In the superposition of Eq.(\ref{eq:Super3}), each component $\psi^\lambda$ is distinguished from the others by the spectral components $\hat{{\cal{H}}}(\lambda)$ such that \begin{equation}{\mbox{if}}\ \ \hat{{\cal{H}}}(\lambda)\psi^{\lambda_1}\neq0\;,\ \ {\mbox{then}}\ \ \hat{{\cal{H}}}(\lambda)\psi^{\lambda_2}=0\;,\end{equation} for $\lambda_1\neq \lambda_2$ and the state space ${V}$ of the system. Concretely, the spectral component $\hat{{\cal{H}}}(\lambda)$ is defined by the restriction of $\hat{{\cal{H}}}$ on the part which lies within the eigenspace $V_\lambda$ for eigenvalue $\lambda$,\begin{equation}\hat{{\cal{H}}}(\lambda)=\hat{{\cal{H}}}|_{V_\lambda}\;,\ \ V=\bigoplus_\lambda V_\lambda\;,\end{equation} which induces a non-unitary action on the wave function within the non-zero quantum variance of the increment of time as an operator of the contraction semigroup in the time evolution.
 We identify the cause of the state reduction with this non-unitary action on the wave function. 
 
 As an easy but a very important remark, due to Eq.(\ref{eq:Est}), the quantum superposition retention time tends to zero for the macroscopic objects.
 
 This Penrose thesis will be reformulated statistically in 
       Chapter 5 
%Section 5
 using hidden time variable.
 \section{Positions in Our Theory}
 The two notions considered in this 
       chapter,
 %  section,
that is, the Wheeler-De Witt equation, especially in its consequent TRpS, and the Penrose thesis on state reductions, play central roles in the description of the quantum mechanical world and their roles naturally intersect at a fundamental level. Here, we briefly indicate the positions of these ideas in our theory, which will be constructed in following 
chapters.
%sections.
As explained in the last
section
%subsection
shortly and in 
Chapter 3,
%Section 3,
 the Penrose thesis can explain the conscious activities of observers, in particular the completion of measurement processes.
In 
Chapter 4,
%Section 4,
 we reinterpret the role of the Penrose thesis on the conscious activities of observers in terms of the Goldstone theorem for the spontaneous breakdown of TRpS in the quantum mechanical world as seen by the observer and governed by the Wheeler-De Witt equation,
  which is caused by the classical and quantum mechanical self-identities of the observer. In 
      Chapter 5,
 %Section 5,
  we reformulate the Penrose thesis itself in the gauge theory of affinized S-duality symmetry as a statistical model of time increments via an infinite number of hidden parameters and the hidden time variable of a linear combination of the Virasoro operators. Here, we identify the Kugo-Ojima physical state condition\cite{KO1,KO2} of this gauge theory with the 
  free part\footnote{
  In the following, we refer to the {\it{free part}} of the Hamiltonian or the Wheeler-De Witt equation in the asymptotic field description of the canonical momenta.} of the
  Wheeler-De Witt equation in type IIB string theory based on the HKK's result. Due to this identification, we call the BRST charge in this gauge theory the Hamiltonian, and the Penrose thesis is reformulated using this Hamiltonian.

\chapter{A Model of Quantum Mechanical Observers}
In this 
chapter,
%section,
 we model human brain-type quantum mechanical observers based on the Penrose thesis.\cite{Konishi}
The quantum properties of observers originate in the existence of the macroscopically coherent quantum ground state and the machinery of the reduction of the classical mechanical dynamical degrees of freedom.
First,
we explain the quantum ground state. 
Second, we explain the neural-glial network model and show the criteria of its reduction of the classical mechanical dynamical degrees of freedom. 
Finally, we define what an observer is and explain the observer's ability to complete measurement.

\section{A Model of the Human Brain}
The fundamental ingredients of the neural system in the human brain are {\it{neurons}} (i.e., neural cells) and {\it{synapses}}.\cite{Kandel} A synapse is a junctional structure between two neurons, and its activity is controlled by electric or chemical signals. Their fundamental activities are described in Appendix A.1.
\subsection{The Superradiative Circuit and Its Ground State}
The main physical subject of our model of observers is a globally coherent quantum field theoretical {\it{non-perturbative}} ground state in the human brain, which describes as {\it{classical fields}} a macroscopically steady electro-magnetic field and the macroscopic electric dipole field of water molecules (i.e., a field describing a macroscopic number of the electric dipoles that have a coherent direction and strength) caused by ferroelectric and hydrophilic materials such as cell membranes and the {\it{dendrites}} (i.e., the branched projections) of neurons.\footnote{This description is due to {{a condensation of quasi-particles induced by the boson transformation}}.\cite{Umezawa}} This ground state has the quantum coherence consistent with the coherent activities of assemblies of neuronal cells\cite{FL,Freeman1,Freeman2} and describes a network of Josephson currents over the whole of the brain where the quantum coherent regions are bridged by microtubules, which are thought to act like dielectric waveguides for photons\cite{Hameroff}, and other cytoskeletal structures as a superradiative circuit through quantum non-coherent regions.\cite{JHHPY,JPY} From this ground state, for $n(\sim 10^{11})$ neurons\cite{Kandel}, we express the excitatory or inhibitory neural states, and their junctions proportional to the expectation values of the polarization currents of the ions between post- and pre-synaptic site junctions, which couple to neural states in the Hopfield type Hamiltonian\cite{H1,H2}.\cite{Konishi}

First, we explain a few details of these statements.\cite{Konishi} The spatial localization of this ground state is the set of normalized quantum field states $|{{C}}_i\rangle$ of the quanta of the electric dipole fields of water molecules, photons and damped Goldstone bosons of the spontaneously broken rotational symmetry of the dipole field as the result of the interactions between dipole field quanta and radiative photons\cite{JPY} in the perimembranous region of the synaptic site of the $i$-th neuron $C_i$, $i=1,2,\ldots,n$,
described in the interaction Hamiltonian as linear couplings of the spin variables of dipole field quanta to the creation and annihilation operators of radiative photons,
 and in the Schr$\ddot{{\rm{o}}}$dinger picture, this ground state has the structure\footnote{This non-perturbative ground state is treated by thermo field dynamics, since the system is open and dissipative (see 
   Chapter 4).\cite{Konishi,Umezawa,U2,Vitiello}}
 %Section 4).\cite{Konishi,Umezawa,U2,Vitiello}}
\begin{equation}|0(\beta)\rangle= \phi_i|{{C}}_i\rangle\ \ {\mb{in}}\ \ C_i\;,\ \ i=1,2,\ldots,n\;,\label{eq:state}\end{equation}
for real-valued coefficients ${{\phi}}_i$ and the body temperature of the system $T=1/(k_B\beta)$. The phase factor of $\phi_i|C_i\rangle$ is absorbed into $|C_i\rangle$.
(There is an ambiguity of a factor $\pm1$ in the phase. This factor plays a very important role in our model.) This ground state $|0(\beta)\rangle$ stores memories as its order parameters in the spontaneous symmetry breakdown. It is the statement of the Ricciardi-Umezawa theory.\cite{Vitiello,RU1,RU2,RU3} Each quantum state $|C_i\rangle$ is macroscopically coherent due to the general collective behavior of the dipole field quanta of water molecules as a laser and the spontaneous breakdown of the rotational symmetry of the ground state of quantum dipole fields around ferroelectric and hydrophilic materials such as cell membranes and dendrites of neurons.\cite{JPY,water}

 Actually, from the exhaustive study by Jibu, Pribram and Yasue (JPY) of the dynamics of water molecules in the perimembranous region of the neurons\cite{JPY}, which is based on the papers by Fr${\ddot{{\rm{o}}}}$hlich and the early papers about the role of dipole wave quanta in living matter\cite{Froehlich1,Froehlich2,GDMV1,GDMV2,GDMV3}, it has been shown theoretically that Bose-Einstein condensates of evanescent photons with high enough critical temperature exist. The evanescent photons are generated by absorbing the Goldstone mode into the longitudinal mode of the radiation field as in the Higgs mechanism.\cite{JPY,GDMV1} Moreover, at each synaptic site the Josephson currents, that produces a quantum tunneling between two such macroscopically quantum coherent regions of dendrites, whose coherence lengths are about $50$ $\mu$m, separated by a thin enough quantum incoherent region, are theoretically predicted.\cite{JPY} Between the pre- and postsynaptic cells is a gap about 20 nm wide\cite{Kandel}, and the unit of the dendritic net falls within these quantum coherence lengths. Consequently, in the brain a global tunneling circuit exists.\cite{JPY} (Its quantum coherent regions are bridged by microtubules and other cytoskeletal structures as a superradiative circuit through quantum non-coherent regions.\cite{JHHPY,JPY}) In this sense the global nature of Eq.(\ref{eq:state}) is satisfied.\cite{GDMV1,GDMSV}

 It will be shown that in the quantum field description, the real parts of the $c$-number coefficients $\phi$ of the localized quantum states $|C\rangle$ in Eq.(\ref{eq:state}) correspond to the neural states, also denoted by $\phi$, in the semi-classical model Eq.(\ref{eq:start}). Because of the reduction of the dynamical degrees of freedom of the activities of {\it{spikes}} (i.e., active membrane potentials) by Eq.(\ref{eq:EK}) shown later, it is obvious that the informational representations of spike activities in the brain depend on the quantum field states $|{{C}}\rangle$ as well as the Shannon representation of the bits of spikes in the brain\cite{Shannon}. It should be noted that since the $\phi$, carrying information about spikes, are not superposition coefficients, the off-diagonal long-range order is insufficient for this consistency.\cite{Konishi}

\subsection{The Neural Network}
Using the ground state Eq.(\ref{eq:state}), we give a quantum field theoretical derivation of the neural network model of the spike activities in the brain.\cite{Konishi} This description depends on the explanations in Appendix A.1.

We construct the neural network Hamiltonian operator from quantum field theory. The {{neural network system}} is a non-linear electric circuit of spikes (see Appendix A.1), whose energy is given by the summation of the products of the expectation value of the electric charge density of the ions mediating spikes and the post synaptic potentials over all of the synaptic sites. (Since the quantum field description by the ground state Eq.(\ref{eq:state}) is effective only in the perimembranous regions of the synaptic sites, we pay attention to the fact that the intracellular and extracellular ions, and the water molecules in the perimembranous regions of the synaptic sites\cite{JPY}, create currents that are proportional to the post synaptic potentials in the brain.) The corresponding Hamiltonian operator of the neural network is given by the product of the electric charge operator of the ions mediating spikes, the electric synaptic resistance and the operator of the polarization current in the perimembranous regions of the synaptic sites:
\begin{equation}\hat{\ch}_{net}=(-)^\sigma\f{1}{2}\sum_\kappa {\hat{q}}({\boldsymbol{n}}\cdot  \kappa) \hat{{\cal{N}}}_\kappa\;,\label{eq:net}\end{equation}
where the electric charge operator of the ions mediating spikes is denoted by $\hat{{q}}$, ${\boldsymbol{n}}$ is the unit vector field of the transverse directions of the axons at the synaptic junctions from one neuron to another, and $\hat{{\cal{N}}}_\kappa$ is the number operator of the dipole field quanta with a mode $\kappa$. The symbol $\sigma$ takes the value $1$ in excitatory synapses and $0$ in inhibitory synapses. Here, we assume two simplifications. First, the density of the polarization dipoles reflects that of the ions mediating spikes. In Eq.(\ref{eq:net}), we simplify this reflection to be an identity relation.
Second, the expectation values of Eq.(\ref{eq:net}) reflect the number of calcium ions (${\rm{Ca}}^{2+}$) flowing in through the voltage-dependent calcium channels at the pre-synaptic sites, and reflect indirectly the synaptic inputs on the other neurons via the neurotransmitters.
In Eq.(\ref{eq:net}), we simplify these reflections to be proportionality relations. For convenience, we select the unit of electric charge and the unit of resistance to be those of the unit dipole and synaptic resistor.

In the neural network model, we define {\it{memories}} in the quantum field description by incorporating the definition in the Hopfield model, that is, its law on the plasticity of the synapses, in the sense that both of them assign the role of memories to the strengths $J$ of neuron junctions in the Hamiltonians.\cite{H1,H2} In the quantum field description, we replace the classical values of $J_{ij}$ by the time dependent expectation values of the quantum neural network Hamiltonians $\hat{{\cal{H}}}_{net}^{j\to i}$ 
\begin{equation}J_{ij}=-2\langle {{C}}^{j\to i}_j|\hat{{\cal{H}}}^{j\to i}_{net}|{{C}}^{j\to i}_i\rangle\;,\label{eq:memory}\end{equation} on the common domains $C^{j\to i}$ of two neurons $C_{i}$ and $C_j$ at the synaptic junction of $i$-th neuron where the expectation values of $\hat{{\cal{H}}}^{j\to i}_{net}$ do not vanish.  Since the variables $|C^{j\to i}\rangle$ and $\phi_i$ are independent of each other, $J_{ij}$ also is independent of $\phi_i$. We note that the definition of memories in Eq.(\ref{eq:memory}) is descended from both the Hopfield theory and the Ricciardi-Umezawa theory.\cite{Konishi}

For the neural states $\phi_i$ in Eq.(\ref{eq:state}) and the junctions $J_{ij}$ between them, with $i,j=1,2,\ldots,n$, the effective Hamiltonian of the neural network has a Hopfield form:\cite{Konishi}
\begin{equation}
\ch_{hop}=-\f{1}{2}\la \phi,J\phi\ra\;,\label{eq:Hop}
\end{equation}
where $\la A,B\ra$ denotes the inner product of $A_i$ and $B_i$ on their index $i=1,2,\ldots,n$, and in analogy with the time development in the Hopfield model, the presence or absence of a spike of the $i$-th neuron is expressed as $\phi_i>0$ or $\phi_i<0$, respectively. Due to Eq.(\ref{eq:state}), the $n$-dimensional vector $\phi_i$ and the quantum state Eq.(\ref{eq:state}) are normalized, i.e., \begin{equation}\sum_i\phi_i^2=1\;.\label{eq:norm}\end{equation}This condition is equivalent to the normalization of the ground state $|0(\beta)\rangle$.

When a set of neurons ${\cal{S}}$ is given, the time developments of the phases $\vartheta_i[k+1]$ of the neural states $\phi_i[k+1]$, whose absolute value is defined by Eq.(\ref{eq:state}), are ruled to be given by the recursion equations:
\begin{equation}\phi_i[k]=e^{i\vartheta_i[k]}\hat{\phi}_i[k]\;,\ \ \vartheta_i[k]\in\{0,\pi\}\;,\ \ \hat{\phi}_i[k]\in{\bs{R}}_{\ge0}\;,
\label{eq:brainwave}\end{equation}with
\begin{eqnarray}e^{i\vartheta_i[k+1]}={\mb{sign}}\Biggl(-\sum_jE_{ij}e^{i\vartheta_j[k]}\Biggr)={\mb{sign}}\Biggl(\sum_j J_{ij}\phi_j[k]\Biggr)\;,\label{eq:ev}\end{eqnarray}due to actually $\hat{\phi}>0$, where $[k]$ with $k=1,2,\ldots,N-1$ represents the temporal steps and $\phi[1]=\phi$ and $E_{ij}$ is the synaptic part of the potential energy of the synaptic junction between two neurons $C_i$ and $C_j$\begin{equation}
E_{ij}=-\f{1}{2}\hat{\phi}_i\hat{\phi}_jJ_{ij}\;.
\end{equation} In Eq.(\ref{eq:ev}), the corresponding threshold potential energy value to produce a spike at the post-synaptic site of $i$-th neuron is given by $-\sum_jE_{ij}$ and may be time dependent. Here, we recall that the number of emitted neurotransmitters is proportional to the concentration of calcium ions in the pre-synaptic site. Since the calcium channel is voltage dependent (here, we must not confuse the active membrane potential with the potential of the synaptic current),\cite{Kandel} we can simplify the model so that the concentrations of calcium ions are common between all pre-synaptic sites of a neuron without losing the physical essence of the model. Due to Eq.(\ref{eq:ev}), we can treat $\phi$ as the dynamical variables.
In Eq.(\ref{eq:brainwave}), $e^{i\vartheta_i}$ expresses whether there is a spike or not by $+1$ or $-1$ respectively and $\hat{\phi}_i$ describes the semi-classical behavior of 
the ions mediating spikes and gives the expectation value of their electric charge density.
We define the vector of the temporal set of neural states
\begin{equation}
(\varphi[k])_i={{\phi}}_i[k]\;,\ k=1,2,\ldots,N\;,
\end{equation} for a total number of temporal steps $N$.\cite{Konishi}

 Regarding the phase factor of the neural states $e^{i\vartheta}$, we assume the following homogeneity criterion on the neural state dynamics.\cite{Konishi}
 \begin{criterion}The $n$ distinct configurations of the sites of neuron $i$ can be represented by the signs of their 
 states in the temporal set.
  Namely, \begin{equation} {\cal{S}}\simeq\{ {\rm{sign}}(\varphi_i[k])|\ i=1,2,\ldots,n\;,\ k=1,2,\ldots,N\}\;,\label{eq:hom}\end{equation}holds.\end{criterion}
 In the neural network (not in the neural-glial network) it may not hold exactly but only in a weaker form. This criterion first requires that in any pair of neurons its elements have different time developments of $e^{i\vartheta}$. Besides this condition, this criterion requires the local periodicity condition on neural dynamics. This criterion will be generalized for generalized neural states (see Eq.(\ref{eq:general})) and its second condition will be expressed as the temporally {\it{local}} integrability of the neural-glial system.\footnote{Here, the term `local' is used to mean that the `conserved' quantities are defined in a temporally local way and are, from a global perspective, temporally variable in general.}
 Thus, by considering the glia's physiological functions that are explained in Appendix A.2, the second condition of this criterion is natural in the neural-glial system. Throughout this 
  chapter,
 %section,
  to simplify our arguments, we assume that the dynamics of the system considered is constrained to satisfy this criterion exactly by initial conditions. Practically, we assume the first condition of this criterion.

 Due to this assumption, the site information of neurons is coded in an $N$-dimensional vector of signs of neuron states. The total number of steps is \begin{equation}N=\lfloor\log_2n\rfloor\;,\label{eq:N}\end{equation}where $n=\dim\phi+1$ and Gauss' symbol is defined by \begin{equation}x-1<\lfloor x\rfloor\le x\;,\ \ \lfloor x\rfloor \in{\boldsymbol{Z}}\;.\end{equation}
We will revisit this criterion after we take into account the glial degrees of freedom.

\subsection{Glial Modulations and Gauge Symmetries}

Next, based on arguments about glial modulation of the neural network given in Appendix A.2, we incorporate the glial network into the neural network Eq.(\ref{eq:Hop}).\cite{Konishi} For reasons that will be explained soon, we assume that the variable of the glial action ${\cal{G}}$ on the neural state $\varphi$ takes its value in the orthogonal Lie algebra $o(N)$.

By keeping in mind {{the temporal decrease of the value of the Hamiltonian}}, under the two-fold structure with Eq.(\ref{eq:Hop}), we define the Hamiltonian of the neural-glial network written using the bilinear form of $\varphi$ to be\cite{Konishi}
\begin{equation}\ch=-\f{1}{2N}\la\la\varphi,\exp(\Delta) \varphi\ra\ra\;,\label{eq:start}\end{equation}
 where $\la\la A,B\ra\ra$ denotes the inner products of $A_i[k]$ and $B_i[k]$ by contracting on both $i=1,2,\ldots,n$ and $k=1,2,\ldots,N-1$. We have also introduced the {\it{temporal}} covariant difference 
\begin{equation}\Delta\varphi=\delta\varphi+{\cal{G}}\varphi\;,\label{eq:gliaren}\end{equation}
with 
\begin{equation} \delta\varphi[k]=\varphi[{k+1}]-\varphi[{k}]\;.\end{equation}
 By analogy with the dynamics of the Ising model at zero temperature, we find from Eq.(\ref{eq:start}) a recursion equation for the generalized neural state vector $\hat{\varphi}$ of the neural-glial system, which no longer satisfies the normalization condition,
 \begin{equation}
\hat{\varphi}[1]=\varphi[1]\;,\ \ \hat{\varphi}[k+1]=\exp(\Delta)\hat{\varphi}[k]\;,\ \ k=1,2,\ldots,N-1\;.\label{eq:general}
 \end{equation}
When we normalize $\hat{\varphi}$, it is interpreted as the same expression of the neural state by $\varphi$ and will be used to classify the non-linearities of the neural-glial system.
  The glial variable ${\cal{G}}$, whose full condition is \begin{equation}{\cal{G}}_{kl}\in o(N)\;,\label{eq:gliadef}\end{equation} has indices $k$ and $l$ representing time values. We assume that the span $I_0$ represents a periodic pattern to retain the homogeneity criterion. Namely, the summation $\sum_l{\cal{G}}_{kl}\varphi[l]$ over $l=1,\ldots,N$ for a resulting index $k$, which is actually $k^\prime+pN$ for a natural number $k^\prime \in (I_0/t_0)$, time interval of spikes $t_0$ (see Appendix A.2), and a natural number $p$, means the summation of past elements over the time range $t_0\times [k^\prime+(p-1)N+1,k^\prime+pN]_{\bs{N}}$. We note that Eqs. (\ref{eq:Hop}) and (\ref{eq:start}) represent different physical systems. The latter system is larger than the former system by the number of degrees of freedom of the astrocytes.\cite{Konishi}

The glia's activity is originally defined in the infinite time span and irrelevantly to the periodic time span $N$ of the neural states. However, due to Eq.(\ref{eq:main}), it is related to $N$ and the constants of motion associated to the basis of $o(N)$ mean that ${\cal{G}}$ is the unit of the non-dense periodic pattern in the $o(\infty)$ matrix.\cite{Konishi}

 We remark on the relation of Eq.(\ref{eq:start}) to the neural network Eq.(\ref{eq:Hop}). For the unitary or non-unitary time promotion operator $\hat{U}$ of the states in Eq.(\ref{eq:state}), using the relation between the orthogonal matrices with the sizes $n$ and $N$,
  the replacement 
  \begin{equation}\delta\to \Delta\;,\label{eq:LTP0}\end{equation} in Eq.(\ref{eq:gliaren}) is recognized as a temporal redefinition of $\hat{U}$ and $e^{i\vartheta}$ that absorbs the new degrees of freedom of the astrocytes. We note that, due to the rule Eq.(\ref{eq:ev}), this replacement reflects the glia's function on the modulation of synaptic transmission and $\exp(\delta)$ is sufficient to represent the coupling of neural states with the $J$ matrix.\cite{Konishi}

Owing to the homogeneity criterion Eq.(\ref{eq:hom}), our model Eq.(\ref{eq:start}) has {{local}} $O(N)$ symmetry transformations ${\cal{O}}_r$, whose spatial variables are defined to be the signs of the $N$-vector $\hat{\varphi}$, on the index $k$ (the temporal steps):\begin{eqnarray}\varphi\to {{\cal{O}}}_r\varphi\;,\ {\cal{G}}\to {\cal{O}}_r{\cal{G}}{\cal{O}}_r^{t}-(\delta {\cal{O}}_r) {\cal{O}}_r^t\;,\  {{\cal{O}}}_r\in O(N)\;,\label{eq:sym}\end{eqnarray}
where $\delta \co_r$ is the variation of the dependence of $\co_r$ on the site of neuron by the variation of the neuron site $\delta \varphi$ defined by Eq.(\ref{eq:hom}). The generators of this gauge symmetry are related to the constants of motion in the glial modulation of the synaptic junctions. Due to the local property of the symmetry, the glial variable ${\cal{G}}$ is recognized as a gauge field.\cite{Konishi}

This gauge symmetry 
 means that, due to the modulation of synaptic junctions, the distinction between the neurons corresponding to the variable sites of the local symmetry transformations $\co_r$ by the time process labeled by the index $[k]$ loses its validity.\cite{Konishi}

The reasons why we model the glial gauge group to be $O(N)$ and construct the symmetry transformations in Eq.(\ref{eq:sym}) are as follows.
 Due to Eq.(\ref{eq:norm}), the neural states $\phi[k]$ are $O(n)$ vectors.
Then, the $O(n)$ global transformation $\co$ on the neural state is also a transformation of the Hamiltonian in Eq.(\ref{eq:start}) for an $O(N)$ matrix $({{\cal{O}}}_r)_i$ with the local index $i=1,2,\ldots,n$, which corresponds to the site of $i$-th neuron, such that the $O(N)$ vector part 
(temporal part) of the variables $\varphi$ satisfies the identity\begin{equation}{\cal{O}}\varphi={{\cal{O}}}_r\varphi\;.\label{eq:vee}\end{equation}
This equation has at least one solution ${{\cal{O}}}_r$ except for the $N=1$ case, since against $n$ equations there are $\frac{1}{2}n{N(N-1)}$ degrees of freedom. The solutions of this equation are distinct if the vectors $\varphi[k]$, for $k=1,2,\ldots,N$, are linearly independent.\cite{Konishi}

It should be noted that the idea of gauging the synaptic connection in neural network models was initially proposed and investigated by two papers\cite{Matsui1,Matsui2}. However, we note that the present context for the gauge symmetry, which was proposed by the author\cite{Konishi}, is independent of theirs and the originality of his approach is 
that he incorporated the glial network into the neural network model Eq.(\ref{eq:Hop}) based on the recently discovered roles and activities of the astrocytes.\cite{Glia1,Glia21,Glia22,Glia23,Glia24}

\section{Properties of the Model}
\subsection{Reduction of Dynamical Degrees of Freedom}

Based on the discussion so far, we introduce the information entropy of neural-glial dynamics and show the criterion of the reduction of dynamical degrees of freedom in the neural-glial networks.\cite{Konishi} (Here, the extension is done by the replacement of $\varphi$ in Eq.(\ref{eq:hom}) by $\hat{\varphi}$ in Eq.(\ref{eq:general}).)
In the neural network model, the spike of the neuron has been considered as a bit of information, and by itself we can consider a closed information structure. This is seen in the Boltzmann machine-type neural network model\cite{Boltzmann}, which is an extension of the Hopfield-type neural network model incorporating stochastic processes. However, when we take into consideration the glial network, it is more natural to define the information entropy using the interaction between the neuron states ${{\phi}}_i$ and the glia state ${{{\cal{G}}}}$. We introduce the classical entropy of the neural and glial networks by\cite{Konishi}
\begin{equation}
H(t)=-\sum_{s\in I(t)}\int\int d\varphi_s d{\cal{G}}(p_{t}(s t_0)\log_2 p_{t}(st_0))\;,\label{eq:cl}
\end{equation}
where $t$ labels the time evolution of the neural states $\varphi$ by a certain threshold structure as seen in the non-linear Hopfield model, $t_0$ is the unit of time (the interval between spikes) and $s$ indexes the time value.
In Eq.(\ref{eq:cl}), we define the set $I(t)$ of $s$, which is associated with the time $t$, as
\begin{equation}
I(t)={{\boldsymbol{Z}}_{\ge0}}\cap\biggl[0,\frac{t}{t_0}\biggr]\;.\label{eq:sequence}
\end{equation}
 The time distribution of the modulation of synaptic transmission by astrocytes is given by the probability
\begin{equation}
p_{t}(st_0)=\frac{1}{Z_{t}} \exp\biggl(\beta\frac{1}{2}\langle{{\varphi}}_s,(\exp(\Delta){{\varphi}})_s\rangle \biggr)\;,\label{eq:p}
\end{equation}
with 
\begin{eqnarray}
\sum_{s\in I(t)}\int\int d\varphi_sd{\cal{G}}(p_{t}(st_0))=1\;.\label{eq:sample}
\end{eqnarray}
The reason why we adopt $p_{t}(st_0)$ as the probability is that $\exp(\beta(1/2)\langle {{\phi}},J{{\phi}}\rangle)$ is the probability of the time evolution of the Boltzmann machine-type neural network\cite{Boltzmann}
 and it is generalized to Eq.(\ref{eq:p}) by the generalization of the synaptic junction $J$ to the glial action $\exp({{\Delta}})$.
In Eq.(\ref{eq:p}), the normalization factor $1/Z_{t}$ is the inverse of the partition function of the neural-glial system.
 The temporal index $k$ is replaced by $s$ in Eq.(\ref{eq:sequence}), and the
 time evolutions 
  are given by
   Eq.(\ref{eq:ev}).\cite{Konishi}

 We classify the non-linear behavior of the neural-glial network using the temporally local classical information entropy $H_{{{loc}}}(t)$. This local entropy is given by Eq.(\ref{eq:cl}) within a time span of $t_0N$. When
\begin{equation}
 H_{{{loc}}}( t)\sim \log_2{ t}\;,\label{eq:period}
\end{equation} the classical system is locally non-random;
 otherwise it is 
  locally random.
    Simultaneously, the extended homogeneity criterion on the neural-glial network holds.
 Here, we use a relation for almost all trajectories of time evolution with time variable $t$ and the probability $p_{t}(st_0)$ on the sample space of the neural and glial states in Eq.(\ref{eq:sample}):\cite{Kl}
 \begin{equation}
 H( t)\simeq K(t)\;,\label{eq:algo}
 \end{equation}
 where $K$ is the algorithmic complexity of a trajectory over a time $t$. $K$ is the length of the smallest program able to reproduce the trajectory on a universal classical Turing machine.\cite{Kl1,Kl2}

 For the property of being an observer, if the system satisfies the criterion given in Eq.(\ref{eq:period}) for the working hypothesis, we define an index $N$ that takes into account the time span of the brain function being considered:
  \begin{equation}N= \biggl(\frac{\Delta t}{t_0}\biggr)\times \lfloor\log_2n\rfloor\;,\end{equation} and if 
 \begin{equation}N\gg1\;,\label{eq:measurement}\end{equation} the dynamical degree of freedom for the brain function, which does not appear till we observe it during a time span $\Delta t$, is reducible by the Eguchi-Kawai large $N$ reduction.\cite{Konishi,EK} 

The statement of the Eguchi-Kawai large $N$ reduction is that, if we assume a large number of local symmetry generators (of course the Eguchi-Kawai large $N$ reduction is not valid for a global symmetry) and the existence of unbroken $U(1)$ phase symmetries between the gauge fields and their Hermite conjugates (in our case this latter assumption is not necessary), then the spatial degrees of freedom can be completely removed from the partition function of the system due to the factorization properties\cite{FAC} of the loop correlation functions. We apply this statement to our model Eq.(\ref{eq:start}). We define the $O(N)$ vector part $\varphi_N$ of the $O(Nn)$ vector ${\varphi}/{\sqrt{N}}$, which is obtained by quenching the other degrees of freedom. Then, the functional integrals over the quenched neural state variable $\varphi_N$ and the glial variable ${\cal{G}}$ in the partition function are reduced, in the large $N$ limit, to matrix integrals over the $O(N)$ matrix $\Phi$ and the glial $O(N)$ matrix $\Gamma$, due to the relation tr$(\varphi^t\co \varphi)$=tr$(\co \varphi\varphi^t)$:
\begin{eqnarray}Z[\beta,J]
=\int\int {{D}}\varphi_N{{D}}{\cal{G}} e^{-\beta{\mathcal{H}}[\varphi_N,J,{\cal{G}}]}
=\int\int d\Phi d\Gamma e^{-\hat{\beta}{\rm{Tr}}{\mathcal{H}}_{mat}[\Phi,J,{\Gamma}]}\;,\label{eq:EK}\end{eqnarray}
where $\ch_{mat}$ is the Hamiltonian of the reduced matrix model corresponding to Eq.(\ref{eq:start}) and $\hat{\beta}=\beta/N$.
 This means that, for the variable $\varphi$, the number of dynamical degrees of freedom is reduced from $O(n^N)$ to $O(Nn)$: the sites of neurons change from being variables to being merely indices. The other degrees of freedom of the $O(Nn)$ vector still survive. In the Eguchi-Kawai large $N$ reduction, the value ${\hat{\beta}}$ is kept constant.\cite{EK}
 Here, we have used the thermal variable $\hat{T}=1/(k_B{\hat{\beta}})$.\cite{Konishi}

The Eguchi-Kawai large $N$ reduction in Eq.(\ref{eq:EK}) leads to the globalization of both the quantum correlations of the operators and the quantum mechanical properties, reflecting the spike activities, of the neurons at their pre-synaptic sites.\cite{Konishi}

\subsection{Human Brain as a Quantum Mechanical Observer}
 Now, we can define what an observer is.\cite{Konishi} Here, we invoke Penrose's state reduction thesis, which claims that the non-unitary processes of measurement result from the quantum variance of the increment of time $\Delta t$ due to quantum fluctuations caused by the effects of quantum gravity.\cite{Penrose} We denote the 
quantum superposition retention time of the neuron's pre-synaptic site and the brain's spatial domain $D$ (not of the superradiative circuit but of the neural network) by $\tau_{ps}$ and $\tau_{br}$, respectively.\footnote{The 
quantum superposition retention time of the pre-synaptic site $\tau_{ps}$ is estimated to be equal to that of the microtubules. The latter is calculated to be in the order of 10ms to 100ms due to the ordering of water around microtubule bundles.\cite{HHT} From this result, we confirm that the quantum theory is relevant to real conscious activities.}
 Then, due to the main result Eq.(\ref{eq:EK}) under the condition Eq.(\ref{eq:measurement}), our scheme for defining an observer is simply\cite{Konishi}
\begin{equation}
\tau_{br}\sim \tau_{ps}\neq0\;,\label{eq:Penrose}
\end{equation}
even though the volume of the domain $D$ belongs to the classical limit of the wave function of each pre-synaptic site. Eq.(\ref{eq:Penrose}) is compatible with the functions of the neural network.
We note that Eq.(\ref{eq:Penrose}) does not always imply that there is a macroscopic superposition of the brain wave functions $\psi(\beta)$. (Here, the brain wave function $\psi(\beta)$ is the coordinate representation of the ground state $|0(\beta)\ra$ in the Heisenberg picture.)
On the basis of the scheme in Eq.(\ref{eq:Penrose}), an observer would become just a quantum system in which the superposition of the wave functions is maintained during a non-zero time span as well as in the microscopic system, and in which memories are the vacuum expectation values of order parameters of its wave function.\cite{Konishi}

In the following, we briefly explain the scenario of the completion of a measurement process by an observer, which is expected from this scheme and the idealized roles of the neural network.\cite{Konishi}

We assume an objective quantum system with a superposition of $l$ wave functions, which are the eigenfunctions of an observable $\hat{\co}$ with eigenvalues $\Lambda_i$ for $i=1,2,\ldots,l$. In contrast to an ordinary quantum system, the real human brain can recognize each eigenvalue of the observable $\Lambda_i$ in the informational database of neural state configurations, denoted by $\Phi_i$ for $i=1,2,\ldots,l$. 

First, by the brain's recognition, the information about $\hat{\co}$ would be translated into the information of the bits of spikes in the neural network. This process needs to be done between the superposed wave functions of the objective quantum system and the observer, since the classical informational mediation in the brain takes too long to make the collapses of the quantum superpositions coincide. The superposition of the brain wave functions $\psi(\beta)$ corresponding to that of the objective quantum system will be generated by a unitary transformation on the brain wave function $\psi(\beta)$ via Eq.(\ref{eq:brainwave}). 

Second, the superposition of the wave functions of the objective quantum system would collapse due to the quantum variance of the time increment. Simultaneously, due to the common quantum variance of the time increment, the superposition of  the brain wave functions $\psi(\beta)$ would collapse within a wide enough time span $\tau_{br}$. (If the time span $\tau_{br}$ were vanishingly short, the coincidence of the collapses of the superpositions of the objective quantum system and the observer would be a rare occurrence.) Here, we recognize that {\it{free will}}, constrained by the probability law of the state reduction, works. Then, the neural state configuration, as a coefficient of the superposition of the brain wave functions $\psi(\beta)$, would be chosen from $\Phi_i$, $i=1,2,\ldots,l$.

 Consequently, within the observer's conscious experience, the observer and the objective quantum system would enter the same world branch. The reason why their world branches are same is that the stochastic variable is not the wave function but the time increment $\delta t$. The results of the measurement processes would be recorded in the memories $J$. Due to its expression in Eq.(\ref{eq:memory}), $J$ is determined by the foliation of the collapsed branches of the brain wave function $\psi(\beta)$. 
 \section{Summary}
 In this 
 chapter,
 %section,
 we give an example of a quantum mechanical observer by modeling the human brain. There are two structures in the human brain: the superradiative circuit and the neural-glial network. The latter has gauge symmetries. Due to the dynamical reduction of the semi-classical degrees of freedom, caused by a large number of gauge symmetries in the neural-glial system with an exponentially large number of elements as in Eq.(\ref{eq:measurement}), the retention time of superposition of the ground state wave functions does not vanish in the neural-glial network. By the combination of the informational processes and the state reduction, this model satisfies the expected properties of observers.
\chapter{The Quantum Mechanical World as Seen by Observers}
In this 
chapter,
%section,
 we clarify the meaning of the existence of an observer in the quantum physical description of nature. By taking the human brain as an example, we show that when an observer exists, time reparametrization symmetry is spontaneously broken in its surrounding quantum mechanical world, although this symmetry is unbroken in the full quantum mechanical world. This description depends on the results in 
  Chapter 3.
 %Section 3.
  The categorical formulation of the quantum mechanical world made in this 
       chapter 
 %section
 is a prototype of the one that will be made in 
      Chapter 6.
 %Section 6.
\section{Quantum Classes and Reparametrization Symmetries}
Based on the Penrose thesis on state reduction, we define the {\it{quantum classes}} of the quantum mechanical wave functions by the identification of pairs of related wave functions. These wave functions are related to each other by a certain compatible complication of the renormalizations of their physical scales, that is, the space-time scale and the scales of physical quantities (e.g., the scales of {\it{time variables}} which are the coefficients in the exponential map of the conserved charge operators of the system, such as time $t$ for the Hamiltonian $\hat{{\cal{H}}}$ and the rotation angle for angular momentum in a central force system, etc.). The renormalization commutes with the quantum mechanical time development as operations on the wave functions. That is, the complication of the renormalizations preserves the non-unitary effects of the quantum variance of the time increment of the wave function when its Hilbert space is transformed. Importantly, since in the Penrose thesis the stochastic variable is not the wave function but time increment, the time developments of the wave functions can transition between the different Hilbert spaces within the common quantum class without losing their physical meaning. Consequently, the wave function pairs are classified by the superposition retention time, and once we define such classes, we can distinguish the quantum mechanically trivial and nontrivial states by this superposition retention time.\cite{Konishi3}

 Here we make three observations about non-unitary processes.\cite{Konishi3} First, we note that discussing non-unitary processes using the notion of quantum classes is meaningful only when each quantum class being considered is related to an open quantum system. In our central arguments of this 
       chapter,
 %section,
  we want to discuss the case where the quantum classes are the brain wave functions. Fortunately, human brains are open quantum systems and their dynamics are, in the context of the Ricciardi-Umezawa theory of the quantum brain\cite{RU1,RU2,RU3}, characterized by dissipation.
  The dissipative quantum model of the human brain was theoretically founded by Vitiello and has recently received experimental support.\cite{FL,Vitiello} 
  The whole construction of this 
              chapter
  %section
   is based on this openness of the brain quantum systems. Second, the non-unitary processes treated in this 
              chapter
  %section 
   are those not in quantum mechanics but in quantum field theory; in the latter the decoherence phenomena are known to be less harmful than in the former. Third, regarding the non-unitary processes of measurement, they may be triggers of symmetry breaking in the system being considered, and place the system in a specific state space unitarily inequivalent to other state spaces.\cite{Cel,Bla}

Here, for convenience, we introduce the concept of the {{category}} of the quantum classes, denoted by $C$.\cite{Konishi3} In mathematics, a {\it{category}} consists of a set or class of objects and the morphisms between each pair of objects, which include the identity map for pairs of the same object and have a composition structure with associativity.\cite{Category} The category $C$ of the quantum classes is mathematically defined as follows. First, its objects are the sequences of the non-unitary temporal developments of the spaces of temporally varying quantum classes of wave functions. Second, the morphisms between objects are defined from the restrictions between the Hilbert spaces of wave functions of objects, which are unitarily inequivalent spaces. Third, the compositions of morphisms are the compositions of the transformations. From its definition, this category is time dependent.
This category will be reformulated as a derived category in 
Chapter 6
%Section 6
 based on the hidden time variable theory developed in 
       Chapter 5.
%Section 5.
There, the morphisms will be treated to have the spatial degrees of freedom.

Next, we introduce the notions of scale and time reparametrization symmetries in the quantum mechanical world.\cite{Konishi3} To simplify the argument, we consider integrable systems only. First, the above definition of quantum classes is scale-intrinsic. Thus, in an obvious argument, it is not affected by arbitrary spatial scale reparametrizations as renormalization group-like changes of the time variables $x_a$ of conserved charge operators $Q_a$ of spatial symmetries:
\begin{equation}
x_a\longrightarrow x_a^\prime\ \ {\mbox{with}}\ \ x_a\equiv f_a(x^\prime)\;,
\end{equation}
where the index $a$ runs over all of the time variables. 
Second, the wavefunction of the Universe is a solution of the Wheeler-De Witt equation that is the result of the canonical quantization of gravity (to quantize we take the spatial-spatial parts of the space-time metric as variables) and matter, in the ADM decomposition of the space-time metric,\cite{ADM} applied for the null classical Hamiltonian constraint.\cite{KT} Thus, the dynamics of the wavefunction has a symmetry under arbitrary monotonic and differentiable time reparametrizations:
\begin{equation}
t\longrightarrow t^\prime\ \ {\mbox{with}}\ \ t\equiv f_t(t^\prime)\;.
\end{equation}
The scale reparametrization invariance of quantum classes and this time reparametrization symmetry mean that the quantum mechanical world does not depend on the choices for the parametrizations of time and scale variables. The time reparametrization invariance requires the absence of a Newtonian external time.\footnote{In the canonical theory of quantum gravity, since we do not separate the observing system corresponding to the coordinate frame and the observed objects, two arbitrary states linked by a diffeomorphism are equivalent to each other.}

On the other hand, when an observer uses its classical mechanical self-identity, which avoids the quantum gravity effects of time, to fix the temporal lapse function in the ADM decomposition of the space-time metric to a particular one, the time reparametrization symmetry is spontaneously broken.\cite{Konishi3} Indeed, even when we write down the matter Schr${\ddot{{\rm{o}}}}$dinger equations, we already assume the spontaneous breakdown of the time reparametrization symmetry. In the following, we consider the spontaneously broken phase of the time reparametrization symmetry, whose existence will be shown in the next 
section.
%subsection.
 In this phase, a time parametrization under the spontaneously broken time reparametrization symmetry is the sum of the vacuum expectation value $\langle t\rangle$ and the Goldstone mode $\tilde{t}^G$:
 \begin{equation}t=\langle t\rangle +\tilde{t}^G\;,\ \ \langle \tilde{t}^G\rangle=0\;,\label{eq:t1}\end{equation}
where the vacuum expectation value and the Goldstone mode of time are defined by those of the temporal lapse function. (The temporal lapse function that is in the temporal-temporal part of the space-time metric\cite{ADM} is not a parameter like time but a physical quantity, and plays the role of a dynamical order parameter of the time reparametrization symmetry.)
On the other hand, the external time increment is the sum of the mean time increment $\widehat{\delta t}=\mu$ and the quantum gravitational fluctuation $\widetilde{\delta t}^Q$, treated as a normal stochastic variable:
\begin{equation}\delta t=\widehat{\delta t}+\widetilde{\delta t}^Q\;,\ \ \widehat{\widetilde{\delta t}^Q}=0\;.\label{eq:t2}\end{equation}
As will be explained, $\mu$ is not constant in time when the time reparametrization symmetry is unbroken.
Then, comparing Eqs.(\ref{eq:t1}) and (\ref{eq:t2}) we make the main statement of this
chapter:\cite{Konishi3}
%section:\cite{Konishi3}
\begin{equation}\delta\tilde{t}^G=\widetilde{\delta t}^Q|_{\mu=\mu_0}\;,\label{eq:prop}\end{equation}
where we fix the mean time increment to be a constant $\mu_0$.
Namely, we state that the degree of freedom of the quantum gravity effects of the time increment, from which non-unitary time processes on wave functions follow in the Penrose thesis, originates in the Goldstone mode $\tilde{t}^G$ of the spontaneously broken time reparametrization symmetry. The vacuum expectation value $\langle t\rangle$ and the Goldstone mode $\tilde{t}^G$ of a time parametrization cause unitary and non-unitary time developments, respectively, in the corresponding system. As will be explained in the next 
section,
%subsection,
 the spontaneous breakdown of the time reparametrization symmetry is due to the fact that, though a quantum mechanical observer is described by a macroscopic quantum state, it retains the classical mechanical self-identity\cite{Konishi}, and is produced or recovered by the birth or death process respectively (see Appendix B).\cite{Konishi3}
The scale of the quantum variance of the time increment is 
determined
 by that of the vacuum expectation value of time in TRpS breaking.

Now, we have the following perspective grounded on the above arguments:  {{The scale and time structures of the world, which we recognize by using our own scales of time variables and clocks (see Eq.(\ref{eq:unbroken2})),
 depend on and are formed via our own broken scale and time reparametrization symmetries.}} {{In particular, when the scale and time reparametrization symmetries of a quantum mechanical observer are broken or unbroken, the scale and time reparametrization symmetries for its perceptible surrounding world are also broken or unbroken, respectively.}} The {\it{surrounding world}} indicates its classical mechanical time development by the constant mean time increment $\widehat{\delta t}$. So, a quantum mechanical system cannot always have its own quantum mechanical world in the above sense.\cite{Konishi3}

\section{Concepts of Observers}
\subsection{Self-identities in Observers}
In the Penrose thesis on the state reduction, the universal definition of a human brain-type quantum mechanical observer, which is modeled in 
Chapter 3,
%Section 3,
 is summarized to be the triple of the nonzero quantum superposition retention time $\tau$
 in the spatially global region due to the gauging control, the information entropy $H$ of neural-glial dynamics and the vacuum expectation values of order parameters $J$ of the ground state:
\begin{equation}(\tau, H,J)_{\hat{{{{\cal{H}}}}}, {{V}}}\;,\label{eq:finals}
\end{equation}
for the Hamiltonian $\hat{{\cal{H}}}$ of the system and the Hilbert space $V$ of macroscopically coherent wave functions.\cite{Konishi3} For example, with respect to a human brain, the ability to perform quantum measurement results from the combination of $\tau$ and $H$ in the case of the learning process of $H$.
For the human brain, the elements $\tau$, $H$ and $J$ represent the primitive free will, the dominant informational brain activities (e.g., the informational processes of recognition, learning and unlearning, etc.) and the dynamical memory stores, respectively. The extension of the primitive free will is due to the Eguchi-Kawai large $N$ reduction by the action of glia cells.\cite{Konishi} The memory stores $J$, that are the quantum field theoretical synaptic couplings and reproduce the corresponding non-unitary changes, have an interrelation to the informational entropy of brain activity $H$ as seen in the Hopfield model of associative memory and learning.\cite{Hopfield} This human brain-type interrelation between $H$ and $J$ is a universal property of the definition of consciousness.\cite{Konishi3}

In this 
section, 
%subsection,
based on the criterion for human brain-type consciousness given in Eq.(\ref{eq:finals}), we treat the conceptual aspects of quantum mechanical observers.\cite{Konishi3}

First, we note that in quantum mechanics, as each quantum with the same quantum numbers has no individuality and is a probability cloud due to the uncertainty principle, each quantum state with its $\tau$ also has no individuality (i.e., it is {\it{none}}). When we consider a set of such quanta, we cannot distinguish among them. So, to follow them in spatially different regions or to discuss where they belong does not make sense unless observers measure their positions, and there is no clear border line between them.
Their non-unitary dynamics are reduced on the fluctuation of the time increment, and via the concept of the fluctuation of the time increment, the only self-identical abstract notions are the quantum classes and their category.

Next, we apply this perspective to our theory. Our brains constitute macroscopically enlarged and nontrivial quantum classes. The fact that {{they have no individuality}} means that quantum recognition, which we associate with {\it{qualia}}, has a universality between brains. On the other hand, the classical mechanical notion $H$ has an individuality and is distinguishable. Namely, it depends on the classical mechanical material properties of the neural-glial network. In the Ricciardi-Umezawa theory, the memories $J$ are self-identical, in a different sense, as the vacuum expectation values of order parameters\cite{RU1}. We identify ourselves, and distinguish ourselves from others, mainly by these notions $H$ and $J$. However, we must not confuse these self-identities with the identity of the quantum class of wave functions with a particular $\tau$. Just like the relation between quantum mechanics and classical mechanics, regarding self-identity, the quantum class of wave functions with a particular $\tau$ differs essentially from these notions, and the concept of {\it{ourselves}} is an approximate classical mechanical concept.

Our consciousness as ourselves plus its none under the criterion of Eq.(\ref{eq:finals}) was derived in the birth process and accompanies the {{spontaneous breakdown of the time reparametrization symmetry}} on the category of none, that is, losing the arbitrariness in the parametrization of time. This is due to the fact that, when we {{a priori}} admit the quantum gravitational effects on the time increment (r.h.s. of Eq.(\ref{eq:prop})) and the matter Schr$\ddot{{\rm{o}}}$dinger equations for systems of observers with a formal time parameter, the count $\nu$ of derived non-unitary processes with constant mean time increment $\mu_0$ (where $\nu$ and $\mu_0$ are due to the fact that our consciousness has the quantum (none) and classical mechanical (ourselves) self-identities, respectively) plays the role of a clock, which does not allow any time reparametrization and introduces the Newtonian external time. Thus, the time reparametrization symmetry is spontaneously broken (l.h.s. of Eq.(\ref{eq:prop})). This argument is compatible with Eq.(\ref{eq:prop}). We note two points. First, of course, every quantum class with a concept of itself, not only human brain-type consciousness states, has such a clock. Second, the constant time increment can be applied to general classical mechanical systems. However, in general, due to their zero 
quantum superposition retention time, they are trivial and out of consideration in the classification of quantum mechanical objects by quantum classes and we consider only the quantum mechanical world by this classification. Furthermore, as easily noted, the birth process also accompanies the spontaneous breakdown of the scale reparametrization symmetry. These two types of broken symmetries are related to each other in Schr$\ddot{{\rm{o}}}$dinger equations but not in the Wheeler-De Witt equation.

As explained here, our conscious activities always contain an immortal, quantum mechanical and macroscopically enlarged characteristic of none in a macroscopically nontrivial quantum class of wave functions besides the living classical mechanical characteristic of ourselves, which produces this quantum class and surrounds it by a classical mechanical level potential barrier. The free will in conscious activities, defined by the fluctuating time increments, and the qualia of perceptions, defined as a high-order role of the free will in the case of the learning process of $H$, are characteristic not of ourselves but of the enlarged none. In general, the free will and the qualia survive as long as the system being considered consists of Eq.(\ref{eq:finals}), otherwise they are too faint to be detected at the classical mechanical scale, as rates of general quantum mechanical effects occur on spatial nanoscales and usually decrease exponentially, as seen in tunneling effects.

\subsection{Pure None States}
After the neural death and before the neural birth, an observer's quantum state is a pure none state, which possesses the exact time reparametrization symmetry.
The time property of pure none states is characterized by
\begin{equation}\widetilde{(\mu,{\cal{T}})}_C\;,\label{eq:unbroken}\end{equation}
where ${\cal{T}}$ denotes the 
quantum superposition retention time of the objects in the quantum mechanical world $C$ along the ensuing world branch ruled by the Penrose thesis, and $\mu=\mu(t)$ is the mean time increment, which is not constant, due to the time reparametrization symmetry. (Here, $\mu>0$.)
Namely, the variable characterizing the time property increases from ${\cal{T}}$ to $\widetilde{(\mu,{\cal{T}})}$. Here, the tilde denotes the equivalence classification under the time reparametrization symmetry. By this increment of the number of variables, the time reparametrization symmetry is retained.\cite{Konishi3}

The most important point of Eq.(\ref{eq:unbroken}) is that the time reparametrization symmetry is a gauge symmetry of time. For a gauge symmetry, under its gauge equivalence, the moduli space of symmetry variables ${\cal{M}}$, which survives the gauge equivalence, contracts to ${\cal{M}}/{\cal{G}}$, where ${\cal{G}}$ is the symmetry group. In a global symmetry, its moduli space is still ${\cal{M}}$. The gauge invariant quantity of this symmetry is the count of the non-unitary changes. From these facts, in Eq.(\ref{eq:unbroken}), the unitary time evolution between two arbitrary non-unitary changes loses its quantitative sense. In other words, the pure none state is unable to recognize these unitary time developments, which are gauge equivalent to each other.\cite{Konishi3}

We note that the same statement holds for the relic of the pure none state (i.e., the Goldstone mode of the broken time reparametrization symmetry) that carries the core role of consciousness as already explained in detail.\cite{Konishi3}

Another important point of Eq.(\ref{eq:unbroken}) is that in the surrounding world of an observer, in whose wave function time reparametrization symmetry is spontaneously broken by using their own clock,  Eq.(\ref{eq:unbroken}) is a property of the quantum mechanical world itself, which contains all of the quantum mechanical phenomena along an ensuing world branch, as described by this observer:
\begin{equation}(\mu_0,{\cal{T}})_C\;,\label{eq:unbroken2}\end{equation} where the mean time increment is fixed to a constant $\mu_0$. We remark that the spatial inclusion relation between the quantum mechanical world and the observer is irrelevant to the issue of time induced by the observer's clock in Eqs. (\ref{eq:unbroken}) and (\ref{eq:unbroken2}). We recall that to describe physical phenomena by quantum mechanics, we always assume the existence of an observer's measurement. But when we describe an observer's quantum mechanical world with Eq.(\ref{eq:unbroken2}), in which time reparametrization symmetry is broken, we usually idealize the description by ignoring them. We must not confuse Eqs.(\ref{eq:unbroken}) and (\ref{eq:unbroken2}).\cite{Konishi3}

\section{Summary}
In this 
chapter,
%section,
 we showed that in any quantum mechanical observer, there are two levels of consciousness. We call them ourselves and the {{relic}} of a pure none state (i.e., the Goldstone mode of the spontaneously broken time reparametrization symmetry around the vacuum expectation valued time parametrization, where the spontaneous breakdown of the time reparametrization symmetry of the pure none state is due to the classical and quantum mechanical self-identities of the observer and produced or recovered by the {{birth}} or death process respectively). The former corresponds to the vacuum expectation valued time increment and its causal dependence on the history of time is unitary, and the latter corresponds to the Goldstone mode of the time increment and its causal dependence on the history of time is non-unitary. Since the time reparametrization symmetry is a gauge symmetry of time, the pure none state with exact time reparametrization symmetry and the relic of it are unable to distinguish unitary time processes between two arbitrary non-unitary changes. We classified the none, that is, the wave functions by their non-unitary temporal behavior for the quantum variance of the time increment under the equivalence produced by spatial rescaling of the spatial time variables and renormalizations of the wave functions.

\chapter{A Model of Quantum Mechanical World I}
The purpose of this 
chapter
%section
 is to construct the model of a temporally statistical theory of non-unitary time processes in a quantum geometrodynamics and derive the statement of the Penrose thesis on state reductions. To do this, we consider the statistics of time increments via an infinite number of hidden parameters and the hidden time variable for a Hamiltonian-form conserved charge.
  Via this charge, the time dependence of each field variable is statistically deformed.
  In particular, we consider string theory, which has been a promising candidate for the self-consistent unifying quantum theory of the fundamental forces of Nature including gravity since its first revolution in 1984 following Green and Schwarz's celebrated discoveries.\cite{GSW,GSW2,GS1,GS2,Polchinski1,Polchinski2,BBS} Concretely, we consider a reducible system of fundamental strings (F-strings) and Dirichlet-branes (D-branes) as a model of type IIB string theory. As the fundamental symmetry of this model, we consider the gauged and affinized S-duality symmetry.\cite{Konishi2}
\section{Gauged and Affinized S-Duality Symmetry in Type IIB String Theory}
To explain the reason why the gauged and affinized S-duality symmetry is fundamental, we start by describing the duality symmetries in string theory.\cite{Konishi2}

Since the second revolution of string theory around 1995, we have seen that the five traditional ten-dimensional string theories (of type I, type IIA, type IIB, heterotic $E_8\times E_8$ and heterotic $SO(32)$) produced by the first revolution can be unified in an eleven-dimensional {\it{M-theory}} which appears as the strong coupling limit of type IIA string theory with its Kaluza-Klein modes of D-particles and each string theory describes a different aspect of the same theory.\cite{WP1,WP2,WP3}

 The links between the five traditional string theories and M-theory are the {\it{string dualities}}\cite{Sen,T1,T2,r2,r3}, which are classified into two kinds. The first kind is the S-duality which relates the strong and weak coupling
 phases of the same theory or of two different theories. Type IIB string theory is an example of the former case;
on the other hand, a familiar example of the latter case is
the heterotic string theory with the $SO(32)$ gauge group which is S-dual to type I string theory with same gauge group in $D=10$.
 The second kind is the target space duality, T-duality, which is examined perturbatively. A simple example is a bosonic closed string theory whose one spatial coordinate is compactified 
on a circle with radius $R$. The 
perturbative spectrum of this theory matches with the one whose 
corresponding spatial coordinate is compactified on a circle with 
radius $1/R$. This is a consequence of the modular invariance of the partition function under the exchange of the temporal and the string-coordinate directions on F-string world sheets (i.e., the exchange of winding and unwinding strings around the circle). The T-duality translates type IIA and type IIB theories into each other in this way and shifts the dimensions of D-branes by plus and minus one. Using combinations of S-duality and T-duality with compactifications, a duality web between all of the five traditional ten-dimensional string theories and the eleven-dimensional M-theory has been conjectured.

Besides the discoveries of M-theory and its duality web, another aspect of the second revolution of string theory is the discovery of the fundamental role of D-branes.\cite{WP2} In particular, evidence has been found for the integrability of string/M-theory: there are as many symmetry charges as degrees of freedom in the theory when we consider string/M-theory as a many-body system of D-branes. The well-known matrix theories are formulations of string/M-theory based on this evidence.\cite{BFSS,IKKT,Taylor} The degrees of freedom of these theories are those in the many-body systems of D-branes (D-particles for type IIA string/M-theory and D-instantons for type IIB string theory) with the parallel and vertical Chan-Paton degrees of freedom of open F-strings attached to them.
Based on this integrability argument, we can reformulate type IIB string theory as a reducible model with gauged and affinized S-duality symmetry as its fundamental symmetry.\cite{Konishi2} This symmetry will be explained in detail as follows.

The notion of {\it{gauged S-duality}}, introduced by the author,\cite{Konishi2,K} has a representation on type IIB string theory vacua parameterized by the complex string coupling constant $\tau_s$. In type IIB supergravity, usually, S-duality symmetry is considered as a non-linear ${SL}(2,{\boldsymbol{R}})_S$ {{global}} symmetry on the Poincar\'e upper half-plane $\fh$ of the complex string coupling constant.\cite{Sen} In contrast, we take the gauge transformation on each vacuum on $\fh$ to be independent of the others and consider the non-linear ${SL}(2,{\boldsymbol{R}})_S$ local symmetry for the axion and dilaton. In string theory, F-strings and D-branes are also the physical states associated with the gauged S-duality. In type IIA string/M-theory, D-particles are the Kaluza-Klein particles of the eleventh dimension whose radius is reciprocal to the coupling constant.\cite{WP1,WP2,WP3,BFSS} Their cousins in type IIB string theory, the D-strings and fundamental open strings, form a doublet of the S-duality symmetry via a coupling to the axion and dilaton.

Here, we explain the necessity of gauging S-duality based on the ideas in Yoneya's paper on D-brane field theory.\cite{Konishi2,Yoneya0}
To do this, we consider an analogy between the duality in string field theories and the Coleman-Mandelstam duality (CM duality) in two-dimensional space-time.\cite{CM1,CM2} CM duality relates between the two-dimensional fermionic system of the massive Thirring model and the solitonic solutions (kinks) of the bosonic sine-Gordon model. Here, we consider the second quantized theory of the sine-Gordon model. In string theory, D-branes are non-trivial classical kink solutions of the supergravity approximation of the closed string field theory. In the analogy between string theory and CM duality, the closed string field theory corresponds to the sine-Gordon field.  CM duality explains the duality between open string field theory, that is, as a massive Thirring model and closed string field theory, that is, as a sine-Gordon model. When we second-quantize the D-brane system, it is recognized to be a second-quantized open string field theory. This open-closed string duality explains the duality between closed string field theory and the open string field degrees of freedom in D-brane field theory. Thus, the dualities between these three string field theories are explained. We shall identify the physical states of the gauge theory of S-duality with those of type IIB string theory. Accordingly, when we consider the second quantized type IIB string theory, we are inevitably led to the gauging of S-duality in order to exclude the artificial distinctions between perturbative and non-perturbative excitations of strings in their field theories.

Now, we consider the reducible model of type IIB string theory as a field theory of gauged and affinized S-duality on the configuration space
(to be defined later) 
by utilizing the S-duality doublet of the axion and dilaton and that of F- and D-strings in type IIB string theory. Here, the {\it{affinization}} of S-duality means the incorporation of the world sheet degrees of freedom of a perturbative string theory into an affine Lie algebra
 without the central extension, that is, a loop algebra 
 based on $sl(2,{\boldsymbol{R}})_S$. This is needed to cover all of the degrees of freedom. The field theory of gauged and affinized S-duality has the enlarged Hilbert space of the third quantized D-brane fields instead of the one of the second quantized string fields. It will be shown that our model can incorporate 
the perturbative modular symmetry
 in the weak string coupling region as the invariance of the vacua under the modular transformations of the modulus parameter.\cite{Konishi2}

 The model of the gauged and affinized S-duality has two distinct structures.

  The first structure is the gauge field theory on the configuration space, obtained by the T-duality of the gauge field theory on the Minkowski space-time container,
such that each vacuum is specified by the Kugo-Ojima physical state condition\cite{KO1,KO2} and is the temporally stable field configuration when we regard the BRST charge as a free Hamiltonian based on the HKK representation of the field variables\cite{HKK}.
At this stage, this gauge field theory is just an infinitesimal local description of the configuration space
 in type IIB string theory. Non-perturbative effects are not yet described and non-perturbative field configurations are fixed and not dynamical. So, the contents of this stage of the modeling are perturbative regarding dynamics.

 The non-perturbative description or dynamics of type IIB string theory, that is, the transition between the stable configurations, is achieved by introducing another non-linear potential that represents a gauged string field, as the second structure of the model, which can describe the configuration space
 globally. Then, we can describe non-perturbative effects, such as an infinite many body effect and the dynamics of D-branes. Our way to introduce the second gauge potential is based on the derived category structure of the state spaces generated by a fixed vacuum in the HKK representation. This derived category structure bases on Eq.(\ref{eq:time2}) and results from the perturbative string symmetry. This second structure will be studied in the next
     chapter.
 %section.

 Due to these structures, the theory studied in this paper can be considered as  a generalization of the standard non-perturbative formulation of type IIB string theory\cite{IKKT} that resolves the non-unitarity issues.
{{Incidentally, we note that $SL(2,{\bs{R}})$ has no finite dimensional unitary representation due to its non-compactness; thus, throughout this paper we consider its infinite dimensional unitary representation on the Hilbert space.}}
\section{Model Setting}
 
Our model is based on the Neveu, Schwarz and Ramond (NSR) model of type IIB string theory that contains the massless and bosonic excitations of the axion ${\chi}$, dilaton ${\Phi}$, graviton ${{{g}}}_{\mu\nu}$, 2-form Neveu-Schwarz-Neveu-Schwarz (NS-NS) and Ramond-Ramond (R-R) potentials (${B}^{(i)}_{\mu\nu}$ for $i=1,2$, respectively) and the R-R 4-form potential with its self-dual field strength. 
The effective action in the Einstein frame is\cite{jhs,hull}
 \begin{eqnarray}
S=\frac{1}{2\kappa^2}\int d^{10}x\sqrt{-{{{g}}}}
\biggl[{R}_{{{{g}}}}+\frac{1}{4}{\rm{Tr}}(\partial_\mu{{\cal{M}}}\partial^\mu{{\cal{M}}}^{-1})-\frac{1}{12}{{{\boldsymbol{H}}}}_{\mu\nu\lambda}^T{{\cal{M}}}{{{\boldsymbol{H}}}}^{\mu\nu\lambda}\biggr]\;, \label{eq:NSR}
\end{eqnarray}
where the axion-dilaton moduli matrix ${{\mathcal{M}}}$ and the vector of $H$-fields are
\begin{equation}{{\mathcal{M}}}=\left(\begin{array}{cc}{\chi}^2e^{{\Phi}}+e^{-{\Phi}}&{\chi} e^{{\Phi}}\\ {\chi} e^{{\Phi}}&e^{{\Phi}}\end{array}\right)\;,\ \ {{{{\boldsymbol{H}}}}}_{\mu\nu\lambda}=\left(\begin{array}{cc}{H}^{(1)}\\ {H}^{(2)}\end{array}\right)_{\mu\nu\lambda}\;,
\end{equation}
and ${H}^{(i)}=d{B}^{(i)}$.
We exclude the R-R 4-form potential from consideration.
The action in Eq.(\ref{eq:NSR}) is manifestly invariant under the S-duality transformations
\begin{eqnarray}{{\cal{M}}}\to\Lambda{{\cal{M}}}\Lambda^T\;,\ \ {{{\boldsymbol{H}}}}\to (\Lambda^T)^{-1}{{{\boldsymbol{H}}}}\;,\ \ 
{{{g}}}_{\mu\nu}\to {{{g}}}_{\mu\nu}\;,\label{eq:Sgroup}
\end{eqnarray}where $\Lambda\in SL(2,{\boldsymbol{R}})_S$.

In the following, we gauge and quantize the S-duality group of Eq.(\ref{eq:Sgroup}).

We regard the pair of the axion and dilaton and that of NS-NS F-strings and R-R D-strings as the gauge bosons of gauged and {{affinized}} S-duality.
(Note that in the NS-NS sector, their field strengths can be unified in a single intrinsic tensor (see Eq.(\ref{eq:Furuta})) by lifting the former by the Kronecker delta.\cite{HKK2}) Since the axion and dilaton parameterize the coset $SL(2,{\bs{R}})/SO(2)\simeq{\fh}$, the {{Chan-Paton}} {{S-duality}} gauge potentials on lifted D-instanton fields and on D-string fields, in the coordinate representation, as the connections on the underlying principle bundle, satisfy
\begin{equation}a_\mu\in d{\fh}\;,\end{equation}
for the tangent space of the Poincar\'e upper half plane $d{\fh}$ at an arbitrary point on $\fh$ and index $\mu$ of the later introduced base space coordinates $s^\mu$. 
However, as will be explained, we assume that the supersymmetry extends the coset $SL(2,{\bs{R}})/SO(2)$ to the gauge group $SL(2,{\bs{R}})$.
 Then, the number of the generators of the gauge symmetry is still three. We denote the generators of $SL(2,{\bs{R}})_S$ by $Q^i$ for $i=0,1,2$. We define the two-dimensional representations\footnote{In the following, we refer to the representations of generators merely as the {\it{generators}}.} of the gauge symmetry, $\Sigma^i$ for $i=0,1,2$, such that the provisional field variables $\psi_r$ satisfy
\begin{equation}[\psi_r,Q^i]=(\Sigma^i)_{rs}\psi_s\;,\ \ r,s=1,2\;.\label{eq:trans1}\end{equation}
 The 
  gauge transformations corresponding to Eq.(\ref{eq:trans1})
\begin{equation}{\cal{U}}(\lambda(s^\mu))=\exp\Biggl(i\sum_{i=0}^{2} \Sigma^i\lambda^i(s^\mu)\Biggr)\;,\end{equation} act on the provisional field variables $\psi$ and the gauge potential $a_\mu$ as
\begin{equation}\psi\to {\cal{U}}(\lambda(s^\mu))\psi\;,\ \ a_\mu\to {\cal{U}}(\lambda(s^\mu)) a_\mu{\cal{U}}(\lambda(s^\mu))^{-1}-\frac{1}{ig}{\cal{U}}(\lambda(s^\mu))\partial_\mu {\cal{U}}(\lambda(s^\mu))^{-1}\;,\label{eq:trans2}\end{equation} where we denote $\partial/\partial s^\mu$ by $\partial_\mu$ and $g$ is the coupling constant, which is proportional to the cube of the string slope parameter.
In type IIB string theory, the S-duality gauge group of Eq.(\ref{eq:Sgroup}) is quantized from $SL(2,{\bs{R}})_S$ to $SL(2,{\bs{Z}})_S$ by imposing Dirac's charge quantization condition on the charges $\Sigma^i\lambda^i$ for $i=0,1,2$. In this 
chapter,
%section,
 our scheme for this quantization process is, first, to build the theory for the continuous family $\lambda(s^\mu)$ (in the BRST transformation this is the ghost field); second, we restrict the coordinates $s^\mu$ to be discrete, so that the transformation operators ${\cal{U}}(\lambda(s^\mu))$ belong to the representation of $SL(2,{\bs{Z}})_S$. However, this restriction of the coordinates $s^\mu$ does not hinder the mathematical structure of our theory before the restriction and it will be sufficient only to refer to the necessity of the restriction here. When we refer to the coordinates $s^\mu$ on the base space in type IIB string theory (not in type IIB supergravity), we regard them as discrete variables.\cite{Konishi2}

In our modeling we, first, introduce the variables $t_0$ and their differentials $q_{0}={\partial}/\partial t_0$ to represent the infinitesimal transformations in the Lie algebra ${\mathfrak{g}}=s\ell(2,{\boldsymbol{R}})_S$ with the infinitesimal generators
\begin{equation}{\mathfrak{g}}=\la\Sigma^i(0);\ i=0,1,2
\ra_{{\bs{R}}}\;.\label{eq:BPS}
\end{equation} 
In addition, we introduce 
the
 time variables ${t}_n$, where $n\in{\boldsymbol{Z}}\backslash\{0\}$, for the open string charges ${{q}}_n$ associated with independent gauge symmetries of the parallel Chan-Paton factors (the degrees of freedom of the coupling edges of open strings to a 1-form on D-strings, for $n>0$, and anti-D-strings\footnote{Here, an {\it{anti-D-string}} is simply a D-string with opposite orientation.}, for $n<0$) and the vertical Chan-Paton factors (the degrees of freedom of the open string fields vertical to D-strings). {\it{The edges of open strings and vibrating open strings represent adjoint matter and gauge connections, respectively.}} Thus the gauge bosons, which couple to the string states with the Chan-Paton charges, correspond to the Chan-Paton gauge bosons and their anti-bosons with the gauge group $\bigoplus_{N\ge1}U(N)$.
{{Here, we note that the unitary group $U(N)$ has $N^2$ generators. For our purpose we focus on the numbers of interacting D-branes in their clusters in the vacua, so we simplify the situation by ignoring the internal degrees of freedom of D-branes, which are those of the bound states of D-branes and open strings, and represent the $N^2$ generators of $U(N)$ by one generator $q_n$, satisfying 
a loop algebra.}}
By introducing this infinite number of open string charges,
 we affinize the ${\mathfrak{g}}$ generators of the gauge transformation in order to create and annihilate F-strings, D-strings and anti-D-strings.\cite{Y1,Y2}
The affinization to the $\hat{\mathfrak{g}}$ algebra is given by the
loop
 algebra of the generators
\begin{equation}
[\Sigma^1(l_1),\Sigma^2(l_2)]_-
=[\Sigma^1,\Sigma^2]_-(l_1+l_2)
\;,\label{eq:affine}
\end{equation}for $l\in{\boldsymbol{Z}}$. 
In the present model, to also incorporate the Bogomol\'nyi-Prasad-Sommerfeld (BPS) supersymmetry\cite{WP2} into the $\hat{{\mathfrak{g}}}$ gauge symmetries,
 we extend the moduli space of S-duality multiplets from the cosets $SL(2,{\bs{R}})/SO(2)$ of the axion and dilaton to $SL(2,{\bs{R}})$ by introducing the dilatino included in the NSR model of type IIB string theory.\cite{Konishi2} We denote the generator belonging to the Cartan subalgebra of Eq.(\ref{eq:BPS}), which survives in the S-duality group $SL(2,{\bs{R}})_S$ under the modulo of its maximal compact subgroup $SO(2)$, between two Cartan subalgebras of Eq.(\ref{eq:BPS}) by $\Sigma^0(0)$. Then, we interpret the $l\neq0$ generators in $\hat{{\mathfrak{g}}}$ for the $i=0,1,2$ parts to be the creation- and annihilation-like operators of R-R D-strings, NS-NS F-strings and bosonized
  NS-R F-strings respectively. The $s\ell(2,{\boldsymbol{R}})_{{{{S}}}}$ symmetries on the complex string coupling constant and the $B$-field parts of NS-NS and R-R states are realized by an $SO(2,1)_{{{{S}}}}$ rotation about the third axis (i.e., NS-R states), and the BPS supersymmetry between NS-NS and bosonized NS-R states is realized by the Weyl symmetry. Here, we consider the finite unitary symmetry transformation.
 Since the R-R D-strings are the BPS saturated states,\cite{WP2} we associate them with the axis fixed except for the sign under the Weyl group action. Here, we treat the bosonization of the NS-R states in the Hilbert space of two-dimensional conformal field theory by truncating their fermionic Klein factor to adjust them to the bosonic 
loop generators.
 Thus, the BPS generators that we consider also are bosonic. In the following, we refer to these two processes merely as {\it{bosonization}}.

Throughout this
chapter,
%section,
 we denote the Lie bracket and the (anti-)commutator by $[,]$ and $[,]_-$ ($[,]_+$), respectively.
\section{BRST Field Theory on the 
Minkowski Space-time Container}
\subsection{Fundamental Ingredients}
Since the critical dimension of superstring theory is ten, we consider a ${\hat{{\mathfrak{g}}}}$-valued covariant Yang-Mills theory of ten Chan-Paton S-duality gauge potentials $a_\mu(s^\mu)$, on lifted D-instanton field ($l=0$) and D-string field ($l\neq0$) in the coordinate representation, on a ten-dimensional base space with coordinates $s^\mu$ and metric $\eta_{\mu\nu}={\mbox{diag}}(-1,+1,\cdots,+1)$.
We call the underlying principle bundle the {\it{configuration space}}. 
 {{{Here, the fiber space of this principle bundle is the corresponding infinite dimensional Lie group of ${\hat{{\mathfrak{g}}}}$, that is, the loop group.}}} We denote the temporal and spatial coordinates by $s:=s^0$ and $x^i:=s^i$ for $i=1,2,\ldots,9$ respectively. The fact that the coordinates $s^\mu$ are those of ten-dimensional Minkowski space-time container relates to the fact that, in the D-instanton model interpretation of type IIB matrix model,\cite{IKKT} the eigenvalues of matrices are distributed over a flat ten-dimensional container space. As will be explained in Section 5.4.1, this interpretation will be revised such that
 {{each set of ten  gauge potentials $a_\mu(s^\mu)$ represents its own curved space-time with coordinates $s^\mu$}} or a higher-spin field fluctuation on it.
 
 For the ${\hat{{\mathfrak{g}}}}$-valued classical fields, the gauge potential $a_\mu$, the Faddeev-Popov ghost field $c$ and the anti-ghost field $\bar{c}$ are\begin{equation}a_\mu=\sum_{i}\sum_{l\in{\boldsymbol{Z}}}{\Sigma}^i(l)a_\mu^{i,l}\;,\ c=\sum_{i}\sum_{l\in{\boldsymbol{Z}}}{\Sigma}^i(l)c^{i,l}\;,\ 
\bar{c}=\sum_{i}\sum_{l\in{\boldsymbol{Z}}}{\Sigma}^i(l)\bar{c}^{i,l}\;.\end{equation}

The Lagrangian is given by\footnote{Our gauge theory is sourceless,
because the NS-R sector is incorporated into the gauge potential.}
\begin{eqnarray}
{\cal{L}}= -\frac{1}{4}F_{\mu \nu}^{\alpha} F^{\mu\nu\ \alpha}+i\partial^{\mu}\bar{{c}}^{\alpha}D_\mu{c}^{\alpha}+\alpha\frac{\bigl(b^{\alpha}\bigr)^2}{2}
+\partial^\mu b^{\alpha}{{a}}_\mu^{\alpha}\;,\label{eq:YM}
\end{eqnarray}
where
$\mu$ is the global Lorentz index, the index $\alpha$ represents $(i,l)$, and repeated indices ($\mu$, $\nu$ and $\alpha$) are contracted. The field strength of the gauge potential is \begin{equation}-igF_{\mu\nu}=[D_\mu,D_\nu]_-\;,\end{equation} with covariant derivative \begin{equation}D_\mu\phi=\partial_\mu\phi-i{{g}}[a_\mu,\phi]\;,\end{equation} and 
the gauge is $\alpha$. 
We introduce an auxiliary field $b$ by\begin{equation}\alpha b=\partial^\mu a_\mu\;.\end{equation}
As in other formulations, the masses are taken into the string excitations.

The 
Euler-Lagrange equations of the fields are
\begin{subequations}
\begin{align}
D^\mu F_{\mu\nu}+\partial_\nu b-i{{g}}[\partial_\nu \bar{c},{c}]&=0\;,\label{eq:cgauge}\\
\partial^\mu D_\mu{c}&=0\;,\label{eq:cghost} \\
D^\mu\partial_\mu\bar{c}&=0\;,\label{eq:antighost}\end{align}
\end{subequations}
for the gauge potential $a_\nu$, the ghost field $c$ and the anti-ghost field $\bar{c}$.

For any of the classical fields, which we generically label $\phi$, we define each field component $\phi^\alpha$ so that the dependency of $\phi$ on the coordinate $s$ is given by the gauge transformation-like exponential map 
\begin{eqnarray}
 \phi(s,x)&=&U(s)\phi(0,x)\;,\nonumber\\
 U(s)&=&\exp(is\cdot ad(L))\;,\ \ L\in Vir\;,\ \ L^\dagger =L\;,\label{eq:exmap}\end{eqnarray} of the adjoint representation of a Hermitian Virasoro operator $L$, that is, a Hermitian linear combination of the non-commutative Virasoro operators $L_n$ coupling to their moduli coefficients and the coordinate $s$ under an infinite number of relations $L_n^\dagger=L_{-n}$ for $n\in{\boldsymbol{Z}}$. Here, the moduli coefficients of $L_n$ in $L$ are a kind of temporal hidden variable. A finite number of them can be detected by measurement together with the initial data of the field variables.

Here, we make an important remark about Eq.(\ref{eq:exmap}).
 In two-dimensional conformal field theory,\cite{BPZ} time, whose definition ensures the locality of the time development of fields, is the time variable of $L_0$ which is the scale generator on the complex plane. However, in the present model, we make an attempt to generalize this `time' concept in the Yang-Mills theory Eq.(\ref{eq:YM}) so that it is the time variable for a general Virasoro operator $L$, which is the provisional `Hamiltonian', by admitting non-locality of the `time' development of fields. 
  After we proceed with the canonical quantization using this provisional `time' $s$ in Eq.(\ref{eq:exmap}), we take the non-unitary factors appearing in the canonical quantization relations into the actual time increment $\delta \tau(s)$ (see Eq.(\ref{eq:time1})) for the actual Hamiltonian ${\cal{H}}$. In this sense, $s$ is the hidden time variable, 
 and the provisional `time' $s$ and 
 the provisional `Hamiltonian' $L$ are the backstage contrivances of non-unitary phenomena in our model.

  Eq.(\ref{eq:exmap}) determines the coordinate dependence of the classical field variables $\phi$.
Then, the following identities hold:
 \begin{subequations}
 \begin{align}
 \partial_\mu \phi(s,x)&=U(s)\partial_\mu\phi(0,x)\;,\label{eq:s2}\\
  [\phi_1,\phi_2](s,x)&=U(s)[\phi_1,\phi_2](0,x)\;.\label{eq:s3}
\end{align}
 \end{subequations}
Actually, Eqs.(\ref{eq:s2}) and (\ref{eq:s3}) are due to the form of $U(s)$ as an exponential map and the Jacobi identity in the Virasoro-loop algebra, respectively.

These field variables $\phi$ are real-valued only in Hermitian linear combinations of components (see Eqs.(\ref{eq:Her1}) and (\ref{eq:Her2})).

 Next, we consider the quantum regime of Eq.(\ref{eq:YM}) in the Heisenberg picture under Eq.(\ref{eq:exmap}). By fixing the gauge, we turn the field variables (i.e., the gauge potential $a_\mu$, the ghost field $c$ and the anti-ghost field $\bar{c}$) into
 operators\footnote{We will suppress the hat indicating the field operators in the equations without further notice.}
\begin{equation}\hat{\phi}(s,x)=\sum_{i}\sum_{{{l}}\in {\boldsymbol{Z}}}{\Sigma}^i({{l}})\hat{\phi}^{i,l}(s,x)
\;.\label{eq:gauge}\end{equation}

Here, the field components of $\hat{\phi}$ and their independent canonically conjugate variables
\begin{equation}
\hat{\pi}_\phi(s,x)=\sum_i\sum_{l\in {\boldsymbol{Z}}}{\Sigma}^i({{l}})\hat{\pi}_\phi^{i,l}(s,x)
\;,\ \ \pi_\phi^{\alpha}=\frac{\partial L}{\partial \bigl(\partial_0\phi^{\alpha}\bigr)}\;,\label{eq:pigauge}
\end{equation} need to satisfy the functional equal-time canonical commutation and anti-commutation relations of Eq.(\ref{eq:YM}) for the time variable $s$, up to the non-unitary factors $\sigma(s)$, which are the same between kinds of field variables:\footnote{Since the interaction terms in the canonical conjugate variables drop from the (anti-)canonical commutation relations, a field variable $\phi^\alpha(s)$ with a negative $\sigma^\alpha(s)$ means $\phi_0^\alpha(-s)$, where $\phi_0^\alpha (s)$ obeys the normal (anti-)canonical commutation relation, and for it, the sign of $\pi_\phi^\alpha(s)$ reverses.}
\begin{subequations}
\begin{align}
[\hat{\breve{a}}_i^{{\alpha}_1}(s,x),\hat{\pi}_{\breve{a}_j}^{\alpha_2}(s,x^\prime)]_-&=i\hbar\delta_{\alpha_1\alpha_2}\delta_{ij}\delta(x-x^\prime)\sigma^{\alpha_1}(s)\;,\ \ i,j=1,2,\ldots,9\;,\label{eq:CCR1}\\
[\hat{\breve{a}}_0^{{\alpha}_1}(s,x),\hat{\breve{b}}^{\alpha_2}(s,x^\prime)]_-&=i\hbar\delta_{\alpha_1\alpha_2}\delta(x-x^\prime)\sigma^{\alpha_1}(s)\;,\label{eq:CCR2}\\
[\hat{\breve{c}}^{{\alpha}_1}(s,x),\hat{\pi}_{\breve{c}}^{\alpha_2}(s,x^\prime)]_+&=[\hat{\breve{\bar{c}}}^{\alpha_1}(s,x),\hat{\pi}_{\breve{\bar{c}}}^{\alpha_2}(s,x^\prime)]_+=i\hbar\delta_{\alpha_1\alpha_2}\delta(x-x^\prime)\sigma^{\alpha_1}(s)\;,\label{eq:CCR3}
\end{align}
\end{subequations}
where we set
\begin{subequations}
\begin{align}
\hat{\breve{\phi}}^{i,0}&=\hat{\phi}^{i,0}\;,\label{eq:Her1}\\ 
\hat{\breve{\phi}}^{i,l}&=\hat{\phi}^{i,l}+\hat{\phi}^{i,-l}\;,\ \ l\in{\boldsymbol{N}}\;,\label{eq:Her2}
\end{align}
\end{subequations}
 and diagonalize the canonical (anti-)commutation relations of the field variables $\hat{\phi}$ by a unitary matrix $\breve{U}(s)$. We also transform the tensor $\delta_{m+n}^l$ appearing in the Lie brackets by the tensor product of this unitary matrix and itself. Other pairs among the field variable operators $\hat{\breve{\phi}}$ and their canonical conjugates $\hat{\pi}_{\breve{\phi}}$ are commutative or anti-commutative.
 These relations can be decomposed into those at a particular time $s=0$, which are Schr${\ddot{{\rm{o}}}}$dinger picture type relations, and the time promotion factors, which result in the factors $\sigma(s)$, via the time promotion operator $U(s)$, owing to Eqs.(\ref{eq:s2}) and (\ref{eq:s3}).
 Our canonical (anti-)commutation relations appear to be different from the conventional ones because $U(s)$ does not adjustably separate the Lie algebra components of the field variables $\phi^\alpha$. Here, we note that the (anti-)canonical commutation relations are independent of the canonical equations (or, equivalently the Euler-Lagrange equations).
In the following, we use the diagonalized field variables, but for simplicity, denote them by the same notation as the undiagonalized field variables.

We make several remarks about the non-unitary factors $\sigma(s)$.

By this definition, a good behavior is that the signs of appropriate $\sigma(s)$ can be definite within their convergence regions, since by setting $\sigma(0)$ to appropriate values, $\sigma(s)$ are proportional to $\sigma(0)$ and the squares of the norms of an arbitrary kind of field variable. However, by this definition, the $\sigma(s)$ diverge by oscillation outside their convergence regions. Moreover, in general, the field variables $\phi(s)$ also diverge by oscillation outside their convergence regions. Due to the form of the time promotion operator $U(s)$, the role of the coordinate $s$ in the field variables $\phi(s)$ is the same as that of the coupling constant in the perturbation expansion series of an S-matrix, and the divergence of the $\phi(s)$ outside their convergence regions is physically the same as that of the perturbation expansion series of an S-matrix in the strong coupling region. So, this kind of divergence is not physically essential and can be removed by analytic continuation, which preserves the local functional equations, or by more general methods\cite{Hardy,Suslov}. Besides this kind of divergence, $\sigma(s)$ have another kind of divergence coming from the infiniteness of the number of field variables. This divergence can be removed by redefining the $\sigma(s)$ as the finite norms of an arbitrary kind of field variable $\phi(s)$ in an infinite dimensional vector space.\footnote{As an illustration, for $L=L_1+L_{-1}$, the squares of the norms of the undiagonalized field variable operators $\phi(s)$ become almost $f(s)(1-(ks)^2+(ks)^4-(ks)^6+\cdots)^2$ for finite constants $k$ with $0\le (ks)^2<1$ and power series $f(s)$ of $(ks)^2$ with the coefficients $\sigma(0)$. The analytic continuation of the second factor to the regions $(ks)^2>1$ is $\Bigl(\frac{1}{1+(ks)^2}\Bigr)^2$. So, when we set $\sigma(0)$ to appropriate values, this result gives well-behaved non-unitary factors $\sigma(s)$ before the diagonalization.
However, our model is, in some aspects, like a toy model; for our model to be a realistic statistical theory of time, we might need some improvement of the time promotion operator $U(s)$.}

Next, if the Virasoro and initial data are trivial, $\sigma^{\alpha}(s)=1$ for all generators $\Sigma^\alpha$ with $l\neq0$ and unitarity for the coordinate $s$ is maintained.
If so, the equal-time canonical (anti-)commutation relations with $l=0$ are 
not well-behaved previous to this.
 As will be seen, the norms of the factors $\sigma(s)$ need to be finely tuned to be around $1$ with the width of the ratio of the scale of the time increment fluctuation to the constant mean time increment $\mu_0$.
Owing to this fine-tuning of $\sigma(s)$, we can assume that $\partial_0{\breve{U}}(s)\approx0$.
Here, we note that Eq.(\ref{eq:exmap}) expresses the non-local development of the field components for $s$ and, due to it, the Virasoro and initial data are the non-local hidden variables via their statistics in the factors $\sigma(s)$.

Besides these canonical quantization relations, there has been still the condition on the field variables $\hat{\phi}$ given by Eq.(\ref{eq:exmap}).
This condition is necessary for two reasons.
First, due to this condition, the coordinate $s$ has the meaning of the time variable of the linear combination of the Virasoro operators in Eq.(\ref{eq:exmap}). Second, this condition is needed to ensure the 
Hermiticity of the field variable operators
$\hat{\breve{\phi}}$ and the Hermiticity of the symmetry charge operator.

\subsection{Symmetry Structures}

\bigskip

\noindent\underline{BRST Symmetry.}

\bigskip

On the basis of the preliminaries so far, we consider a
 global symmetry of Eq.(\ref{eq:YM}) after the gauge fixing. The BRST symmetry is defined by the following infinitesimal global transformation with the gauge function $c$.\cite{BRST1,BRST2,BRST3}
\begin{equation}
\delta a_\mu=\epsilon D_\mu c\;,\ \ \delta c=-\epsilon\frac{1}{2}g[c,c]\;,\ \ \delta\bar{c}=i\epsilon b\;,\ \ \delta b=0\;,\label{eq:BRST}
\end{equation}
where the parameter $\epsilon$ is an anti-commuting $c$-number.
The transformation $\delta$ is nilpotent.

As is well known, this transformation 
 has a clear meaning in the geometry of the principle bundle.\cite{YM,ghost} Once the gauge degree of freedom is fixed, the gauge potential $a_\mu$ and the ghost field $c_{\alpha}$ can be regarded as contravariant components of the Ehresmann vertical connection $\nabla_v$ on the section of the principle bundle:
\begin{equation}
\nabla_v =a_\mu ds^\mu +c_{\alpha} dy^{\alpha}\;,
\end{equation}
where $y^{\alpha}$ are coordinates of the internal fiber space. For later discussion, we put $dy^l=\sum_i dy^{i,l}$. Since any system of coordinates could be used without affecting the definitions of field operators, we distinguish between the index of these coordinates and the index of the ${\hat{{\mathfrak{g}}}}$ generators (i.e., the cotangent space index).\cite{ghost}
The BRST transformation is constructed from the Maurer-Cartan equations for the curvatures of $ig\nabla_v$\cite{ghost} \begin{equation}R_{{\alpha}{\beta}}(c,s,y)=R_{{\alpha} \mu}(a,c,s,y)=0\;,\label{eq:MCeq}\end{equation}
 as
\begin{equation}\delta=d y^{{\alpha}} \partial_{{\alpha}}\;,\ \ \delta^2=0\;,\label{eq:nilpotency}\end{equation}where we put $\partial_{{\alpha}}=\partial/\partial y^{{\alpha}}$ and two $dy^{{\alpha}}$ anti-commute.

\bigskip

\noindent\underline{Kugo-Ojima Physical State Condition and Its Solutions.}

\bigskip

From the Yang-Mills theory of Eq. (\ref{eq:YM}), the charge operator $Q$ of the BRST transformation $\delta$ for the time variable $s$ is
\begin{equation} \epsilon {{Q}}= \int d^9x
\Bigl({\pi}_{\breve{a}_\mu}^{{\alpha}}(\delta \breve{a}_\mu)^{\breve{\alpha}}+{\pi}_{\breve{c}}^{{\alpha}}(\delta \breve{c})^{\breve{\alpha}}+{\pi}_{\breve{\bar{c}}}^{{\alpha}}(\delta\breve{\bar{c}})^{\breve{\alpha}}
\Bigr)
\;,\end{equation}
where the repeated indices, $\mu$ and ${\alpha}$ with $l\ge0$ are contracted. In the case of the $\alpha$ indices, the $\breve{}$ is ignored for the contraction.
Remarkably, due to the nilpotency of the BRST transformation, the {\it{normalized}} BRST charge, denoted by ${\boldsymbol{Q}}$, is invariant under the BRST transformation.\cite{KO1,KO2}

The Kugo-Ojima physical state condition\cite{KO1,KO2} on 
the wave functions
$\Psi[{{{g}}},s,\phi]$ of the coupling constant ${{{g}}}$, the coordinate $s$ and the field variables $\phi$ that we adopt in the coherent-state representation about $\phi$ is
\footnote{Later we will impose the condition that these wave functions $\Psi[g,s,\phi]$ are independent of the variable $s$: namely, $\Psi=\Psi[g,\phi]$ due to
Eq.(\ref{eq:Diff}).
 However, at present, they
  depend on $s$ and we denote them by $\Psi[g,s,\phi]$.}
\begin{equation}({\boldsymbol{Q}}+\zeta)\Psi[{{{g}}},s,\phi]=0\;,\label{eq:brst}\end{equation}
where the normalized BRST charge ${\boldsymbol{{Q}}}$ is normally ordered and we add Grassmann odd frozen Casimir $\zeta$\cite{KM2} to the normalized BRST charge ${\boldsymbol{Q}}$ owing to \begin{equation}[\zeta,\hat{\phi}]_\pm=0\;.\end{equation} For simplicity, we will denote ${\boldsymbol{Q}}+\zeta$ just by ${\boldsymbol{Q}}$.

 In the absence of $\zeta$, an illustrative example of the 
  solutions to Eq.(\ref{eq:brst}) written in the form of a theta function as an additional requirement is
\begin{subequations}
\begin{align}
\Psi&=\sum_{n\in \bz}
\sum_{\Delta}
c_{n,\Delta}\Psi_{n,\Delta
}\;,\label{eq:sol3}\\ \Psi_{n,\Delta
}&=
\sum_{p\in{\boldsymbol{Z}}_{\ge0}}\epsilon_{p,n}\Delta_{pn}
O(\hat{y})
\Biggl(\sum_{k\in{\bs{Z}}}
\exp
\biggl(\frac{i\pi{\cal{T}}}{\hbar}
 \hat{{\bs{\cal{{H}}}}}_{k\hat{\phi}}
\biggr)
\Biggr)\Delta
\psi_v\;,\label{eq:sol2}
\end{align}
\end{subequations}
 for the ground state wave function $\psi_v$, operators 
 \begin{equation} \Delta_n=\hat{\phi}_0^n\ \ (\Delta_0=id)\;,\ \ n\ge0\;,\ \ \delta\phi_0=0\;, \end{equation}
  operator ${{O}}(\hat{y})$
  and
   Grassmann numbers $\epsilon_{p,n}$ 
  that will be defined later,
   the complex modulus parameter ${\cal{T}}$ on the upper half plane,
the normalized Hamiltonian operator $\hat{{\bs{\cal{H}}}}$ of Eq.(\ref{eq:YM}) in the weak coupling region where the world sheet Hamiltonian is effective,
a functional $\Delta$ of the operators $\Delta_n$,
  and $c$-numbered coefficients  $c_{n,\Delta}$.
  We denote the operator part in $\Psi$ which acts on $\psi_v$ by $\hat{\Psi}$. 
  We restrict the Grassmann number degree of each solution to one.

 Each solution $\Psi_{n,\Delta
 }$ with $n\ge2$ in Eq.(\ref{eq:sol3}) has 
 broken symmetries. We denote the (un)broken BRST charge by ${{Q}}_{br}$.
As will be explained shortly, the 
symmetry breakdown is caused by the non-zero term $\zeta$ in Eq.(\ref{eq:brst}).

To complete the definition of Eq.(\ref{eq:sol2}), we introduce the normalized charge operators ${\boldsymbol{Q}}_n$ and their commutative dual coordinate operators $\hat{y}^n$ by 
 \begin{equation}
 {\boldsymbol{Q}}_n={\boldsymbol{Q}}|_{d{{y}}^{n}}\;,\ \ [{\boldsymbol{Q}}_m,\hat{y}^n]_-=\epsilon_m\delta_{mn}\;,\ \ [\epsilon_m,\epsilon_n]_+=0\;,\ \ m,n\in{\boldsymbol{Z}}\;,\label{eq:y}
 \end{equation}
 which satisfy an infinite number of relations
 \begin{equation}
[{\boldsymbol{Q}}_m,{\boldsymbol{Q}}_n]_+=0\;,\ \ [\Delta_{m},{\boldsymbol{Q}}^{(n)}]_-=0\;,\ \ {\boldsymbol{Q}}^{(n)}=\sum_{q\in{\bs{Z}}}{\boldsymbol{Q}}_{qn}\;.
 \end{equation}
 The operators $\hat{y}^n$ are explicitly written as 
 \begin{equation}
 \hat{y}^n=-\frac{\delta}{\delta \sqrt{{\bs{Q}}_n{\bs{Q}}_n^\star+{\bs{Q}}^\star_n{\bs{Q}}_n}}\;,
 \end{equation}
 for the normalized co-BRST charge operators ${\bs{Q}}_n^\star$.\cite{co-BRST}
 
 In Eq.(\ref{eq:sol2}), the operator ${{O}}(\hat{y})$ is defined by
 \begin{eqnarray}
  {{O}}(\hat{y})=\exp\Biggl(\sum_{n\in \bz} \alpha_{n}\hat{y}^{n} \Delta_{i_n}\Biggr)\;,\label{eq:expO}
 \end{eqnarray}
with certain constants $\alpha_{n}$. $i_n$ are non-negative integers; for example, we choose $i_n=n$ for $n\ge1$ and $i_n=-n$ for $n\le-1$. Eq.(\ref{eq:expO})
is consistent, since an infinite number of relations
\begin{equation}
[\Delta_m,\Delta_n]_-=0\;,
\end{equation}
holds.
 They satisfy an infinite number of relations
\begin{equation}[{\boldsymbol{Q}}^{(n)},{{O}}(\hat{y})]_-=\sum_{qn\in \bz}\alpha_{qn}\epsilon_{qn}\Delta_{|q|n}{{O}}(\hat{y})\;.
\label{eq:alpha}\end{equation}
Due to Eq.(\ref{eq:alpha}), we obtain \begin{eqnarray}{{\boldsymbol{Q}}^{(n)}}\epsilon_{p,n}\Delta_{pn}O(\hat{y})
 \hat{{\bs{\cal{{H}}}}}^r\Delta\psi_v=
 &&
 -\epsilon_{p,n}\Biggl(\sum_{qn\in \bz}\alpha_{qn}\epsilon_{qn}\Delta_{(p+|q|)n}O(\hat{y})
 \hat{{\bs{\cal{{H}}}}}^r\Delta
 \nonumber\\&&
 +\Delta_{pn}O(\hat{y})
 \hat{{\bs{\cal{H}}}}^r\Delta{\boldsymbol{Q}}^{(n)}\Biggr)\psi_v\;,\ \ r\in{\bs{Z}}_{\ge 0}\;,\end{eqnarray}
 where 
 we use 
\begin{equation}
[{\boldsymbol{Q}}^{(n)}, \hat{{\bs{\cal{H}}}}^r\Delta]_-=0\;,\ \ r\in{\bs{Z}}_{\ge0}\;.\label{eq:GSOres}
\end{equation} 
  ${\boldsymbol{Q}}^{(n)}\psi_v$ is zero, due to the uniqueness of the ground state wave function $\psi_v$ and the symmetry charge property of ${\boldsymbol{Q}}^{(n)}$.\cite{KO2}
This is\begin{equation}{{\boldsymbol{Q}}^{(n)}}\psi_v=0\;.\end{equation}
 The solution $\Psi_{n,\Delta
 }$, which satisfies \begin{equation}\sum_{p+|q|=N}\alpha_{qn}\epsilon_{qn}\epsilon_{p,n}=0\;,\ \ qn\in{\boldsymbol{Z}}\;,\ \ N=0,1,2,\ldots\;,\label{eq:qn}\end{equation} with $n\ge1$ can be checked by acting with ${{{{\boldsymbol{Q}}}_{br}}}$ on Eq.(\ref{eq:sol2}).

\bigskip

\noindent\underline{Symmetry Properties of the Solutions.}

\bigskip

 As in Eq.(\ref{eq:sol2}), we impose  the modular form property of the wave function $\Psi$ via the linear fractional action of a discrete modular group $\Gamma$ on the complex modulus parameter ${\cal{T}}$
  \begin{equation}\gamma{{
  {\cal{T}}}}= \frac{a{{
  {\cal{T}}}}+b}{c{{
  {\cal{T}}}}+d}\;,\ \ \gamma=\left(\begin{array}{cc}a&b\\c&d\end{array}\right)\in\Gamma\;.\end{equation} 
  The modular form property is imposed up to an operator factor $\Lambda^\gamma$, which acts on the theta function part of $\Psi$ and comes from the eigenstructure of the operator 
  \begin{equation}\hat{{\cal{B}}}=k\textstyle{\sqrt{{ \hat{{\bs{\cal{{H}}}}}/\hbar
  }}}\;.\end{equation}
   The summation over $k$ in Eq.(\ref{eq:sol2}) reflects this constraint.
   
Actually, when we assume that the symmetry is unbroken, the modular form property in the case $\gamma=\left(\begin{array}{cc}0&-1\\1&0\end{array}\right)$ can be confirmed by using the Poisson summation formula for the wave function $\Psi$ and its Fourier transform about the absolute values of the eigenvalues of $\hat{{\cal{B}}}$ performed by completing the square for the absolute value of each eigenvalue of $\hat{{\cal{B}}}$ in the exponent and using the Gauss integral formula by substitution.

The summation over $k$ in Eq.(\ref{eq:sol2}) is taken within the eight-fold root lattice of the S-duality algebra $s\ell(2,{\bs{R}})_S$.\footnote{Due to the Jacobi identity between the Neveu-Schwarz partition function and the Ramond partition function, the results of the Gliozzi, Scherk, and Olive (GSO) projection in superstring theory\cite{Modular1} allow us to introduce consistently the eight independent physical transverse directions of string vibrations and the eight-fold total degeneracy of the states at the lowest excitation level in the NSR model of type IIB string theory.} When we modularly transform the wave function to $\Psi(-1/{
{\cal{T}}})$, we retake this summation within the dual lattice. The mathematical content of Eq.(\ref{eq:sol2}) is the same as that in the theta function, which takes the form ${\mbox{tr}}(\exp(-{\cal{T}} \hat{{\cal{H}}}/\hbar))$ for the modulus parameter of F-string world sheets ${\cal{T}}$ and the Hamiltonian $\hat{{\cal{H}}}$, in the context of the toroidal compactifications. Here, the eigenvalues of the Hamiltonian $\hat{{\cal{H}}}$ are given by the summations of the squares of masses associated with the discrete momenta conjugate to the coordinates of the compactified space dimensions and 
the winding modes. In this analogy, the theta function is defined via the root lattice of the gauge algebra of Kaluza-Klein modes. From this analogy for the open F-string and D-brane system, we see that the discrete modular symmetry of Eq.(\ref{eq:sol2}) can be interpreted as a perturbative symmetry.

On the other hand, the character of the representation of the central extension of $\hat{{\mathfrak{g}}}$,
 as another procedure to quantize the same model,
 produces the algebra of the Kac-Peterson type theta functions.\cite{KP} These theta functions are defined on the affine Weyl lattice of the root system of the central extension of ${\mathfrak{g}}$ and must be distinguished from the above theta function. The complex variable of the character is non-trivially identified with the complex string coupling constant that is the S-duality variable. Thus, the discrete modular symmetry of the Kac-Peterson type theta functions is non-perturbative and is interpreted as the quantized S-duality symmetry with a discrete modular symmetry group.

In the perturbative interpretation of the modular symmetry property of $\Psi$ about ${{\cal{T}}}$, in the weak string coupling region (if necessary by taking S-duality), we interpret the complex modulus parameter ${{{\cal{T}}}}$ as that of F-string world sheets. In the following, we consider only the solutions which possess this perturbative discrete modular symmetry.

In the presence of the term $\zeta$ in Eq.(\ref{eq:brst}), the affinized symmetries are broken
 by the generators whose central extension parts are indexed by ${\boldsymbol{Z}}_{{{\boldsymbol{N}}}}=\bigoplus_N{\boldsymbol{Z}}_N$ with
\begin{equation}\sum_{i}\sum_{l_M=0}^{N_M(\lambda)-1}\cdots\sum_{l_1=0}^{N_1(\lambda)-1}\Biggl(\bigotimes_{k=1}^M
{\boldsymbol{Q}}^{i,l_k}\Biggr)\Psi=-\zeta^{\otimes M}\Psi\;,\label{eq:zeta}
\end{equation}
where we 
 put $\dim {\boldsymbol{N}}=M$.\cite{Konishi2}\footnote{Here, we apply the formula $Q_{G\times H}=Q_G\otimes 1_H  +1_G\otimes Q_H$ to the case in which $G=N{\boldsymbol{Z}}$ and $H={\boldsymbol{Z}}_N$.} This type equation holds independently for each class of degenerate transverse directions of string vibrations.
 For the unbroken parts, the quantity corresponding to the l.h.s. of Eq.(\ref{eq:zeta}) is zero.
Eq.(\ref{eq:zeta}) means that the r.h.s., which is the number of the strings with all of numbers of members of clusters possessed by $\Psi$, is equal to the l.h.s., which is the number of the $l$-interacting strings where $l$ belongs to ${\boldsymbol{Z}}_{{\boldsymbol{N}}}$.\footnote{As will be explained later, we reinterpret the Kugo-Ojima physical state condition as the Wheeler-De Witt equation in the asymptotic field description of the canonical momenta. In this reinterpretation, the BRST charge ${Q}$ plays the role of the free Hamiltonian.}
Thus the representation on $\Psi$ satisfies that in Eq.(\ref{eq:zeta}) the elements of the algebra, which is generated by the field operators corresponding to the generators of the ${\boldsymbol{Z}}_{{\boldsymbol{N}}}$ part of the affinized symmetry, multiplicatively generate the non-zero states labeled by the weights of the universal enveloping algebra of the loop algebra $\hat{{\mathfrak{g}}}$.
The physical meaning of this statement is that, owing to the existence of the Chan-Paton charges,
{{for each class of degenerate transverse directions of string vibrations, its
 low energy effective theory of type IIB string theory vacua is the Yang-Mills theory with the gauge group $\bigoplus_{N} U(N-1)$ of the Chan-Paton fields (i.e., the adjoint matter fields and the gauge fields) on the 
clusters of the interacting
 D-strings with ${\boldsymbol{N-1}}$ 
 numbers.}}

If the affinized symmetries are broken 
 as Eq.(\ref{eq:zeta}), the 
 loop part of their eight-fold root system is deformed. The quantized S-duality symmetry in the character of the equivalent representation of the central extension of ${\hat{\mathfrak{g}}}$ is therefore also broken, like as for products of the Dedekind $\eta$ functions. Secondary, the eight-fold root lattice of $s\ell(2,{\bs{R}})_S$ is also deformed and consequently the 
perturbative discrete modular symmetry, defined by using this root lattice, may be also broken.
\section{Space-time Structures}
\subsection{T-dual Representation of the Field Variables}
In this
section,
%subsection,
 we study the space-time structures of our model.
 
In the last
 section,
%subsection,
 as already pointed out, we considered the model on the Minkowski space-time container by invoking the old interpretation of the actual space-time, in type IIB matrix model, as the eigenvalue distribution of the D-instanton matrices over a flat container. This is because our model and type IIB matrix model do not directly contain the graviton modes and describe space-time points by the positions of D-branes which are solitons of closed F-strings. On the basis of this formalism, we have introduced several important quantities (e.g., $L$ and $\sigma^\alpha(s)$) which are characteristic of our model, and studied the symmetry properties of the vacua by using the BRST field theory.

However, this formalism is not Einstein's metric description of gravity and is thus not convenient for studying the space-time structure of the theory purely mathematically without relying on simulation methods. So, in the following, keeping the solutions of the equations of motion of the model, we switch to a new equivalent formalism, which is based on the interpretation of the space-time in type IIB matrix model proposed by Hanada, Kawai and Kimura\cite{HKK}. It is, in our context, a reversible mapping of the field variables $\phi(s^\mu)$, described in the last 
section,
%subsection,
 to the variables $\phi^\ast(s^\mu)$ appearing in the covariant derivative of a curved space-time ${\boldsymbol{M}}^{9,1}$ via the variables $\Phi$ of the gauge-fixed matrix model:
\begin{equation}
\phi(s^\mu)\ \; {\mbox{on}}\ \; {\boldsymbol{R}}^{9,1}\longrightarrow \Phi\ \; {\mbox{in}}\ \; \hat{{\mathfrak{g}}}\longrightarrow \phi^\ast(s^\mu)\ \; {\mbox{on}}\ \; {\boldsymbol{M}}^{9,1}\;,\label{eq:HKKRep}\end{equation} in the classical regime. Even though the two formalisms are equivalent, in general, corresponding variables $\phi(s^\mu)$ and $\phi^\ast(s^\mu)$ seem to have different physical meanings. On this point, in the HKK interpretation, we interpret matrices as covariant derivatives on a background space-time. Namely, matrices represent momenta in contrast to the old interpretation in which they represent space-time coordinates. In this sense, the new formalism can be considered as just the {{T-dual}} of the old formalism. To do the mapping in Eq.(\ref{eq:HKKRep}), we need the following preliminary mathematical result.

Hanada, Kawai and Kimura showed that there is a correspondence between space-time covariant derivatives $\nabla^\ast_{(\mu)}$ and infinite rank matrices $\Phi:=A_\mu$:
\begin{equation}
A_\mu=i\nabla^\ast_{(\mu)}\;,\ \ \nabla^\ast_{(\mu)}f(s^\mu,\gamma):=R_{(\mu)}^\nu(\gamma^{-1}) \nabla_\nu^\ast f(s^\mu,\gamma)\;,\label{eq:HKK}
\end{equation} for any smooth function $f$ on a principle $Spin(10)$ bundle on a space-time manifold ${\boldsymbol{M}}^{9,1}$ with space-time coordinates $s^\mu$ and elements $\gamma$ in $Spin(10)$. Here, $Spin(10)$ is the ten dimensional Lorentz group and $R_{(\mu)}^\nu$ is the vector representation of $Spin(10)$. In the representation matrix $R_{(\mu)}^\nu$, by redefining the base of the $Spin(10)$ representation vector space, $(\mu)$ can be chosen to be not the local Lorentz index but just a label of the redefined base of the representation vector space. 
Eq.(\ref{eq:HKK}) is consistent with coordinate patch gluing on space-time manifolds.
It was also shown that when $A_\mu$ are the variables of type IIB matrix model, the equations of motion of the HKK covariant derivatives are the vacuum Einstein field equations of general relativity.\cite{HKK}

In Eq.(\ref{eq:HKKRep}) we invoke this result in the following way.
 First, we re-express the covariant derivatives of the $\hat{{\mathfrak{g}}}$ gauge potentials $a_\mu(s^\mu)$ as matrices $A_\mu$ acting on a representation vector space different from above. Second, we re-interpret these matrices $A_\mu$ as space-time covariant derivatives $\nabla_{(\mu)}^\ast$ with coordinates $s^\mu$ by Eq.(\ref{eq:HKK}). Then, each set of ten 
gauge potentials $a_\mu(s^\mu)$ represents its own curved space-time with coordinates $s^\mu$ or a higher-spin field fluctuation on it.
The cases of the other field variables are the same.
According to the above result, this re-expression of the gauge potentials $a_\mu(s^\mu)$ is a background-independent description of space-time.

Here, we make an important remark. In the same way that we reduced the $U(N)$ Chan-Paton internal degrees of freedom of clusters of $N$ D-strings to just a loop algebra index $N$ of the field variables $\phi$, now we have reduced the local structure of space-time represented by infinitesimal general coordinate transformations in the form $x^{n+1}\cdot{d}/{dx}$ for $n\in{\boldsymbol{Z}}$ that obey the diffeomorphism algebra, to just loop algebra elements $z^n\otimes d/dx$, corresponding to the T-dual description of clusters of D-strings, for a multiplicable and dimensionless symbol $z$. Due to this reduction, the following descriptions of the local space-time degrees of freedom are very simplified.

The components of the covariant derivative $\nabla_\mu^\ast$ can be recognized as $Spin(10)$ gauge fields. Due to this fact, we extract the translation gauge fields, that is, the vierbeins $e_\mu^{a\ast}$, as 
\begin{equation}
e_\mu^{a\ast}=\langle \nabla_\mu^\ast,\partial_a \rangle\;.\end{equation} Then, the metric tensor is defined using the inner product over the index $a$ of the vierbeins:
\begin{equation}
g^\ast_{\mu\nu}=\langle e_\mu^\ast,e_\nu^\ast\rangle_a\;.
\end{equation}

The following combinations of the vierbeins and the space-time derivatives are the Utiyama gravitational gauge fields $\Gamma^\ast$\cite{U}:
\begin{equation}
\Gamma_{\mu\nu}^{\lambda\ast}=\frac{1}{2}\la {\partial}^{\lambda}{{{e}}}_{\mu}^\ast- {\partial}_{\mu}{{{e}}}^{\lambda\ast},{{{e}}}_{\nu}^\ast\ra_a\;.\label{eq:gravity1}
\end{equation}

As found from Eq.(\ref{eq:gravity1}), we assume, for purely physical reasons, that the indices $\mu$ and $\nu$ are commutative in $\Gamma_{\mu\nu}^{\lambda\ast}$ in order to consider the Riemannian space
\begin{equation}
\Gamma_{\mu\nu}^{\lambda\ast}=\Gamma_{\nu\mu}^{\lambda\ast}\;.
\end{equation}

 In the Utiyama's formalism, gravity is represented not by the metric but by the Utiyama field (i.e., the affine connection).\cite{U} We can take the vierbein, the metric or the affine connection as the variable of gravity. They have their own representations of the Einstein equation and can be translated into each other.

 This construction of the space-time structure is done in the classical regime. That is, it is an approximation of the quantum reality. In the quantum regime, the gauge potentials $a_\mu$ are operators. Thus, Eq.(\ref{eq:HKK}) makes sense only in terms of expectation values for the wave function $\Psi^\ast$.

As an aspect of the quantum regime, now we explain how the dimensionality of space-time is linked to the symmetries of the wave function $\Psi$ before the HKK mapping.

 Based on the GSO result, the trivial form of the dimensionality of $\hat{a}_\lambda$ in terms of the vector $(\lambda_i)_i$ of the eight independent physical transverse directions of string vibrations is
\begin{equation}
[\hat{a}_{\lambda_i},\hat{\pi}_{a_{\lambda_j}}]_-\propto \delta_{ij}\;,\ \ i,j=1,2,\ldots,8\;,\label{eq:pol}
\end{equation}
where the full eight-dimensional spatial rotational symmetry is unbroken.
Here, we note that in general, this formula can be locally satisfied by taking, at every $s^\mu$, the certain local coordinate frame for the spatial coordinates. The non-trivial dimensionality formula is given by using the variables which have spatial rotational symmetries and are reduced by these symmetries due to the dimensional degeneracy.
  
In $\Psi$ the cusps represent the transverse degrees of freedom, since they do so in the NSR partition function.
Then, the dimensionality formula of the field operators is reflected by the transverse factorization of $\hat{\Psi}$ such that
\begin{equation}
\Psi\sim \hat{{\cal{O}}}_1\cdots \hat{{\cal{O}}}_n\psi_{v_1}\otimes\cdots\otimes\psi_{v_n}\;,\end{equation}for operators $\hat{{\cal{O}}}_i$, representing $n$ independent spatial rotational symmetries, and their corresponding ground state wave functions $\psi_{v_i}$. So it is natural to consider that, if another decomposition of $\hat{\Psi}$, written as a summation within the eight-fold root system of $\hat{{\mathfrak{g}}}$, is deformed, then this dimensionality formula may also be deformed.

This is confirmed from the perturbative discrete modular symmetry of the wave function. To see this, we need the following preliminaries. 
Except for one case,
 if the perturbative discrete modular symmetry of the wave function is broken,
  the dimensionality formula also needs to be non-trivial.

In the last
section,
%subsection,
we found that, in the presence of the non-zero frozen Casimir $\zeta$ in Eq.(\ref{eq:brst}), the 
loop symmetry $\hat{{\mathfrak{g}}}$ is broken and its eight-fold root lattice is deformed. Consequently, the perturbative discrete modular symmetry of $\Psi$ may be also broken. From the result obtained in the paragraph above, the dimensionality may be also deformed and non-trivial.

Next, we make an important point. In this subsection, so far only the space-time variables have appeared. However, of course, the matter variables also need to be incorporated into the model. To address this point, Furuta, Hanada, Kawai and Kimura generalized HKK's argument in Eq.(\ref{eq:HKK}) to covariant derivatives with torsion:\begin{equation}
A_\mu=iR_{(\mu)}^\nu (\gamma^{-1})(e_\nu^{a\ast}(s^\mu)\nabla_a^\ast+S_{\nu}^{\lambda\rho\ast}(s^\mu)\co_{\lambda\rho}^\ast)\;,\end{equation}
where $e_\nu^{a\ast}$, $\co_{\mu\nu}^\ast$, $\nabla_a^\ast$ and $S_{\mu}^{\nu\lambda\ast}$ are the vierbeins, the Lorentz generators of the $Spin(10)$ group, the torsionless covariant derivatives and the contorsions, respectively.
Then, some components of the torsion field \begin{equation}T^\ast_{\mu\nu\lambda}=S^\ast_{\mu\nu\lambda}-S^\ast_{\nu\mu\lambda}\;,\label{eq:Furuta}\end{equation} can be identified with the matter fields in type II string theories, that is, in the bosonic sector, the dilaton field and the $B$-fields.\cite{HKK2}

 Here, all of the local fields are unified in each loop algebra component of the generalized Utiyama fields with torsion $a_\mu^\ast(s^\mu)$, which is considered to represent the T-dual description of the field of clusters of D-strings, and are distinguished by their geometrical roles in $a^\ast_\mu(s^\mu)$. This distinguishability is compatible with the loop algebra structure of the gauge potentials $a_\mu^\ast(s^\mu)$, as remarked before, just as it is in the context of type IIB matrix model.
 
  In the following, we invoke HKK's argument in this generalized form.
 
\subsection{Cosmic Time in the Quantum Regime}

The cosmic time $\tau$ is the affine parameter assigned on spatial hypersurfaces sliced from space-time. As will be seen in the next 
section,
%subsection,
our formulation of type IIB string theory vacua is close to that of the wave functions of the Universe. Then, to quantize the Universe, the treatment of the cosmic time needs to be trivial, in other words, the physical quantities do not depend on the choice of the cosmic time.\cite{Konishi2}
Actually, to quantize the Universe conventionally, we decompose the space-time metric by the ADM formalism and after the variation of the action by the lapse function we set the lapse function at unit. So, the wave function of the Universe has no information about the potential of the increment of the cosmic time that is the temporal-temporal part of the gravitational potential (i.e., the space-time metric).
Our context of the quantization after TRpS breaking is in the same situation. Thus, our definition of the increment of the cosmic time needs to be done not by the space-time metric but by a clock of the string excitations.\cite{Konishi2}

{{In the Heisenberg picture, before doing the HKK representation,
we define the increments of the {{cosmic time}} $\delta{\tau}(s)$ for the canonical variables, to describe the change of the system only, in units of the Planck time as an operator-valued function of the hidden time variable $s$:
\begin{equation}\delta{\tau}(s)=\int_s^{s+\delta s} ds({\cal{O}}_{id}(s))\;,\label{eq:time1}\end{equation}
where $\delta s$ is a constant due to TRpS breaking.
Here, ${\cal{O}}_{id}(s)$ is given by
\begin{equation}
{\cal{O}}_{id}(s)=\sum_\phi\frac{1}{i\hbar} \int d^9x \Bigl(\pi_{\breve{\phi}}^{{\alpha}}ad\Bigl(\breve{\phi}^{{\alpha}}\Bigr)-\breve{\phi}^{{\alpha}}ad\Bigl(\pi_{\breve{\phi}}^{{\alpha}}\Bigr)\Bigr)\;,
\end{equation}
where the summation is taken over all of the field variables and the repeated indices $\breve{\alpha}$ with $l\ge0$ are contracted.
${\cal{O}}_{id}(s)$ is not a temporally constant operator
and it generates $\sigma^{{\alpha}}(s)$, which are the non-unitary factors of the uncertainties of the string excitations in Eqs. (\ref{eq:CCR1}), (\ref{eq:CCR2}) and (\ref{eq:CCR3}) and have an infinite number of independent parameters.
As already mentioned, due to the non-locality of the development of field components for $s$, the Virasoro and initial data appear in $\sigma(s)$ as the non-local hidden variables.

When we assume TRpS breaking, the Heisenberg equations of field operators ${\cal{O}}(\breve{\phi},\pi_{\breve{\phi}})$ are, using the normalized Hamiltonian operator ${\bs{\ch}}$ of Eq.(\ref{eq:YM}), given by
\begin{equation}
-i\hbar\frac{\partial{\co}}{\partial {\tau}}=[\co,{\bs{\ch}}]_-\;.\label{eq:Heisenberg}
\end{equation}
These are consistent with the definition of the time increment given in Eq.(\ref{eq:time1}). Here, we 
use 
$\partial_0\sigma(s)\approx 0$
for Eq.(\ref{eq:Heisenberg})
 when ${{\cal{O}}}$ is the canonical (anti-)commutator of a field variable $\phi$. We note two points. First, Eq.(\ref{eq:Heisenberg}) does not logically lead to the canonical (anti-)commutation relations of the field variables $\phi$. Second, 
the combination of Eq.(\ref{eq:Heisenberg}) and the canonical (anti-)commutation relations leads to the Euler-Lagrange equations (or, equivalently the canonical equations) of the field variables ${\phi}$ using the time $\tau$. This time $\tau$ is the coordinate $s$ corrected by quantum effects. Thus, we identify $\tau$ with $s$ as a variable in $\co$.

Next, we consider the HKK map $\ast$ from the set of field operators ${\cal{O}}$, defined by Eq.(\ref{eq:Heisenberg}), on the Minkowski space-time to the set of field operators ${\cal{O}}^\ast$ on the space-time manifold ${\bs{M}}^{9,1}$ given by the representation, by the HKK method, of the gauge potential $a_\mu(s^\mu)$.
Then, on the HKK theater ${\bs{M}}^{9,1}$, the free part $(\ch^{(0)})^\ast$ of the Hamiltonian operator $\ch^\ast$ is given by the BRST charge $Q^\ast$ in the asymptotic field description of the canonical momenta.

 The grounds for this statement are as follows. First, in our model $Q^\ast$ is the generator of the unique symmetry transformation, which is of the gauge transformation form, on the gauge potentials $a_\mu(s^\mu)$ in the Lagrangian ${\cal{L}}$. So, on the HKK theater, $Q^\ast$ is translated into the generator of a general coordinate transformation. Second, $Q^\ast\Psi^\ast=0$ gives the coordinate representation of single string states and, thus, corresponds to the Schr${\ddot{{\rm{o}}}}$dinger equation 
of the asymptotic behavior.
As will be explained, in our model, we assume that $\Psi^\ast$ does not depend on the time also in the 
coherent-state representation. Thus, this Schr${\ddot{{\rm{o}}}}$dinger equation is just $\ch^\ast\Psi^\ast=0$. The generator of general coordinate transformations which defines this Schr${\ddot{{\rm{o}}}}$dinger equation is $\ch^\ast$ only. From these two facts, we conclude that the free part of the Hamiltonian $(\ch^{(0)})^\ast$ is identified with the BRST charge $Q^\ast$ on the HKK theater.

More precisely, our statement is
\begin{equation}
(\ch^{(0)})^\ast=\Upsilon Q^\ast\;.\label{eq:Diff}
\end{equation}
This is similar to the relation between the Virasoro charges and the BRST charge in two-dimensional conformal field theory.\cite{GSW}
Here, $\Upsilon$ is defined by
\begin{equation}
\Upsilon=d\upsilon^\alpha \Lambda^\alpha\;,
\end{equation}where the repeated indices $\alpha$ are contracted, $d\upsilon^{\alpha}$ is the linear dual basis of the Grassmann number basis $dy^{\alpha}$, defined by the inner product, and $\Lambda^\alpha$ are the $\hat{{\mathfrak{g}}}$-endomorphisms which decompose the BRST transformation of fields into a linear combination of the Grassmann number basis vectors.
 Consequently, we have
\begin{equation}
\delta{\tau}^\ast(s)=\Upsilon\int_s^{s+\delta s} ds({\cal{O}}_{id}(s))\;.\label{eq:time2}
\end{equation}
Since the HKK map preserves the equations of motion,
the canonical (anti-)commutation relations 
among $\breve{\phi}^\ast$ and $\pi_{\breve{\phi}^\ast}$ are 
locally
the same as those 
among $\breve{\phi}$ and $\pi_{\breve{\phi}}$.

When we assume TRpS breaking, we write the Heisenberg equations of the field variables $\breve{\phi}^\ast$ on the HKK theater as
\begin{equation}
-i\hbar\frac{\partial{\breve{\phi}}^\ast}{\partial \tau^\ast}=[\breve{\phi}^\ast,{\bs{Q}}^\ast]_\pm\;.\label{eq:Heisenberg2}
\end{equation}
The consistency condition $\ddot{\breve{\phi}^\ast}=0$ holds due to the nilpotency of the time $\tau^\ast$.

Here, we make important remarks.
The field variables and the wave functions are determined by the Heisenberg equations and Schr${\ddot{{\mbox{o}}}}$dinger equations, which are defined on ${\boldsymbol{R}}^{9,1}$ before the HKK mapping. Of course, the fields $\phi^\ast$ have interactions and Eq.(\ref{eq:Heisenberg2}) is just a supplementary consistency condition on $\phi^\ast$. Actually, the HKK map is a classical mechanical notion and the mapping is done in the Lagrangian formalism. So, Eq.(\ref{eq:Heisenberg2}) and the later introduced Eq.(\ref{eq:Sch}) are not the fundamental equations which determine the model, but are tools in the formulation that is used for describing the asymptotic mechanics of the model.

In the following, we limit the space-time model to be the HKK theater ${\bs{M}}^{9,1}$. So, for simplicity, we denote the quantities on the HKK theater without an asterisk.

As will be explained in the next 
section,
%subsection,
the difference between our time increment and the conventional one is that the former induces non-unitary temporal developments 
of the wave functions
 as well as the unitary one but the latter cannot induce the non-unitary temporal development but only the unitary one. Since the time variable $s$ and the Virasoro and initial data are just hidden, this non-unitarity is just an interpretation of the behavior of wave functions, in terms of the history of time $\tau$, at a particular value of $\tau$. This interpretation is determined by the non-unitary factors $\sigma(s)$, which appear from the BRST charge operator ${Q}$ and are absorbed by $\delta\tau(s)$, via the matter Schr${\ddot{{\mbox{o}}}}$dinger equations.

The cosmic time makes sense only in terms of its change in the Schr$\ddot{{\rm{o}}}$dinger equations (see Eqs.(\ref{eq:Sch})) obtained by assuming TRpS breaking. The issue of quantum mechanical simultaneity will be resolved by an intrinsic way in the derived categorical formulation as the quasi-equivalence of $Q$-complexes that analyze 
cosmic time evolutions.

\section{Cosmic Time Developments on the HKK Theater}
\subsection{The Universe: A Conjecture}
In this 
section,
%subsection,
 we physically interpret the wave function, which corresponds to a type IIB string vacuum.\cite{Konishi2}
As will be seen later, it is consistent that our formulation of type IIB string theory vacua by Eq.(\ref{eq:brst}) does not depend on changes of the cosmic time also in the 
coherent-state representation. This is the same as the formulation of the quantum behavior of the early Universe by the Wheeler-De Witt equation.\cite{WDW1,WDW2,WDW3,WDW4,WDW5} The wave function of the Universe is defined on superspace, that is, the moduli space of spatial metrics and field configurations under the moduli of spatial diffeomorphisms.

The Universe is canonically quantized by the spatial metrics $h_{\hat{i}\hat{j}}$ (with spatial indices $\hat{i},\hat{j}=1,2,\ldots,9$) on the spatial hypersurfaces sliced from space-time by values of the cosmic time and their momenta. It is described by a wave function of the nine-dimensional spatial metric $h$, the axion and dilaton ${\cal{M}}$, the spatial parts of the 2-form NS-NS and R-R potentials ${{{B}}}^{(i)}_{\hat{i}\hat{j}}$ for $i=1,2$, and the cosmological constant appearing in Eq.(\ref{eq:brst}) as $\zeta$. This wave function is the solution of the Wheeler-De Witt equation \begin{equation}\hat{{\cal{H}}}\psi[h_{\hat{i}\hat{j}},{\cal{M}},{{{B}}}^{(i)}_{\hat{i}\hat{j}}]=0\;,\label{eq:WDW}\end{equation} for the quantum mechanical Hamiltonian operator $\hat{{\cal{H}}}$ of type IIB supergravity as the low energy effective field theory of type IIB string theory.

 The Hilbert space of the fields of gauged S-duality on the configuration space
  contains all of the excitations of strings except for gravitons and space-time has been described in the last subsection.

 Therefore, the consistency of our model requires that the wave function of the Universe is equivalent to the wave function, in the 
 coherent-state representation, defined by the generalized Kugo-Ojima physical state condition, that will be introduced in the next
chapter,
% section,
  in its local field description
  \begin{equation}\psi=\Psi\;.\label{eq:Universe}
 \end{equation}
Here, the r.h.s.
is based on a double structure. The first structure is quantum mechanics before the HKK map. Based on this first structure, the second structure is the generalized Kugo-Ojima physical state condition, after the HKK map, that represents a TRpS in the presence of the interactions.

As already referred to, this conjecture is supported by the following arguments. In Section 5.4.1, we invoked HKK's formulation of space-time. They showed that diffeomorphism invariance is realized as a unitary symmetry of the matrix variables in the context of type IIB matrix model.\cite{HKK} The Wheeler-De Witt equation expresses the quantum version of diffeomorphism invariance, which is a gauge symmetry, and the Kugo-Ojima physical state condition
or its generalization
 implies the same gauge symmetry by invoking their result.

Here, we make two remarks about the time aspect of the conjecture in Eq.(\ref{eq:Universe}).

First, the reader may find our conjecture to be strange, since although due to this conjecture the wave function $\Psi$ does not depend on the cosmic time, the Universe which is defined by $\Psi$ and is the solution of the equations of motion of type IIB string action depends non-trivially on the development of the cosmic time. As already seen, this paradox is resolved by the following argument. In our model, in the quantum regime, the 
space-time structure is defined by using the wave function $\Psi$, and $\Psi$ itself does not depend on the cosmic time, but the matter states depend on the cosmic time.  So, by recognizing a matter state as the clock, Eq.(\ref{eq:gravity1}) etc depend non-trivially on the cosmic time.

Second, as already explained in
Chapter 2,
%Section 2,
 in the l.h.s. of Eq.(\ref{eq:Universe}), the cosmic time and the dependence on it of the matter fields have been introduced for the semiclassical phase of the wave function of the Universe by substituting the definition of the momentum of the Universe (in the minisuperspace model, the time rate of the dynamical changes of the scale factor of the Universe) into the Hamilton-Jacobi equations of the matter fields obtained from the Wheeler-De Witt equation Eq.(\ref{eq:WDW}). As a result, we obtain the Schr${\ddot{{\rm{o}}}}$dinger equations of the matter fields with TRpS. 

\subsection{Matter and Hamiltonians}

In the 
coherent-state representation, a matter system of microscopic or macroscopically coherent quantum fields is described by the wave function ${\psi}_\Lambda$, depending on the order parameters $\Lambda$, which describe the symmetry and its breakdown in the effective vacuum.\cite{Konishi2}

 The uniqueness of the bare Hamiltonian of our model requires that the wave function ${\psi}_\Lambda$ of any such system takes the form\begin{equation}
{\psi}_\Lambda=R\psi_{\hat{v}}\;,\ \ \psi_{\hat{v}}=\varrho(\hat{v})\Psi\;,\ \ \hat{v}\in V\;,\label{eq:brain}
\end{equation} where $V$ is the Hilbert space of the system, $\varrho(V)$ is the representation of $V$, and $R$ is a renormalization transformation on the vacuum $\Psi$, in which $\psi_{\Lambda}$ keeps the degrees of freedom of the symmetry of the system.

To reformulate the full Hamiltonian in a form compatible with the free part $Q$ and the time $\tau$, we need the non-linear potentials that will be introduced in the next 
chapter.
%section.
Here, we consider the simpler asymptotic mechanics of the model because, in spite of its simplicity, it contains manifestly non-unitary processes in the particle picture.

At the start, the BRST transformation is
\begin{equation}
\delta \psi_{\Lambda}={{{\boldsymbol{Q}}}}\psi_{\Lambda}\;.
\end{equation}
As well as Eq.(\ref{eq:Diff}), on the HKK theater and in the asymptotic field description, we non-trivially identify the cosmic time with the time variable of the normalized BRST charge
\begin{equation}
\delta_{\tau_R}\psi^{as}_{\Lambda}=\delta \psi^{as}_{\Lambda}\;,\ \ [\delta_{\tau_R},\tau_R]_-=i\hbar\;,\label{eq:postulation}
\end{equation}
where $\tau_R$ is an effectively scaled cosmic time of the renormalized asymptotic matter wave functions $\psi_\Lambda^{as}$.

{{When we assume TRpS breaking, the equation of the asymptotic matter wave function $\psi^{as}_\Lambda$ is\begin{equation}i\hbar\frac{\partial\psi^{as}_\Lambda}{\partial \tau_R}={{{\boldsymbol{Q}}}}{\psi^{as}_\Lambda}\;,\label{eq:Sch}\end{equation}
where we identify $\tau_R$ with $s$ as a variable in $\psi^{as}_\Lambda$.
Due to the restriction of the Hilbert space on $V$, the r.h.s of Eq.(\ref{eq:Sch}) needs not to vanish
(refer to Chapter 2.1).
%(refer to Section 2.1).

The functional form of the increment of the cosmic time $\delta\tau_R(s)$ is given by Eq.(\ref{eq:time2}). Eqs. (\ref{eq:Sch}) specify the form of the derivative $\partial/\partial \tau_R$ in terms of the coordinate $s$.
The derivative $\partial\psi^{as}_\Lambda/\partial \tau_R$ is between functions of the coordinate $s$. Since $s$ is hidden, without Eqs.(\ref{eq:Sch}), when $\tau_R$ changes its value, we do not know how the hidden coordinate $s$ has changed to cause this shift; Eqs.(\ref{eq:Sch}) specify it.
So, the variation of the wave function with respect to the cosmic time is also determined by Eqs. (\ref{eq:Sch}).

Due to the infiniteness of the number of the Virasoro and initial data,
to know the exact form of the 
derivative $\partial/\partial \tau_R$ in terms of the hidden coordinate $s$, we need Eqs.(\ref{eq:Sch}) for the full Hilbert space. Thus the description of $\delta\tau_R(s)$ is stochastic. }}
{{Namely, the unpredictability caused by the infiniteness of the number of the non-local hidden variables, that is, the Virasoro and initial data introduces the stochastic processes. 

Due to Eq.(\ref{eq:brst}), it is consistent that the 
vacuum $\Psi$ does not depend on changes of the cosmic time
\begin{equation}{\delta_{\tau} \Psi}=0\;.\label{eq:SchUniverse}\end{equation} 
The 
free Hamiltonian operator, whose factor $\Upsilon$ is truncated,
  of the Hilbert space of the system $V$ takes the form
\begin{equation}
\hat{{\cal{H}}}_{{\bs{Q}}}=({{{\boldsymbol{Q}}}}R)|_{V}\;.
\end{equation}
Here, it is to be noted that particle states 
are eigenstates of the free Hamiltonian $\hat{{\cal{H}}}$,
 which is written by using asymptotic field operators. 

Now, using this free Hamiltonian $\hat{{\cal{H}}}_{{\bs{Q}}}$, we rewrite the 
derivative by the cosmic time in Eq.(\ref{eq:Sch}) as an average over the 
functional $\delta\tau_R(s)$:
\begin{equation}
\langle\psi^{as}_{\Lambda}(\tau_R)\rangle\approx\exp\biggl(-\frac{i\tau_R}{\hbar}\hat{ {\cal{H}}}_{{\bs{Q}}} -\frac{\sigma_R\tau_R}{2\hbar^2}\hat{{\cal{H}}}^2_{{\bs{Q}}}\biggr)\langle\psi^{as}_{\Lambda}(0)\rangle\;,\label{eq:est}
\end{equation}
where the average is defined by the following recursion equation
\begin{subequations}
\begin{align}
\langle \psi^{as}_{\Lambda}(\mu_R)\rangle&=\int {\cal{D}}\tau_R^\prime(s) \exp\biggl(-\frac{i\delta\tau_R^\prime(s)}{\hbar}\hat{{\cal{H}}}_{{\bs{Q}}}\biggr)\langle\psi^{as}_{\Lambda}(0)\rangle\label{eq:ave}\\
&\approx\int d\tau_R^\prime\exp\biggl(-\frac{i\delta\tau_R^\prime}{\hbar}\hat{{\cal{H}}}_{{\bs{Q}}}\biggr)f(\delta\tau_R^\prime)\langle\psi^{as}_{\Lambda}(0)\rangle\;.\label{eq:ave2}
\end{align}
\end{subequations}
Here, we use the identification of each $\tau^\alpha_R$ and $s$ as a variable in $\psi^{as}_\Lambda$ for the ansatz in Eq.(\ref{eq:ave}).
The exponential map in Eq.(\ref{eq:ave}) is well-defined, because two $\partial/\partial y$ commute.
In Eq.(\ref{eq:ave2}) we rewrite the functional integral with respect to $\delta\tau_R^\prime(s)$ as an average over a normal stochastic variable $\delta\tau_R$ with mean $\mu_R$, variance $\sigma_R \mu_R$ and distribution function $f(\delta\tau_R^\prime)$.
We note that, since
the time increment has the Grassmann number basis $d\upsilon^\alpha$
 which is dual to that of the BRST charge, 
 in general, the combination $\sigma_R\tau_R\hat{{\cal{H}}}^2_{{\bs{Q}}}$ does not vanish.

The free Hamiltonian $\hat{{\cal{H}}}$ is a Hermitian operator. Thus for the eigenvalues $\{\lambda\}$ of $\hat{{\cal{H}}}$, there exists a unique spectral family $\{d{\hat{\cal{H}}}(\lambda)\}$, and the spectral decomposition is\begin{equation}
\hat{{\cal{H}}}=\int \lambda d\hat{{\cal{H}}}(\lambda)\;.\label{eq:Sch2}
\end{equation}

From the elementary property of the spectral components $\hat{{\cal{H}}}(\lambda)$ in Eq.(\ref{eq:Sch2}),
\begin{equation}
\hat{{\cal{H}}}({\lambda_1})\hat{{\cal{H}}}({\lambda_2})=\delta_{\lambda_1\lambda_2}\hat{{\cal{H}}}({\lambda_1})\;,
\end{equation}
it follows that,
\begin{equation}
\hat{{\cal{H}}}^2=\int \lambda^2d\hat{{\cal{H}}}(\lambda)\;,
\end{equation}
and the cosmic time development in Eq.(\ref{eq:est}) satisfies the properties of a contraction semigroup in the parameter $\tau_R$.

The degree of freedom of collapses of the superposition of wave functions \begin{equation}R\Psi=\sum_\lambda c_\lambda R\Psi^\lambda\;,\label{eq:super3}\end{equation} is the spectral component $\hat{{\cal{H}}}(\lambda)$. In the superposition of Eq.(\ref{eq:super3}), each component $R\Psi^\lambda$ is distinguished from the others by the spectral components $\hat{{\cal{H}}}(\lambda)$ such that \begin{equation}{\mbox{if}}\ \ \hat{{\cal{H}}}(\lambda)(\psi^{as}_{\hat{v}})^{\lambda_1}\neq0\;,\ \ {\mbox{then}}\ \ \hat{{\cal{H}}}(\lambda)(\psi^{as}_{\hat{v}})^{\lambda_2}=0\;,\end{equation} for $\lambda_1\neq \lambda_2$ and elements $\hat{v}$ of the Hilbert space ${V}$ of the system. Concretely, the spectral component $\hat{{\cal{H}}}(\lambda)$ is defined by the restriction of $\hat{{\cal{H}}}$ on the part which lies within the eigenspace $V_\lambda$ for eigenvalue $\lambda$,\begin{equation}\hat{{\cal{H}}}(\lambda)=\hat{{\cal{H}}}|_{V_\lambda}\;,\ \ V=\bigoplus_\lambda V_\lambda\;,\end{equation} which induces a non-unitary action on the wave function within the non-zero variance of the increment of the effectively scaled cosmic time as an operator of the contraction semigroup in the cosmic time evolution:\cite{Konishi2}\begin{equation}\Delta(\tau_R)\approx\exp\biggl(-\frac{i\tau_R}{\hbar}\hat{{\cal{H}}}_{{\bs{Q}}}-\frac{\sigma_R\tau_R}{2\hbar^2}\hat{{{\cal{H}}}}^2_{{\bs{Q}}}\biggr)\;.\label{eq:timeexact}\end{equation}

We regard the inference of $\delta\tau_R(s)$ from the incomplete knowledge of it (i.e., Eq.(\ref{eq:Sch})) as a normal stochastic process of the variable $\delta\tau_R$ via Eq.(\ref{eq:ave2}) such that the probability $P_\lambda$ of the collapse into the branch $R\Psi^\lambda$ is\cite{Konishi2}
\begin{equation}
P_\lambda=\langle R\psi^{as}_{\hat{v}},\hat{{\cal{H}}}(\lambda)\psi^{as}_{\hat{v}}\rangle=|c_\lambda|^2\;,\ \ \sum_\lambda P_\lambda=1\;.
\end{equation}

 If we knew the exact form of $\partial/\partial \tau_R(s)$, all of the degrees of freedom of both the non-unitary and unitary cosmic time developments of Eq.(\ref{eq:brain}) would be reducible to an infinite number of 
 Virasoro and initial data.
However, this is fundamentally impossible due to the infiniteness of the numbers of the Virasoro and initial data in the cosmic time increment $\delta \tau_R(s)$.
\section{Summary}
Based on the idea in the Penrose thesis, in the context of type IIB string theory, we have given a temporally statistical background for the phenomenon of state reduction in superposed wave functions by introducing the hidden time variable of a linear combination of the Virasoro operators in the gauged and affinized S-duality symmetry
 with an infinite number of independent Virasoro and initial data. 
 The descriptions are based on the BRST formalism for this gauge symmetry with the Kugo-Ojima physical state condition, which is identified with the 
 free part of the
 Wheeler-De Witt equation of the wave function of the Universe in type IIB string theory based on the HKK representation of the field variables. In this statistical model of time increments via the hidden time variable and the non-local hidden variables, both of the spatial expanses of the network of D-strings and matter are described by using the wave function of the Universe. When we assume TRpS breaking, both of the unitary and non-unitary time developments of matter systems are described the Schr${\ddot{{\rm{o}}}}$dinger equations, which are generalized by incorporating the degrees of freedom of  the
  hidden time variable and the non-local hidden variables.
\chapter{A Model of Quantum Mechanical World II}

\section{Non-perturbative Description of the Vacua Using the Non-linear Potential}
\subsection{Derived Category Structure Using Wave Functions}
As the continuance of 
Chapter 5,
%Section 5,
 in this 
chapter,
%section,
 we reformulate the category of the quantum mechanical world given in 
      Chapter 4 
 %Section 4
 as a derived category and realize a temporally statistical quantum geometrodynamics with a hidden time variable and an infinite number of Virasoro and initial data.\cite{Konishi2} 
   In our modeling of type IIB string theory vacua given in 
      Chapter 5,
 %Section 5,
  due to Eqs.(\ref{eq:SchUniverse}) and (\ref{eq:est}), the perfect description of the Universe is independent of changes in the cosmic time, and non-trivial cosmic time processes can be applied only to closed systems with imperfect, partial descriptions and a non-zero quantum superposition retention time. {Due to Eq.(\ref{eq:est}), the quantum superposition retention time tends to zero for the macroscopic objects.} Systems which lose the quantum superposition retention time
 have a 
 classical cosmic time evolution and are essentially removable objects, whereas systems with a non-zero 
 quantum superposition retention time 
 genuinely constitute a quantum mechanical world with common cosmic time processes such as quantum mechanical branching. That is, for the system with the non-zero quantum superposition retention time, the variance of the increment of the cosmic time induces the non-unitary time development of a system. By Eq.(\ref{eq:brain}) a system is a state space $RV_s$, composed of a state space
  $V_s$ with a certain renormalization $R$, and its time development on the HKK theater is 
  analyzed by considering
  the 
  temporally continuous $Q$-complex in the Heisenberg picture,\footnote{In this
      chapter,
  %section,
for simplicity, we denote the normalized BRST charge by $Q$ and refer to it as just `the BRST charge'.
  } where $Q=Q_\tau$ develops unitarily and non-unitarily as the product of the free Hamiltonian and the nilpotent factor and
probes the non-unitary events and the product of the interaction Hamiltonian and the nilpotent factor will be given by the non-linear potential, 
\begin{equation}
\xymatrix{\cdots& RV_n \ar@{->}
[l]_{{\mathscr{Q}}_{n-1}}& 
 RV_{n+1}\ar@{->}
 [l]_{{\mathscr{Q}}_n}&\ar@{->}
 [l]_{{\mathscr{Q}}_{n+1}}\cdots}\;,\ \ {\mathscr{Q}}_n=\{Q_\tau\;|\; \tau_n<\tau\le \tau_{n+1}\}\;.\label{eq:exact}
 \end{equation}
 In the 
  $n$-th element of Eq.(\ref{eq:exact}) 
  counted from a reference element
   we 
   restrict
    both of $QR$ and
    $\Psi$ 
    to 
    the 
    same
     state space with the fixed $n$-th cosmic time value $\tau_{n}$ counted by the events of non-unitary processes. The non-reversible direction of the $Q$-complex arises from this non-unitarity.\footnote{Here, for the full state space $V(\tau)$ at any time $\tau$, we need to show 
\begin{equation}(QV)(\tau+d\tau)\subseteq V(\tau)\;.\label{eq:consistency}\end{equation} 
Due to $Q\Psi=0$, Eq.(\ref{eq:consistency}) follows if 
 $[Q,\hat{\Phi}](\tau+d\tau)\in O(\tau)$ holds for the full space of operators $O(\tau)$ at any time $\tau$ and any operator $\hat{\Phi}(\tau)\in O(\tau)$. 
  Due to the Heisenberg equation, $O(\tau+d\tau)\subseteq O(\tau)$ holds. Also $[Q,\hat{\Phi}](\tau+d\tau)\in O(\tau+d\tau)$, so Eq.(\ref{eq:consistency}) follows.}

In the Heisenberg picture, due to the Heisenberg equation, the BRST charge $Q$ develops non-unitarily as well as Eq.(\ref{eq:est}) does. These developments preserve the nilpotency of $Q$ at every time. In the $Q$-complex, 
the space of non-zero-norm vacuum eigenvectors\footnote{Here, we invoke the derivation of the Virasoro constraints from the Kugo-Ojima physical state condition in two-dimensional conformal field theory.\cite{GSW}} (and their representations of the state vectors), as the $Q$-cohomology, changes non-trivially due to the unitary and non-unitary time processes for the BRST charge $Q$.

In our formulation, by the $Q$-cohomology content in Eq.(\ref{eq:est}), we specify each system and a part of the cohomological classification is done due to the non-unitary second factor in Eq.(\ref{eq:est}). For a macroscopic physical object, we can interpret this as a collection of microscopic quantum states with non-trivial effects of time variances or as a large-scale macroscopic quantum state with trivial effect of time variance. As already explained in
Chapter 4,
%Section 4,
these interpretations need to be unified.
These observations lead us to the {\it{derived category}} description of the quantum mechanical world, on the HKK theater, under the moduli of quasi-isomorphism equivalences of the BRST complexes. We denote by $D(C)$ this derived category of the BRST complexes of the base abelian category $C$. 
 {The objects of $D(C)$ correspond to the quantum classes introduced in 
Chapter 4.}
%Section 4.}
   Due to Eq.(\ref{eq:Whitham}), the quasi-isomorphisms, which commute with 
   the $Q$ operation, are given by renormalizations. 
Here, the derived category $D(C)$ of a base abelian category $C$ is defined by restriction of the homotopy category $K(C)$ on a closed system of the products of quasi-isomorphisms in $K(C)$.\cite{GM}
 The objects of our base abelian category $C$ are the vector spaces of states of wave functions created from a given vacuum $\Psi$ by the actions $R\varrho$. The morphisms are the transformations compatible with the differential $Q$ or the  covariant derivative $\nabla$ on states (namely, where $Q$ or $\nabla$ is a vector). We denote the base abelian categories in these two cases by $C^Q$ and $C^\nabla$, respectively.
 In particular, the morphisms of base abelian category $C^Q$ are defined 
 by the homomorphisms
 compatible with the differential $Q$. 
 The difference between $C^Q$ and $C^\nabla$ becomes clear when the topology of the base space is non-trivial.

\subsection{Non-linear Potential (Gauged String Field Operator)}
Based on this derived category structure $D(C)$ of the quantum mechanical world description, we generalize the results in the last 
chapter
%section
by a substantially different method. We introduce a single master equation as the generalization of the Kugo-Ojima physical state condition for a non-linear potential, denoted by $\aleph$ (representing the symbol `{{$A_\mu$}}' of a gauge potential), which represents a gauged string field {{operator}} and can describe the non-peturbative effects and dynamics, according to the following three guiding principles. As the concrete form of the equation, we adopt a single vanishing curvature.

\begin{enumerate}
\item
{{The local principle.}} In our modeling, it is gauged S-duality.
\item The generalized BRST covariance using the covariant derivatives.
\item 
The equation vanishes under the adjoint action of the covariant derivative as the generalized BRST invariance.
\end{enumerate}

As the result, the 
 equation for the non-perturbative dynamics
 is regarding 
$\hat{{\mathfrak{g}}}$-connection $\aleph$ on the fiber space with the dual coordinates $\hat{y}^{\alpha}$ of $Q|_{dy^{\alpha}}$:\footnote{Eq.(\ref{eq:master}) has the gauge symmetries, which can transform an arbitrary solution to its corresponding locally trivial solution. We note, however, that the topology of our loop space is non-trivial.}
\begin{equation}\Omega=0\;,\label{eq:master}\end{equation}
where $\Omega$ is the curvature form
\begin{equation}
\Omega=[\na,\na]\;,
\end{equation} and $\nabla$ is the covariant derivative defined on the fiber space
\begin{equation}\nabla{\cal{O}}=\delta{\cal{O}}+[\aleph,{\cal{O}}]\;,\end{equation}for an arbitrary $\hat{{\mathfrak{g}}}$-valued form ${\cal{O}}$ and $[,]$ is the ${\boldsymbol{Z}}$-graded 
commutator for an arbitrary pair of a $d_a$-form ${\cal{O}}_a$ and $d_b$-form ${\cal{O}}_b$:
\begin{equation}[{\cal{O}}_a,{\cal{O}}_b]={\cal{O}}_a\wedge {\cal{O}}_b-(-)^{d_ad_b}{\cal{O}}_b\wedge {\cal{O}}_a\;,\end{equation}
which satisfies the super Jacobi identity
\begin{eqnarray}
(-)^{d_ad_c}[{\cal{O}}_a,[{\cal{O}}_b,{\cal{O}}_c]]+(-)^{d_bd_c}[{\cal{O}}_c,[{\cal{O}}_a,{\cal{O}}_b]]+
(-)^{d_ad_b}[{\cal{O}}_b,[{\cal{O}}_c,{\cal{O}}_a]]
=0\;.\label{eq:Jacobi}
\end{eqnarray}
Here, 
the degree $d$ of the element ${\cal{O}}$ is its ghost number.
We note that in general, ${\cal{O}}_a\wedge {\cal{O}}_b\neq-{\cal{O}}_b\wedge{\cal{O}}_a$ for $\hat{{\mathfrak{g}}}$-valued $1$-forms, since we treat the product of matrices in $\hat{{\mathfrak{g}}}$ 
and the outer product of forms simultaneously. 
Since the BRST differential has ghost number one, the ghost number coincides with the degree of the element as a form. The space of the forms $O$ splits into $\oplus_{i\ge0} O^i$ labeled by the ghost number $i$ with $[O^i,O^j]\subset O^{i+j}$. The BRST differential shifts $O^i$ to $O^{i+1}$ and acts on commutators of forms as \begin{equation}\delta[{\cal{O}}_a,{\cal{O}}_b]=[\delta{\cal{O}}_a, {\cal{O}}_b]+(-)^{d_a}[{\cal{O}}_a, \delta{\cal{O}}_b]\;.\label{eq:Leibniz}\end{equation} 

We check the requirements of the three principles in Eq.(\ref{eq:master}).
The first principle requires that infinitesimal approximations $\Psi$ of the wave function parallel to $\nabla$ obey the linearized equation Eq.(\ref{eq:brst}). (Here, the adjoint action of $\aleph$ on the components of the field variable operators $\hat{{\phi}}^{{\alpha}}$ is defined by $[\aleph,\hat{{\phi}}]^{{\alpha}}$. This non-commutativity indicates non-commutative space-time.\cite{CDS}) The principle of covariance requires that the non-linear potential $\aleph$ obey the equation, whose curvature part is written only using $\nabla$. Eq.(\ref{eq:master}) satisfies these requirements. Finally, to show the third principle on Eq.(\ref{eq:master}), we use the super Jacobi identity in Eq.(\ref{eq:Jacobi}). For $\nabla$, we have\begin{eqnarray}
 [\nabla,[\nabla,\nabla]]=0\;,
\end{eqnarray}where we use the fact that $\nabla$ has ghost number one. This is the third principle.

We note that the equation for the non-linear potential is the same as that of the gauged string field. This point is crucial when we consider the $A_\infty$ equivalence principle.

Based on Eq.(\ref{eq:master}), we define each morphism of the base abelian category $C^{\nabla}$ to be the non-linear\footnote{Of course, this non-linearity is about the element of $V$, 
and each morphism acts on the object linearly.} transformation operator compatible to the non-linear potential $\aleph$ (namely, where the covariant derivative $\nabla$ is an infinite dimensional vector for this transformation just like the situation such that, in the general theory of relativity, the covariant derivative is a vector on the curved space-time, that is, in our case the parallel wave function $\Psi^\nabla$ of $\nabla$ such that 
the generalized Kugo-Ojima physical state condition $\nabla \Psi^\nabla=0$ holds). 
The objects of $C^\nabla$ are redefined to be compatible to the morphisms and do not need a vacuum, which is an infinitesimal approximation of the parallel wave function $\Psi^\nabla$ of $\nabla$.

 {{We change the formulation so that the 
 dynamical contents of wave functions result from the morphisms. In this new vision, the role of the given linear potential $\Psi$ in the $Q$-complexes is substantially taken by the non-linear potential $\aleph$ (in the general theory of relativity, they correspond to Newton potential and the space-time metric respectively).}}
The non-linear potential $\aleph$ describes the dynamics of the morphisms of the derived category $D(C^\nabla)$ which is the morphism structure of base abelian category $C^\nabla$. This description is global. Consequently, the non-linear potential $\aleph$ can describe the transition between the stable configurations. In contrast, the linear wave function $\Psi$ is an infinitesimal local description of $D(C^\nabla)$ and $\aleph$, and cannot describe the non-perturbative dynamics of the morphisms of $D(C^\nabla)$.

This derived category $D(C^\nabla)$ is the conclusive formulation of our model of type IIB string theory vacua via gauged S-duality. In the next 
section,
%subsection,
 we will refine it in the language of the $A_\infty$ category.

\section{$A_\infty$ Refinement: Overview}
In the rest of this 
chapter,
%section,
on the HKK theater, we describe the geometry of the motion of D-branes and open strings in a Universe with wave function $\Psi$. (In particular, to simplify their internal degrees of freedom, we treat D-particles in type IIA string/M-theory as D-branes.) We do it in the language of the second quantized generalization of the quantum cohomology of the field theory of fundamental strings in the sense of the second quantization of D-particle fields.
 In the following geometrical formulation, we identify the D-particles with geodesics on the upper half plane. (The basis for this identification will be explained in the next 
       section.)
 %subsection.)
 We study the homotopy structure of the second quantized D-particle field theory that includes the degrees of freedom of 
 D-particles, interacting via open F-strings, with the parallel and vertical Chan-Paton factors on D-particles, and we construct the $A_\infty$ category $C^Q({\fh})$ of background independent D-particle fields by considering the dynamics of  clusters of interacting
 D-particles on the upper half plane ${\fh}$ of the complex string coupling constant $\tau_s$. $SL(2,{\bs{R}})$ transformations promote the geodesics to other geodesics on the upper half plane ${\fh}$, since {{the Poincar\'e metric on the upper half plane is $SL(2,{\bs{R}})$ invariant}}.
    Then, our new viewpoint is to consider the dynamics of 
    clusters of interacting
     D-particles, which are mapped onto a set of geodesics on the upper half plane of ${g_s}$ deformed by the coordinate.
      Here, as detailed in next 
                            section,
    %subsection,
     we rule that D-particles form a cluster only if their corresponding geodesics intersect on $\fh$.
The Feynman diagram of 
clusters of interacting
 D-particles with bounded open string fields on ${\fh}$ contains the intersection of each set of geodesics on the upper half plane, for example, $n$-number of interacting
 D-particles and $m$-number of interacting
  D-particles fuse into $m+n$-number of interacting
  D-particles on ${\fh}$.
The proper language of the homotopy product structure for bounded open string fields is an $A_\infty$ structure. 
We invoke the minimal model theorem which ensures the existence of the quasi-isomorphism $A_\infty$ functor ${\cal{U}}$ between two $A_\infty$ categories, $C^Q({\fh})$ and its minimal model $H(C^Q(\fh))$, which secondarily results in the connection between the D-particle field Hilbert spaces of $\Psi^\nabla$ for systems $S_1$ and $S_2$ defined in terms of local dual coordinates ${{{{{y^\alpha}}}}}$ of the BRST differential $\delta$:
\begin{equation}
{\cal{U}}:V^\nabla|_{S_1}\to V^\nabla|_{S_2}\;,\label{eq:trf}
\end{equation} 
because the flatness of $\Psi^\nabla$ is defined by that of morphisms in $C^\nabla$. On the basis of this $A_\infty$ functor, the equivalence principle for all the quantized {{conventional}} Chan-Paton gauge interactions from the point of view of gauged S-duality is formulated by invoking the minimal model theorem,\cite{Kad,Kajiura} and we suggest that $A_\infty$ covariance is a principle of the theory of gauged S-duality.

\section{D-particle Field Category: A Toy Model}
An {\it{$A_\infty$ category}} $C$ is the triple of a set of objects $C=\{{O_i}\}_i$, the morphism spaces between two arbitrary objects ${\cal{H}}({O_i},{O_j})$, which are ${\bs{Z}}$-graded vector spaces, and the product structures $m_n$ between $n$ morphism spaces among $n+1$ arbitrary objects ${O_i}$ for $1\le i\le n+1$. The product structures are degree $(2-n)$ multi-linear maps
\begin{equation}
m_n:\bigotimes_{i=1}^n{\cal{H}}({O_i},{O_{i+1}})\to{\cal{H}}({O_1},{O_{n+1}})\;,\end{equation}
for $n=1,2,\ldots$, which satisfy the $A_\infty$ conditions (see Eq.(\ref{eq:l})).\cite{Fukaya}

We define an $A_\infty$ category such that its object set is the set of the direct sums of the $\bc-$vector spaces based by the sets of geodesics $C_1,C_2,\cdots$ promoted to other geodesics by the coordinate ${s}$ 
via the cosmic time increments on the Poincar\'e upper half plane ${\fh}$:
\begin{equation}{C^Q({\fh})}=\biggl\{ \bigoplus_aV_{(\bc_i)_a}\biggr\}\;,\ \ {(\bc_i)_a}=\{(C_{i_j})_a\}_{j_a}\;.\label{eq:CQ}\end{equation}

These sets of geodesics $\bc$ represent the background independent 
(interacting)
D-particle eigenvectors of the D-particle number operator. (When S-duality is gauged, the distinction between D-particles and fundamental strings is removed. On this point, we assign R-R and NS-NS parts of states on the basis of $\hat{\mathfrak{g}}$ generators.) We introduce the morphism structure between the objects by invoking that of the Fukaya category.\cite{Fukaya} We note that geodesics in the upper half plane represent parts of the moduli spaces of S-duality doublets obtained by fixing the R-R sectors, which run along the real axis of the upper half plane, and by keeping the degrees of freedom of NS-NS sectors. Thus, the intersections between geodesics represent the situation in which fundamental strings (an NS-NS sector) connect D-particles (an R-R sector). We define D-particle numbers
 based on these arguments using geodesics. We recognize two geodesics that do not intersect in $\fh$ as non-intersecting D-branes; in the D-particle case these are two isolated 
 (not interacting)
 D-particles. We introduce the class of geodesics by the condition that when $C_1$ and $C_2$ intersect on $\fh$ we consider them to be equivalent, $C_1\sim C_2$. This equivalence relationship gives a class by the fact that if $C_1\sim C_2$ and $C_2\sim C_3$ then $C_1\sim C_3$ (i.e., transitivity). We consider the number of geodesics $n$ that belong to the same class as interacting D-particles with number $n$. 
 The morphism space between two objects of the vector spaces of eigenvectors of the D-particle number operator is defined by the ${\boldsymbol{Z}}$-graded vector space 
 (where vector spaces with degrees less than $1$ are zero-dimensional)
 \begin{subequations}
 \begin{align}
\ch(V_{\bc_i},V_{\bc_j})&=\bigoplus_{k}({\cal{MC}})^{n_{k}}\cap\left\{\begin{array}{c|c}
\co_k& 
\co_{k}\ {\mbox{has\ the\ form-basis\ of}}\ (\bc_i\cap\bc_j)_{k}
\end{array}\right\}\;,\label{eq:mor1}\\
 \bc_i\cap \bc_j&=\bigoplus_{k}\co_{k}\;,\ \ \deg \co_{k}=n_{k}\;,
 \label{eq:mor}
 \end{align}
 \end{subequations}
 for $n_k=1,2,\ldots$. ${\cal{MC}}$ is the space of the solutions of the Maurer-Cartan equation, and each morphism $\co$ is {{a $c$-number form}} on the fiber space
  and represents a gauged string field state as the state of the open string field within the cluster of interacting D-particles. The degree of each morphism for systems of $n$
 interacting D-particles is $n-1$ and the concrete form of the Maurer-Cartan equation for (the equation of motion of) $\co$ will be given later (see Eq.(\ref{eq:MC})). The morphism space between two objects of the vector spaces of composite vectors are defined by the direct sum of Eq.(\ref{eq:mor1}). Since each morphism carries two D-particle cluster indices of its objects, it has $\hat{{\mathfrak{g}}}$-matrix representation. Each morphism depends on 
  the dual coordinates $y^\alpha$ of the BRST differential $\delta$
 whose arrangement will be given later. 
 These 
  represent the degrees of freedom of interacting D-particles.
  For $n=1,2,\ldots$, we represent degree $n$ morphisms by $n$-forms on 
  the fiber space.\footnote{Although our model is based on field theory, we regard the physical states as the ghost number zero components of string fields.}
   The condition in Eq.(\ref{eq:mor1}) implies that this definition is topological.

Here, we make two remarks about this definition of the morphisms. First, D-particles do not intersect with each other. Only the interactions between D-particles promote the open string fields. So $\co$ includes the degrees of freedom of D-particles as well as those of the open string fields. Second, in the previous 
     section,
 %subsection,
  we defined the objects of the category $C^Q$ by the complete set of the subspaces in the full Hilbert space.\cite{Konishi2} These subspaces correspond to the vector spaces in our definition of the objects of $C^Q(\fh)$.

It is natural from a physical point of view to assume that the higher product structures of $\co_{i,j}$ in ${\cal{H}}(V_{\bc_i},V_{\bc_j})$ between $n+1$ ($n=2,3,\ldots$) objects are defined to be the transition rates between open string fields. The degree two product structure is the star product $\star$. In our toy model, the product structures on {$C^Q(\fh)$} are defined by
\begin{subequations}
\begin{align}{m}^Q_1(\co)&=\delta \co\;,\label{eq:low}\\
m^Q_n(\co_1,\cdots,\co_n)&=(-)^{\sum_{i=1}^{\lfloor\frac{n}{2}\rfloor}(\deg \co_{n+1-2i}-1)}\varepsilon^n\int_{\cm^{(n+1)}_{{a^\prime},1,\ldots,n}}S_{{a^\prime},a}\wedge \co_1\wedge \co_2\wedge\cdots \co_n\;,\label{eq:l1}
\end{align}
\end{subequations}where we put $\co_i=\co_{i,i+1}$ for $1\le i\le n$, $\co_{a}=\co_{1,n+1}$, $\lfloor x\rfloor$ is the integer part of $x$ (Gauss' symbol), $\varepsilon^n$ ($n=2,3,\ldots$) are Grassmann numbers\footnote{The product structures without $\varepsilon^n$ have the ghost number anomaly. This issue was fixed by Witten\cite{Witten1} by inserting constant ghost operators $\varepsilon^n$.}, and $\cm^{(n)}$ $(n=3,4,\ldots)$ are the moduli spaces weighted by the complex number valued weight functions. The norm of each weight function is defined as the Hermitian inner product of the function with itself and the weight functions are normalized as well as each morphism. 
The arrangement of the coordinates $y^\alpha$ is chosen according to the following definitions of the moduli spaces $\cm^{(n)}$ ($n=1,2,\ldots$).
 $\cm^{(1)}_{\co}$ are one-dimensional and are defined by
\begin{equation}
\int_{\cm^{(1)}_\co}\co=1\;,\ \ \co\in{\cal{MC}}\;.
\end{equation}
$S_{1,2}$ is the inverse reflector\cite{SFT} that satisfies
\begin{subequations}
\begin{align}
\int_{\cm^{(1)}_{(\co)_1}}S_{1,2}&=(\co)_2\;,\ \ (\co)_1\in({\cal{MC}})_1\;,\ \ (\co)_2\in({\cal{MC}})_2\;,\\
 \delta S_{1,2}&=0\;,
 \end{align}
\end{subequations}where $\co$ is an arbitrary element of ${\cal{MC}}$. $\cm^{(2)}_{1,2}$ is two-dimensional and is defined by
\begin{equation}
\int_{\cm^{(2)}_{1,2}}S_{1,2}=1\;.
\end{equation}
$\cm^{(n)}_{1,2,\ldots,n}$ satisfies the cyclic symmetry
\begin{equation}
{{\cm}}^{(n)}_{1,2,\ldots,n}=(-)^{n+1} {{\cm}}^{(n)}_{{2},{3},\ldots ,{1}}\;.\label{eq:cyc1}
\end{equation}
 We define the coboundary operation by
\begin{subequations}
\begin{align}
\partial \cm^{(2)}_{1,2}=&\emptyset\;,\\
\partial\cm_{1,2,\ldots,n}^{(n)}=&-\frac{1}{2}\sum_{k=1}^n\sum_{l=1}^{n-3}\Bigl((-)^{(n+1)(k+l+1)}\times\nonumber\\ &\Biggl(\int_{\cm^{(2)}_{{a^\prime},a}}\Bigl(\cm_{{k},\ldots,{k+l},a}^{(l+2)}\Bigr)^\vee\wedge\Bigl(\cm^{(n-l)}_{{a^\prime},{k+l+1},\ldots,{k+n-1}}\Bigr)^\vee\Biggr)^\vee\Bigr)\;,\label{eq:cyc2}\end{align}\end{subequations}
for $n\ge 3$, where $\cm^{(n)}_{1,2,\ldots,n}$ for $n\ge3$ is $n$-dimensional and we have introduced the dual transformation $\vee$ from the moduli space weighted by the weight function $\cm$ to the form $\cm^\vee$ by
\begin{equation}\int_{\cm}\cm^\vee=1\;.\end{equation} The operation $\partial$ satisfies $\partial^2=0$\cite{Nakatsu} in the dual concept, that is, the integral domain with the weight function, of the forms.

By using these definitions of the product structures, the Maurer-Cartan equation for degree one morphism $\co$ is
\begin{equation}
\sum_{n=1}^\infty m_n^Q(\co,\ldots,\co)=0\;.\label{eq:MC}
\end{equation}

Here, we remark on the single states. In the definition of the morphisms $\co$, we choose the single open F-string states as the single states in ${\cal{MC}}$. So, each composed state $\co$ describes a state of open F-strings within a cluster of D-particles. On the other hand, in the definition of the wave functions of the Universe $\Psi$, we choose the cluster states of D-particles as the single states. So, each composed state of $\Psi$ describes a cluster state of clusters of D-particles, which reflects the degrees of freedom of the Chan-Paton factors.

Now, we show the nilpotency for $C^Q(\fh)$ in the following way.
 When we apply the BRST differential 
 $\delta$ to the product structure, from the BRST invariance of the inverse reflector, two terms emerge such that $\delta$ acts on the forms $\co_i$ ($i=1,2,\ldots,n$) or on the weight function of $\cm$ (as the coboundary operator $\partial$). Then, using the Leibniz rule of $\delta$ for the former term and using the cyclic symmetry and the coboundary formula for $\cm$, that is, Eqs. (\ref{eq:cyc1}) and (\ref{eq:cyc2}) for the latter term, the following $A_\infty$ conditions hold.\cite{Nakatsu}
\begin{eqnarray}
\sum_{1\le k,n\le m}(-)^\ast{m}^Q_{m-n}(\co_1,\cdots,{m}^Q_{n}(\co_k,\cdots,\co_{k+n-1}),\cdots,\co_{m})=0\;,\label{eq:l}
\end{eqnarray}
where $\ast={\mbox{deg}}(\co_1)+\cdots +{\mbox{deg}}(\co_{k-1})-k+1$, for all $m\ge1$.

We remark that
the morphisms $\co=\co(s^\mu)$ and the product structures $m_n^Q=m_n^Q({s}^\mu)$ ($n=1,2,\ldots$) of $C^Q(\fh)$ also depend on the coordinates ${s^\mu}$,
as the objects do.
\section{Equivalence Principle and $A_\infty$ Covariance Principle}
An {\it{$A_\infty$ functor}} between an $A_\infty$ category $C_1$ and another $A_\infty$ category $C_2$---which, in our sense, may give the {\it{coordinate transformation}} between Hilbert spaces $V^\nabla|_{S_1}$ and $V^\nabla|_{S_2}$---is defined by the set of an infinite number of maps $\{\cu,\cu_1,\cu_2,\ldots\}$. That is, a map between objects 
\begin{equation}
\cu:C_1\to C_2\;,
\end{equation}
 and the multilinear maps $\cu_n$ with degree $(1-n)$ for all objects ${O_1},{O_2},\ldots,{O_{n+1}} \in C_1$ with $n=1,2,\ldots$
\begin{equation}
\cu_n:\bigotimes_{i=1}^n \ch_1({O_{i}},{O_{i+1}})\to \ch_2(\cu({O_1}),\cu({O_{n+1}}))\;,
\end{equation}
 such that for all subsets of the objects $S\subset C_1$ with finite elements
\begin{equation}
\{\cu_n\}_{n\ge1}:\bigoplus_{{O_1},{O_2}\in S}\ch_1({O_1},{O_2})\to \bigoplus_{{O_1^\prime},{O_2^\prime}\in \cu(S)}\ch_2({O_1^\prime},{O_2^\prime})\;,
\end{equation}
is an $A_\infty$ morphism between $A_\infty$ algebras of ${\boldsymbol{Z}}$-graded vector spaces with different product structures. 
An $A_\infty$ functor is called an $A_\infty$-{\it{quasi}}-{\it{isomorphism}} when $\cu_1$ gives the quasi-isomorphism between complexes.

In this 
section,
%subsection,
 we invoke the minimal model theorem.\cite{Kad,Kajiura} (A {\it{minimal}} $A_\infty$ category is 
the $A_\infty$ category in which the gauge symmetries of the string fields are fixed
in such a way its product structure ${m}^Q_1$ is trivial: $m_1^Q=0$.)
\begin{Theorem}
For any $A_\infty$ category $C$, there exists a quasi-equivalence $A_\infty$ functor $\cu$ from a minimal $A_\infty$ category $H(C)$ (called the minimal model of $C$) to it. The morphism space of $H(C)$ is the cohomology of the lowest product structure, as defined by Eq.(\ref{eq:low}), of the original $A_\infty$ category $C$.
 \end{Theorem}
 From this statement, the morphism space of the minimal model of $C^Q({\fh})$ consists of closed forms for the BRST differential $\delta$ of the gauged S-duality. 

From the minimal model theorem, the $A_\infty$ functor ${\cal{U}}$ plays a crucial role in revealing the geometrical structure produced by introducing the nonlinear potential to glue together different BRST invariant wave functions $\Psi$ to form the generalized BRST invariant wave function $\Psi^\nabla$. Namely, the {{equivalence principle for all the quantized gauge interactions as Chan-Paton gauge interactions}} holds. In the following, we consider this in more detail. Here, we note that the perturbative structures of the Chan-Paton gauge interactions appear in the {{product structures}} 
\begin{equation}
{m}^Q_n(\co_1,\ldots,\co_n)\;,\ \ n=2,3,\ldots\;,
\end{equation}
of the $A_\infty$ category, similar to the interactions in the Fukaya category\cite{Fukaya} and quantum cohomology. The statement of this principle is very simple: {\it{the BRST invariant wave function $\Psi$ with nontrivial Chan-Paton interactions (D-particle scattering) cannot be distinguished from the wave function $\Psi^\nabla$ with {{no}} Chan-Paton interaction obtained from $\Psi$ via a certain $A_\infty$ functor.}} That is, {{for given Chan-Paton gauge interactions ${m}^Q_n$ ($n=2,3,\ldots$), describing the D-particle scattering, and for any vacuum $\Psi^\nabla$ for a locality around the dual coordinates ${{{{{y^\alpha}}}}}$ of the BRST differential ${\delta}$, by taking a certain $A_\infty$ functor ${\cal{U}}$, we can completely eliminate these interactions ({{as a whole not in each perturbative degree}}), obtaining the free motion
\begin{equation}
{m}^Q_1({\cal{U}}_1(\co))=0\;,\label{eq:null2}
\end{equation} under this vacuum in its lowest product structure Eq.(\ref{eq:low}). (Here, we recall that the BRST open string field theory requires the Maurer-Cartan equation Eq.(\ref{eq:MC}) for degree one open string field $\co$\cite{Kajiura},
and in our claim the summation of all of the interaction terms in it vanishes.
We note that from its definition any $A_\infty$ functor commutes with $m^Q=\sum_{n\ge1}m^Q_n$ of $H(C)$ and $C$.\footnote{In Eq.(\ref{eq:null2}) we use the fact that ${\cal{U}}_1$ commutes with $m_1^Q$ of $H(C)$ and $C$. This fact follows from this definition of an $A_\infty$ functor.\cite{Kajiura}} Thus the quasi-equivalence $A_\infty$ functor in Theorem 1 maps the solutions of Eq.(\ref{eq:MC}) in $H(C)$ to those in $C$.) The existence of such the $A_\infty$ functor $\cu$ is ensured by Theorem 1.

We need to point out two mathematical facts regarding the statement of this equivalence principle. First, the minimal model of an $A_\infty$ category is unique under $A_\infty$ equivalence.\cite{Kad,Kajiura} Second, we assume only one $A_\infty$ product structure $m_n^Q$ ($n=1,2,\ldots$) of $C^Q(\fh)$ defined in the previous
section.
%subsection.
 Due to these two facts, this equivalence principle is mathematically consistent.

 The locality, around the dual coordinates ${{{{{y^\alpha}}}}}$ of the BRST differential ${\delta}$, of $\Psi^\nabla$ is an object in $C^Q({\fh})$. Hence, by using a certain coordinate frame of ${{{{{{y^\alpha}}}}}}$, we can locally eliminate the lowest product structure ${m}^Q_1$, which results in the renormalizations of higher structures ${m}^Q_n$ ($n=2,3,\ldots$) of the vacuum. (Here, we note again that $m_n^Q$ ($n=2,3,\ldots$) indicate not the dynamics but the structure of the interactions.)
 Besides this, the $A_\infty$ covariance principle for the physical substance of the interactions, that is, the quantum mechanical product structures ${m}^Q_n$ ($n=1,2,\ldots$) in $C^Q({\fh})$ for $A_\infty$ functors, is important for the categorification of the theory of gauged S-duality after introducing the nonlinear potential. The well-definedness of this covariance principle is based on the equivalence principle.

The geometry of the ``${\cal{U}}$-manifold'' (referring to Eq.(\ref{eq:trf})), for an arbitrary number $\ell=1,2,\ldots,\infty$,\begin{eqnarray}
\Psi^\nabla|_S&=&\Biggl(\cdots\Biggl(\Biggl(\Psi|_{S_1}\bigcup_{{\cal{U}}_{(1)}} \Psi|_{S_2}\Biggr)\bigcup_{{\cal{U}}_{(2)}}\Psi|_{S_3}\Biggr)\cdots\bigcup_{{\cal{U}}_{(\ell-1)}}\Psi|_{S_{\ell}}\Biggr)\;,\nonumber\\ S&=&\bigcup_{i=1}^{\ell}S_i\;,\label{eq:Uman}
\end{eqnarray}obtained by gluing the restricted vacua via $A_\infty$ functors gives the geometrization of all of the quantized gauge interactions that exist under the vacuum to be considered. In Eq.(\ref{eq:Uman}), the Chan-Paton interactions are created by the glues ${\cal{U}}_{(i)}$ ($i=1,2,\ldots,\ell-1$) due to the existence of the nonlinear second gauge potential in the covariant derivative $\nabla$. We define a locally $A_\infty$ category $C^\nabla(\fh)$ with product structures $m_n^\nabla$ ($n=1,2,\ldots$) such that \begin{eqnarray}{{\bs{m}}^{Q}}&=&(m_n^{Q}({s}))_n\;,\ \ {{\bs{m}}^{\nabla}}=(m_n^{\nabla}({s}))_n\;,\nonumber\\ {\bs{m}}^\nabla&=&\Biggl(\cdots\Biggl(\Biggl({\bs{m}}^Q|_{S_1}\bigcup_{{\cal{U}}_{(1)}} {\bs{m}}^Q|_{S_2}\Biggr)\bigcup_{{\cal{U}}_{(2)}}{\bs{m}}^Q|_{S_3}\Biggr)\cdots\bigcup_{{\cal{U}}_{(\ell-1)}}{\bs{m}}^Q|_{S_{\ell}}\Biggr)\;,\label{eq:Cnabla}\end{eqnarray} correspond to the wave function $\Psi^\nabla$ by gluing together the $A_\infty$ categories $C^Q(\fh)$ on each patch via a combination of the gauge transformations and the $A_\infty$ functors according to Eq.(\ref{eq:Uman}). Although $C^\nabla({\mathfrak{H}})$ is a locally $A_\infty$ category, in the following we refer to it just as an `$A_\infty$ category' because $\nabla^2=0$.

 In the previous 
       section,
 %subsection,
  by regarding the BRST charge $Q$ as an $\hat{{\mathfrak{g}}}$ invariant free Hamiltonian of a system, we showed that for the local wave function $\Psi$, the temporal non-unitary evolution of strings induced by variance of the increment of time (in the sense of a quantum gravity effect), is  represented 
  by a quasi-equivalence class of 
  temporally continuous $Q$-complexes
   in the derived category with the BRST charge $Q$. By taking the quasi-equivalence class, we remove the ambiguity in the definitions of the states with same variance of the increment of time. Due to our two new principles of equivalence and $A_\infty$ covariance, the refined temporal non-unitary evolution of strings via quantized Chan-Paton interactions is given by an $A_\infty$ quasi-equivalence class of temporally continuous `{{complexes}}'
   on the $A_\infty$ geometry. That is, a {\it{quasi-equivalence class of 
   temporally continuous 
   twisted $A_\infty$ complexes in the derived category $D(C^\nabla(\fh))$ (with a triangulated structure) of the glued $A_\infty$ category $C^\nabla(\fh)$}}. (The difference between the old and the refined model is in their morphism spaces. In the present refined model, by extending the morphism space, we introduce the new degrees of freedom of interaction data via product structures.  Namely, the refined model of the quantum mechanical world is a generalization of the old model in that the former has many body systems (i.e., systems with internal degrees of freedom) as elements.
 When we define the quasi-equivalence class, the refined model has more precision than the old model on this point.)

 Here, a {\it{twisted $A_\infty$ complex}} is a pair $({\cal{C}},{\cal{Q}})$. ${\cal{C}}$ is an object---a finite 
 formal direct sum of objects of $C^\nabla(\fh)$ with $\bs{Z}$ numbers sliding the object degrees,
  which are added to adjust to those of morphism spaces\cite{FOOO}---of
 the additive enlargement $S(C^\nabla(\fh))$ of the original glued $A_\infty$ category $C^\nabla(\fh)$, by which we consider the relation systems of 
 interacting D-particle vectors and transitions between them. ${\cal{Q}}$ is an element of the degree one endomorphism space of the object ${\cal{C}}$
 such that 
  ${\cal{Q}}$ is strictly upper triangular in its matrix representation (as explained in the next paragraph) and satisfies the condition for the product structures ${m^{\nabla,S}_n}$ ($n=1,2,\ldots$) of $S(C^\nabla(\fh))$,
\begin{equation}\sum_{n=1}^\infty {m^{\nabla,S}_n}({\cal{Q}},\ldots,{\cal{Q}})=0\;,\label{eq:tw}\end{equation}to which the nilpotency condition, as seen in that on ${m^{\nabla,S}_1}$, is generalized.\cite{Seidel,FOOO}

 We now explain the details of Eq.(\ref{eq:tw}). ${\cal{Q}}$ has a matrix representation labeled by two indices of the slid ${\bs{Z}}$-grading numbers and Eq.(\ref{eq:tw}) is a simultaneous equation for the matrix elements of ${\cal{Q}}$. Then, the graded morphism structure between an object ${\cal{C}}$ of $S(C^\nabla(\fh))$ and itself, as the direct formal sum of the objects in $C^\nabla(\fh)$ with grading, appearing in the product of matrices in ${\cal{Q}}$ produces the higher terms ${m^{\nabla,S}_k}$ ($k\ge2$). These higher terms, especially $m_2^{\nabla,S}$, have the meaning that is essentially the same as that of the composition $\circ$ of morphisms in an associative category.
 Actually, on the associative model, the condition in Eq.(\ref{eq:tw}) is just `${\cal{Q}}\circ {\cal{Q}}=0$'.
 Upper triangularity of ${\cal{Q}}$ means that Eq.(\ref{eq:tw}) is a finite sum.

 The category of twisted $A_\infty$ complexes with the morphism spaces in $S(C^\nabla(\fh))$ has an $A_\infty$ structure. The lowest product structure of this $A_\infty$ category is the generalization of $\delta$ to the relation system of D-particle vectors. We can define the derived category of the original $A_\infty$ category by the zero-th cohomology of the lowest product structure of the $A_\infty$ category of twisted $A_\infty$ complexes.\cite{Seidel,FOOO} As noted from Eq.(\ref{eq:tw}),  ${\cal{Q}}$ reflects the interaction structures of D-particle vectors, in the given relation system of D-particle vectors ${\cal{C}}$ with its graded morphism structure, that is, the product structures ${m^{\nabla,S}_n}$ ($n=1,2,\ldots$) in $S(C^\nabla(\fh))$ generated by the nonlinear second gauge potential. Regarding this point, we note that our $A_\infty$ category is {{background independent}} and all of space-time and all matter consist of its objects. There is no isolated and completed object, and any object has morphisms between other objects and belongs to the (nontrivial) relation systems, which are explained above. 
 
 \section{Summary}

   In this 
       chapter,
 %section
 we first reformulated the results of 
 time-dependent processes on the HKK theater given in 
       Chapter 5
 %Section 5
 using the language of the derived category. Then, based on this derived category reformulation, 
  by introducing the non-linear potential that represents a gauged string field operator, we performed the globalization of the description over the configuration space.
  Finally,
we reformulated the conventional Chan-Paton gauge interactions, whose non-perturbative distortions are given by the non-linear potential, in the language of the $A_\infty$ category of the D-brane fields (in particular, the D-particle fields), by invoking the minimal model theorem of $A_\infty$ categories as the equivalence principle. We concluded that their non-unitary time evolutions are represented by the objects in the derived category of the $A_\infty$ category, that is, the quasi-equivalence classes of 
the temporally continuous 
twisted $A_\infty$ complexes.

The essential part of this derived category is 
its base $A_\infty$ category. Here, we give an overview of the basic physical interpretation of the base $A_\infty$ category. This category is based on the ground state, which corresponds to a
 wave function parallel to the covariant derivative and represents the 
 configuration of a network of D-branes (i.e., the vacuum configuration of 
(interacting) D-branes with open F-strings vertically attached to them) in a vacuum determined by a list of values of the field variables 
 $\phi$. The objects are the vector spaces of the excited states of D-brane network from this ground state. The morphisms between objects, that is, the relationships are the transition rates between objects as the fields of open F-strings vertically attached to D-branes in the D-brane network. (In the base Abelian category before the $A_\infty$ refinement, the morphisms between objects are also the transition rates.)
In particular, the quasi-isomorphisms, which are morphisms, are spatial and dynamical objects, so the borders between quasi-equivalence classes of complexes (i.e., quantum classes, including the observer's self, discussed in 
Chapter 4)
%Section 4)
are not assumed {{a priori}} but are spatially and dynamically formed.
Although, before introducing the non-linear potential, the base $A_\infty$ category cannot intrinsically describe the non-perturbative transitions between objects, after introducing the non-linear potential, it can do so. In this sense, the non-linear potential means the D-brane network itself that is equivalent to the totality of the objects.

\chapter{Conclusion}
In this final 
chapter,
%section,
 we summarize the time concept, ingredients and geometrical form of the model of our theory and make some conclusions regarding them.

First, we summarize the time processes in the model of our theory, which have the following two stages.
\begin{enumerate}
\item Every non-unitary time process is caused by the statistical variance of
 time increments determined by the values of the hidden time variable via
  non-local hidden variables according to the Penrose thesis. (See 
Chapter 5.)
%Section 5.)
\item In the conclusive form, time developments are geometrized as {{the objects in the derived category of the $A_\infty$ category of the D-brane fields}} affected by the product structure of this $A_\infty$ category and a non-linear potential. The non-perturbative dynamics in the category is intrinsically controlled by the relationship between objects as morphisms determined by the non-linear potential. There are principles of covariance and equivalence. (See 
Chapter 6.)
%Section 6.)
\end{enumerate}

Second, we consider the following (broken) fundamental symmetries of time for the Universe and the observers.
\begin{enumerate}
\item S(trong)TRpS and TRS are broken due to the birth of the Universe, which possesses particular Virasoro and initial data.
\item TRpS is broken due to the existence of the observers which have classical and quantum self-identities simultaneously. (See 
Chapter 4.)
%Section 4.)
\end{enumerate}
Here, the STRpS is the {{reparametrization symmetry of all of the temporal hidden variables (i.e., the hidden time variable and the temporal non-local hidden variables).}} We postulate that STRpS or equivalently TRS is unbroken before the birth of the Universe.
The most important assumption in the construction of any physical theory is that {\it{it is based on the existence of the observers and, for observers only, the TRpS is spontaneously broken. }}
Since, TRpS and STRpS are {{gauge symmetries}} for time and temporal hidden variables
 respectively, the breakdown of TRpS makes the unitary processes distinguishable and the breakdown of STRpS makes the non-unitary processes distinguishable:
\begin{subequations}
\begin{align}
{\mbox{Unitary\ Processes}}&\equiv 0 \ \ {\mbox{Mod\ TRpS}}\;,\\
{\mbox{Non-Unitary\ Processes}}&\equiv 0\ \  {\mbox{Mod\ STRpS}}\;.\label{eq:STRpS}
\end{align}
\end{subequations}
If STRpS is unbroken, due to Eq.(\ref{eq:STRpS}) there is no stock of past events and the concepts of past and future make no sense. STRpS is broken when initial conditions with non-zero statistical variance 
of time increments 
are given.
The {\it{arrow of time}} is generated due to the existence of this non-zero statistical variance: since the non-unitary processes are non-reversible, they produce the arrow of time. If the statistical variance is zero, there are no non-unitary processes and, thus, there is no arrow of time.
We consider this breakdown is a spontaneous breakdown. That is, we regard the Virasoro data and the inverse of the gradients of the time increment by the Virasoro data as the vacuum expectation values of scalar fields $\Theta$ and their Goldstone modes, respectively:
\begin{subequations}
\begin{align}
\langle{\Theta}\rangle&=\{{\mbox{The\ Set\ of\ Non-Zero\ Variance\ Statistics\ of\ Time\ Increments}}\}\;,\label{eq:Theta1}\\
\tilde{\Theta}&=({\nabla}(\delta \tau))^{-1}\;,
\end{align}
\end{subequations}
where $\nabla$ represents the gradients along the Virasoro data, that is, the inverse of the variations of time variable for a given time increment. As seen in Eq.(\ref{eq:Theta1}), the particles $\Theta$ are related to an integrable hierarchy:\begin{equation}
\Theta=(\Theta_n)_{n\in{\boldsymbol{Z}}}\;,
\end{equation} since the Virasoro data are. These particles $\Theta$ accompany every kind of particle in the excitations of fundamental strings.
We note that these particles are {{not time itself}} but objects which {{convert}} the reversible time variable ${s}$ into the non-reversible time $\tau$.
The spontaneous symmetry breakdown of STRpS is considered as a phase transition of the vacuum below a certain high energy scale. Here, we note that the energy concept is superior to time $\tau$ since the energy can correspond to the time variable ${s}$ before the breakdown of STRpS; this breakdown generates $\tau$.

Our results about the concept of time are summarized as follows.
\begin{enumerate}
\item First, there exist a STRpS for all of the temporal hidden variables 
and the $\Theta$-particles. 
The time increment has {{zero}}-statistical variance. In this stage, all of time-dependent processes are indistinguishable and are reversible, thus past and future have no sense. There is no actual measurement or observer, thus there is no arrow of time.
\item This STRpS is spontaneously broken by giving the vacuum expectation values of $\Theta$, which produce particular 
Virasoro data, as a kind of phase transition of the vacuum. Then the Goldstone modes of $\Theta$ become the inverse of the gradients of the time increment by the Virasoro data
\begin{equation}
\tilde{\Theta}=({\nabla}(\delta \tau))^{-1}\;.
\end{equation} These generate the non-unitary processes, break TRS and generate the arrow of time.
\item Moreover the TRpS can be broken by the existence of an observer. Then the Goldstone mode of TRpS is, in our proposal, regarded as the quantum fluctuation of the time increment 
\begin{equation}
\delta\tilde{\tau}^G=\widetilde{\delta \tau}^Q\;,
\end{equation}and it produces non-unitary processes under the broken TRpS and the observers can distinguish between unitary and non-unitary processes. Here, the Goldstone mode of time is given by the Goldstone mode of the temporal lapse function in the ADM space-time metric. 
The scale of the statistical variance of time increments is 
determined by that of the vacuum expectation value of time in TRpS breaking.\end{enumerate}

Regarding the concept of time in the model of our theory, we conclude that the vacuum expectation values of $\Theta$ particles in the spontaneous breakdown of STRpS due to the birth of the Universe cause the contents of the Goldstone modes of TRpS induced by the observer's consciousness and the non-unitary time processes.

Next, we summarize all of the ingredients in the model of our theory in the four following arguments.
\begin{enumerate}
\item There are only two postulates (after STRpS breaking) as primary laws.
      \begin{enumerate}
\item {{{Geometric symmetries}}}: They are 
loop algebra gauge symmetry and the consequent non-perturbative discrete modular symmetry of its quantization.

\item {{TRpS}}: This leads to the Wheeler-De Witt equation that corresponds to the 
generalized
Kugo-Ojima physical state condition.

\end{enumerate}
\item
All of the {{fixed}} data 
with secondary laws, which are arbitrarily produced, are as follows.  
\begin{enumerate}
\item $\langle{{\Theta}}\rangle$: These randomly produced fixed numbers break STRpS and TRS, generate the arrow of time and determine all of the contents of `now', that is, the non-unitary time processes of all of the matter systems. 
\item ${{\zeta}}$: This leads to the fundamental structures of the model, that is, the dimensionality of space-time and the kinds of fundamental gauge interactions. It is the breaker of the discrete modular symmetries. The qualitative statements of the physical laws are determined by it.
\item ${{\aleph}}$: The set of these is a solution of the model. The 
 distortions of the fundamental structures of
 vacua $\Psi$
 given by ${{\zeta}}$ are completely and intrinsically determined by them like as the distortions of space-time are given by the metric potentials in the general theory of relativity.
\end{enumerate}
\item
An observer's consciousness is identified with a macroscopically coherent quantum ground state with non-zero 
superposition retention time and classical and quantum mechanical self-identities.
This leads to TRpS breaking in the observer's quantum mechanical world; and the Goldstone modes give its hard activities (i.e., qualia and free will)\footnote{Of course, these hard activities are processed by classical mechanical information dynamics. We need to distinguish clearly between these two parts of consciousness. In particular, the former is an immortal concept.}, which are determined by $\langle{{\Theta}}\rangle$. {\it{The border of the observer's self is not assumed {{a priori}} but is spatially and dynamically formed (here, we note the TRpS breaking and recovering phenomena).}}
\item
The independent variables appearing in the wave functions of the Universe and matter systems, and the metric fields, given by the HKK method, in the classical regime
\begin{subequations}
\begin{align}
{{\Psi}}&={{\Psi[{{{g}}},{\phi}]}}\;,\\ {{\psi}}&={{\psi[{{{g}}},s,{\phi};\tau]}}\;,
\\
g_{\mu\nu}&=g_{\mu\nu}[s,x;\phi]\;,
\end{align}
\end{subequations}
 are the following. 
\begin{enumerate}
\item ${{{{{g}}}}}$ is the ten-dimensional Yang-Mills coupling constant of the geometric symmetry and is proportional to the cube of the string slope parameter.
\item ${{s}}$ is the hidden time variable of a linear combination of the Hamiltonian-form charges for the geometric symmetry.
\item ${{x}}$ are the spatial coordinates on the space-time represented by
the field variables.
\item ${{\tau}}$ is the history of the cosmic time. $s$ and $\tau$ have quantitative meaning only for matter systems and when TRpS is broken.
\item $\phi$ are the field variables. A list of their values determines all of the  configurations of elements in their network in $\Psi$. 
  In the  metric fields, they appear implicitly.
\end{enumerate}
As noted in 
Chapter 5,
%Section 5,
 the time variable $s$ is hidden behind the cosmic time behavior of matter wave functions. We can only incompletely estimate $s$ and an infinite number of non-local hidden numbers, that is, the Virasoro and initial data, from a finite number of non-unitary quantum mechanical events. Here, the results of all of the non-unitary quantum mechanical events are mapped into and governed by the complete set of the temporal hidden variables $\{s,\langle\Theta\rangle\}$ where $s$ is dynamical.
This
 complete set of the temporal hidden variables 
 is surprisingly common among the elements in the network in $\Psi$
  including consciousness of the observers, and is thus unitary and immortal. 
 This complete set is represented by the respective matter wave functions.
 {\it{In this sense, the fluctuating time increment is substitutable for, and is the most natural generalization of the concept of the observer's consciousness.}} We have proposed that time itself has two reparametrization symmetries, on whose breaking its existence relies, and a statistical structure.
\end{enumerate}

To conclude this paper, besides these four arguments, we need to argue again that the quantum mechanical world with non-unitary time processes is formulated as a derived category as mentioned in the beginning of this 
            chapter 
  %section
  and studied in
      Chapter 6
  %      Section 6
  in detail.
  The main points of this argument are as follows. The objects of the base category of this derived category are the vector spaces of the excited states above the ground state of the D-brane network, whose 
 configuration is determined by a list of values of the field variables $\phi$. The morphisms between objects are the transition rates between them as the fields of open F-strings vertically attached to the D-brane network. Owing to the non-perturbative distortions caused by the non-linear potential, we can intrinsically treat the non-perturbative transitions between D-brane networks by making the morphisms dynamical. The temporal developments of the matter systems are described by the objects in this derived category. 
  We also argue that the basis of the argument of the derived category formulation is the postulation of TRpS, since as mentioned in the above list, TRpS enables us to identify the 
  generalized 
  Kugo-Ojima physical state condition of the geometric symmetry in our model with the Wheeler-De Witt equation in type IIB string theory. This derived category is the conclusive form of our quantum geometrodynamics, and it represents curved, empty space and nothing more.

\begin{appendix}
\chapter{Physiological Elements of the Neural-Glial System}
 
In this appendix, we present minimal accounts of the physiological elements of the neural-glial system in the human brain, which are required in 
Chapter 3.
%Section 3.

\section{The Neural System}
We give three accounts of the neural system in the human brain.\cite{Konishi}

First, today, the classical physical definite formulation of {{spike activities}} in the brain is based on the Hodgkin-Huxley model.\cite{HH} This models the cell membrane and ion-channels of a neuron by the condenser and dynamical registers in an electric circuit. The voltage-dependent sodium (Na$^{+}$) and potassium (K$^{+}$) ion channels are embedded in neuronal cell membranes and keep the equilibrium electric potential by adjusting the ion concentrations inside and outside of the neurons. Each voltage-dependent ion channel has probability factor for its opening. This circuit obeys simple non-linear differential equations for the conservation of electric currents via the electric potential and inflowing currents. This depolarization of membrane potentials induced by a sodium ion current greater than a threshold value is termed a spike. After the generation of spikes, the membrane potential repolarizes and returns to the resting state by the inactivation of the sodium channel and the activation of the potassium channel.

Second, the synaptic signal transmission is mediated by the {\it{neurotransmitters}}. The mechanism of the emission of the neurotransmitters is as follows.\cite{Kandel} When a spike 
 arrives at the pre-synaptic site, the voltage-gated calcium channels open. Then, the outer calcium ions (Ca$^{2+}$) flow into the pre-synaptic site, and due to the action of these calcium ions, a vesicle will couple to the pre-synaptic membrane. Then, the neurotransmitters in the vesicle are emitted into the synaptic gap. The number of emitted neurotransmitters is proportional to the concentrations of calcium ions in the pre-synaptic site and the time span of the opening of the voltage-gated calcium channels.

Third, the dynamics of spikes in the Hodgkin-Huxley theory essentially consists of non-linear oscillations. This activity is compatible with the Hopfield model\cite{H1,H2} of the neural network in a statistical mechanical fashion and can be encoded in its discrete variables. The neural network models are recognized as models of associative memory and learning. Here, {\it{associative memory}} indicates that the system will settle down to stable patterns of the excitatory neurons (i.e., the memories in these models) which are determined by the types of the inputs. In the {{Hopfield model}}, {\it{learning}} processes quadratically strengthen the excitatory couplings of neurons by the synaptic plasticity for the stable memory patterns quadratically embedded in these couplings, according to the Hebbian learning rule. {\it{Unlearning}} processes strengthen the inhibitory couplings by random patterns.\cite{REM1,Hopfield}

\section{Glial Modulation}
We give a minimal account of the astrocyte's role in glial modulation.\cite{Konishi}

During the past two dacades, a revolution has occurred in the recognition of the functions of astrocytes.\cite{Glia1,Glia21,Glia22,Glia23,Glia24} We explain the physiological elements of the neural and glial network\cite{Konishi}, which takes into account this new view.

The astrocytes have recognized to have mainly three functions from the physiological view point:\cite{Glia24} the {\it{modulation of the synaptic transmission}}, the {\it{neural synchronization}} and the {\it{regulation of cerebral blood flow}}. Between them, the one of interest here is focused on the first function including the maintenance of the homeostasis of the concentrations of ions, neurotransmitters and water. We explain it through the following three processes.\cite{Glia22,Glia24}

 \renewcommand{\theenumi}{\alph{enumi}}
\renewcommand{\labelenumi}{(\theenumi)}

\begin{enumerate}
\item As has been recently discovered, each astrocyte communicates with the others through gap junctions via the calcium ion (Ca$^{2+}$) wave produced by the calcium-induced calcium release from the intracellular calcium stores of astrocytes.\cite{Glia1}
 Consequently, the activation of a glial receptor by release of neurotransmitters from the pre-synaptic site results in the modulation of distant synapses by release of neurotransmitters from other astrocytes via the calcium ion waves. This means that the modulation of synaptic junctions (see (b)) among different synapses is done globally.
\item The neurotransmitters flowing at the synaptic sites are modulated by the astrocytes.\cite{Glia1}
 The activation of glial receptors of astrocytes by the release of neurotransmitters from the pre-synaptic site, where this release is evoked by every spike event, inputs to the intracellular concentrations ${{In}}(t)$ into the calcium store at a time $t$, and the output to the extracellular concentrations of the neurotransmitters of astrocytes ${{Out}}(t)$, that is, the activation of the pre- and post-synaptic receptors with the regulation of the synaptic transmitter release can be modeled to satisfy
\begin{equation}
{{Out}}(t)=C^{(1)}{{In}}(t)\;,\ \ {{Out}}(t)=\sum_{k=1}^MC^{(2)}_k(t){{Out}}(t-(k-1)t_0)\;,
\end{equation} for constant $C^{(1)}$, time interval of spikes $t_0$\footnote{For the simplicity of the model, we model the time intervals between spikes to be constant.} in the order of 1ms to 100ms\cite{Kandel} and time span $Mt_0$ of the modulation, without losing the physical essence of the model. $In(t)$ and $Out(t)$ are vectors at every time, with indices corresponding to the synapses, and $C^{(1)}$ and $C^{(2)}_k(t)$ are matrices. These modulations maintain the homeostasis of the concentrations of ions, neurotransmitters and water in the synaptic gaps.\cite{Glia1}
So, there are constraints on $C^{(2)}_k(t)$.
\item
When the astrocytes bridge different synapses via their calcium ion waves and modulate them, the glial action describes the feedforward and feedback properties of the regulation of pre-synaptic junctions. These properties are due to the cyclic activation, via the modulation by astrocytes, of more than one synaptic junction, such as heterosynaptic depression and the potentiating of inhibitory synapses etc.\cite{Glia1}
\end{enumerate}

To define the model of the astrocytes mathematically, we consider two points.\cite{Konishi} First, the energy of the system (see Eq.(\ref{eq:start})) always tends to decrease towards the minimum. Second, the glia's function of the maintenance of the homeostasis (a) and (b) and the feedback or feedforward type of their modulations (c), denoted by ${\cal{G}}$, means that the glial outputs ${{Out}}(t)$, which are temporally accumulated in the synaptic gaps by the temporal changes of the neural states, are linearly averaged in a time range\footnote{This is because the glial network has no threshold structure.\cite{Glia1}} \begin{equation}I_0=t_0\times [1,M]_{\bs{N}}\;,\end{equation} by the linear transformation $\exp({\cal{G}})$ on the corresponding $M$ inputs via the temporal neural state vector $\varphi$. We note that this linear transformation should be recognized not as the modulation of the neural states at different times but the modulation of the glial outputs ${{Out}}(t)$ to the synapse accumulated in the synaptic gaps at different times. Due to (c), these outputs reflect the future inputs by the feedback or feedforward property of the glial modulations. (Here, based on the glia's function of the maintenance of homeostasis (a), that is, the globality of the astrocyte action on the synapses and (b), the glial actions ${\cal{G}}$ are defined to be Lie algebra valued, with a linear basis that is related 
 to the modes of their modulations as constants of motion in the modulation of the synaptic junctions.) The time range $I_0$ determines the time span of the maintenance of homeostasis. In this paper, to simplify the case, we assume that the time range is unique. Our modeling is more advanced than averaging by adding a kinetic term to the Hamiltonian since the latter approach does not incorporate such a time span. Let us assume this time range $I_0$ is equal to the one of the periodic cycles of the neural states in the neural-glial system (see Eq.(\ref{eq:general})), that is, \begin{equation}M=N\;.\label{eq:main}\end{equation}
This assumption is under the following logic. First, if the periodic cycles of the neural states in the neural-glial system $N$ exist, their unit time span is longer than the time span of the maintenance of homeostasis due to the definition of the latter, $M\le N$. Second, for the latter time span $M$, since such the gilal action makes the neural system tend to be linear, approximately $N\le M$. Then, at least under this approximation for the second logic, Eq.(\ref{eq:main}) holds.

Based on these two points, in 
Chapter 3
%Section 3
we incorporate the glial network into the neural network.

\chapter{Transition Rates of Observer's Quantum State}

After neural death of an observer, its aspect of self is lost and only none survives. 
(For a discussion of the terminology `self' and `none', see
Chapter 4.)
%Section 4.)
 Even though the self is lost, there may exist transitions from none to self. We note that the ending self and beginning self put a pure none state between them; so they do not correspond and cannot be connected. Thus, we cannot identify the self in neural birth and the self in neural death.\cite{Konishi3}

In this appendix, we describe the transition from pure none to self (neural birth) according to Ref.5. To simplify the explanation, we take the neural-glial network in the brain as the self.
The birth process, that is, the process of the creation of self (including the relic of pure none) from a pure none state, is described by a non-space-time tunneling into the potential barrier of ourselves. Regarding this point, we note that the neural network, corresponding to ourselves, and the superradiative circuit, corresponding to the none, are different physical entities, though they are highly correlated with each other. So, although this birth process is induced by classical mechanical activation on the neural network (i.e., ourselves) via neurotransmitters, the birth process for the superradiative circuit (i.e., the none) to activate it is quantum tunneling (recall the definition of the neural states $\phi$).

 To describe the activation of this none,\footnote{Here, {\it{activation}} means to coincide two asymmetries of directions of evolutions in that the system evolves as its potential energy decreases and that the system promotes its state only when neurons are fired (i.e., for positive valued $\phi$).} we use the null-Hamiltonian constraint on its wave function in the canonical quantum theory of the variables $\varphi$, not in the language of operator formalism, which leads to the exact time reparametrization invariance of the initial pure none state,
 \begin{equation}(-{\hbar^2}\nabla^2_\phi+2\mu_\phi\ch_{int}(\phi))\psi(\phi)=0\;,\label{eq:Schr}\end{equation}
 where the arbitrariness of the constant additivity of the Hamiltonian 
 $\ch_{int}$ (i.e., $\ch_0$ in Eqs. (\ref{eq:H}) and (\ref{eq:H2})) is fixed to adjust a particular non-spike state (e.g., all $\phi$ are $-\frac{1}{\sqrt{n}}$) to be a zero-energy state. 
Here, in the quantum regime, since the variables of the brain wave function are the neural states $\phi$, the glial states ${\cal{G}}$ can be recognized as a temporally constant background.
So, in Eq.(\ref{eq:Schr}), we quench the glial states ${\cal{G}}$ temporally and focus only on the neural states.

When we need to describe the activation process in a simplified situation, choosing the constraint on the none states to be the null-Hamiltonian constraint is appropriate, since the none state identified with a quantum system itself before any activation of the quantum system is the zero-energy state, as explained above. After the tunneling into activation, the quantum system temporally develops according to the rule in the Hopfield type quantum neural-glial network model (i.e., nonlinearly rolling down the potential energy in the higher dimensional space of variables, $\varphi$ and ${\cal{G}}$).
We remark that in general exact time reparametrization invariant quantum systems, we identify the solutions of Schr${\ddot{{\rm{o}}}}$dinger equations written with equivalent time parametrizations by keeping count of non-unitary processes. When we use a general pure none state, which is assumed to be a maximally closed system, as the initial state, the process of neural birth is also described by tunneling. This is because, if it is not tunneling, the initial state is already self and contradicts the above assumption.

The definitions of the elements in the total Hamiltonian in Eq.(\ref{eq:Schr}) are as follows.
 First, $\nabla^2_\phi$ is the Laplacian.
Second, $\mu_\phi$ is the inertia of neural states measured by the kinetic responses to energy inputs. Third, when the glial states ${\cal{G}}$ are not temporally quenched,
the interaction Hamiltonian of the quantum neural-glial network, which has the {{activation}} potential barrier for none states, is the positively shifted version of Eq.(\ref{eq:start}):
\begin{equation}
\ch_{int}(\varphi,{\cal{G}})=-\f{1}{2N}\la\la \varphi,\exp(\Delta)\varphi\ra\ra+\ch_0\;.\label{eq:H}\end{equation}
When the glial states ${\cal{G}}$ are temporally quenched, the interaction Hamiltonian $\ch_{int}$ can be written as
\begin{equation}
\ch_{int}(\phi)=-\frac{1}{2}\la \phi,J\phi\ra+\ch_0\;,\label{eq:H2}
\end{equation}
where the {{effective}} synaptic couplings are denoted by $J$ and assumed to be temporally quenched with respect to the quantum regime.

In the activation process of the none, the wave function begins from a particular pure none state $\psi(\phi_0)$ with negative-valued $\phi_0$, just like Vilenkin's scenario of the birth of the Universe from nothing. This kind of scenario leads to the absence of an initial singularity.  
 In $\psi$, $\phi$ is not a function but just a number. So, the potential is not a function of space-time coordinates $(x,t)$, reflecting the scale and time reparametrization invariance of the initial pure none state. 
This is the meaning of {{non-space-time}} in `non-space-time tunneling'.

 Now, using Eq.(\ref{eq:Schr}), we derive the formula for the activation rate (i.e., the tunneling rate) for the potential barrier $\ch_{int}$ in the following simplified situation.
When we discuss the tunneling process of the brain wave function, in terms of the behavior of the neural states, the radial part is dominant.\footnote{Here, we relax the normalization condition on the vector $\phi$. After we obtain the vector $\phi=\langle\phi\rangle$ by using $\psi(\phi)$, we normalize it so we can compare the results of $\psi(\phi)$ with those of the wave function $\psi(\beta)$.} So, to simplify the argument, we consider the case in which the system depends only on the radius of the space of the neural states, denoted by $\phi_r$, and the system is reduced to a one-dimensional one. We denote the strength of the reduced effective synaptic couplings by $J_0$. Then, Eq.(\ref{eq:Schr}) is reduced to
 \begin{equation}
 (-{\hbar^2}\nabla_{\phi_r}^2+2\mu_\phi\ch_{int}(\phi_r))\psi(\phi_r)=0\;,\ \ \ch_{int}(\phi_r)=-J_0\phi_r^2+\ch_0\;.\label{eq:Sch4}
 \end{equation}
 When we apply the WKB approximation to the tunneling process for the potential $\ch_{int}$ in Eq.(\ref{eq:Sch4}), the well-known results in the general setting for the in-coming wave function $\psi_{in}$ from the state $(\phi_0)_r$, the transmitting wave function $\psi_{tr}$ and the out-going wave function $\psi_{out}$ are\cite{Landau}
 \begin{subequations}
 \begin{align}
 \psi_{in}(\phi_r)&=e^\Lambda\frac{(-ic)}{\sqrt{p(\phi_r)}}\exp\Biggl[i\Biggl(\frac{1}{\hbar}\int_{\phi_r}^{a}p(\phi^\prime_r)d\phi^\prime_r-\frac{\pi}{4}\Biggr)\Biggr]\;,\\
 \psi_{tr}(\phi_r)&=\frac{-ic}{\sqrt{\varrho(\phi_r)}}\exp\Biggl(-\frac{1}{\hbar}\int_{b}^{\phi_r}\varrho(\phi_r^\prime)d\phi_r^\prime\Biggr)\;,\\
 \psi_{out}(\phi_r)&=\frac{c}{\sqrt{p(\phi_r)}}\exp\Biggl[i\Biggl(\frac{1}{\hbar}\int_{b}^{\phi_r} p(\phi_r^\prime)d\phi_r^\prime-\frac{\pi}{4}\Biggr)\Biggr]\;,\label{eq:out}
 \end{align}
 \end{subequations}for $a=-\sqrt{\ch_0/J_0}$, $b=\sqrt{\ch_0/J_0}$, a complex number $c$, and
 \begin{equation}
 p(\phi_r)={\textstyle{\sqrt{-2\mu_\phi\ch_{int}(\phi_r)}}}\;,\ \ \varrho(\phi_r)={\textstyle{\sqrt{2\mu_\phi \ch_{int}(\phi_r)}}}\;,\ \ \Lambda=\frac{1}{\hbar}\int_{a}^{b}\varrho(\phi_r^\prime)d\phi_r^\prime\;.
 \end{equation}
 Then, for the currents of probability densities $j_{in}$ and $j_{out}$, the formula for the activation rate $T$ is
 \begin{eqnarray}
 T&=&\frac{j_{out}}{j_{in}}=e^{-2\Lambda}\nonumber\\
 &=&\exp\biggl(-\frac{2}{\hbar}\int_{a}^{b}{\textstyle{\sqrt{2\mu_\phi(-J_0(\phi^\prime_r)^2+\ch_0)}}}d\phi^\prime_r\biggr)\nonumber\\
 &=&\exp\biggl(-\frac{\ch_0}{\hbar}\sqrt{\frac{8\mu_\phi}{J_0}}\int_{-1}^1\textstyle{\sqrt{1-(\phi^\prime_r)^2}}d\phi_r^\prime\biggr)\nonumber\\
 &=&\exp\biggl(-\frac{\pi\ch_0}{\hbar}\sqrt{\frac{2\mu_\phi}{J_0}}\biggr)\;.\label{eq:tr}
 \end{eqnarray}
 (However, this is just a theoretical result, and to obtain the numerical value of $T$ we need experimental measurement of three parameters for the human brain: $\mu_\phi$, $J_0$ and $\ch_0$.)
Here we make a remark.
After the activation, the mechanism of the Eguchi-Kawai large $N$ reduction and the presence of a Bose-Einstein condensate, which ensures the existence of the off-diagonal orders, are the necessary conditions for the new-born human brain-like consciousness to satisfy the criterion in Eq.(\ref{eq:finals}). On this point, we note that the quantum mechanical properties of conscious activities, such as the dynamics of their quantum classes, {{all depend}} on whether there is a Bose-Einstein condensate. Actually, based on the JPY model, it has been hypothesized by Jibu that, during anesthesia, the order structure of water molecules in the perimembranous regions of neural cells is broken by the anesthesia molecules and the critical temperature of the Bose-Einstein condensate falls to less than the living body temperature.\cite{Jibu} When we are using the standpoint of JPY model, where the Bose-Einstein condensate is considered to be directly connected with the physical substance of consciousness, this hypothesis explains the pressure reversal of the potency of anesthesia molecules\cite{Rev}. The experience of anesthesia mediates the pure none. Only in this case the selves mediating a pure none correspond.

Here we consider a time paradox.\cite{Konishi3}
In the category of pure none states there is no Newtonian external time as is in the Wheeler-De Witt equation\cite{KT}. In the above scenario of neural birth, the brain wave functions $\psi(\phi)$ correspond to this situation. This paradox is resolved for quantum mechanical observers, as already mentioned, when we admit the matter Schr$\ddot{{\rm{o}}}$dinger equations with a formal time parameter, by adopting the counting of non-unitary processes as a clock after the spontaneous breakdown of the time reparametrization symmetry. The matter Schr$\ddot{{\rm{o}}}$dinger equations are derived from the Wheeler-De Witt equation in the semiclassical regime of the expanding Universe by adopting the growing scale factor of the Universe as a formal time parameter (see 
Chapter 2).\cite{Vilenkin4}
%Section 2).\cite{Vilenkin4}

From our perspective, we make a remark about the sleeping brain state.\cite{Konishi3}
The spontaneous breakdown of time reparametrization symmetry is accompanied by non-unitary processes within a self, which has a fixed mean time increment, that is, {{measurement}}. The absence of measurement recovers the broken time reparametrization symmetry. Since sleeping state is considered to have no measurement process, due to the blocking of sensory inputs, that is, real world data\cite{Dream}, except for reproduction from the memory stores, in it the time reparametrization symmetry may be recovered partially. Thus, under the unlearning activities of REM sleep, viewed as internal random inputs to forget and stabilize the memories $J$\cite{REM1,Hopfield}, the time concept in a sleep state may be recognized as that in a pure none state satisfying Eq.(\ref{eq:unbroken}) partially.
 
\chapter{Renormalization of the Vacua}
This short appendix provides a brief account of the renormalization method of the vacua relating to the symmetry and its breaking structures of the vacua.\cite{Konishi2} The contents of this appendix are about the vacua before we impose Eq.(\ref{eq:Diff}). So, in the coherent-state representation, the vacua depend on the coordinate $s$.

For a given order parameter $\Lambda$ of a symmetry structure of a vacuum $\Psi$, we define its scale constants as the ratios of the scales of the critical points of the order parameter $\Lambda$ to the bare scale of the vacuum:
\begin{equation}1>\epsilon^{(1)}>\epsilon^{(2)}>\cdots\;,\label{eq:spec}\end{equation}
where the scale of $\epsilon^{(i)}$ gives the critical value $\Lambda^{(i)}$ of the order parameter $\Lambda$.

The renormalization transformation $R$ between the vacuum $\Psi$ and an effective vacuum is formulated by the following {\it{multi-phase average method}} (the {\it{Whitham method}}). It introduces a positive valued relative cut-off scale $\epsilon^{(i)}$ into the effective vacuum $R\Psi$.

 In the Whitham method,\cite{Wh1,Wh2} we redefine our variables of the coordinate $s$ and  the field variables $\phi$ as the fast variables, which are distinguished from the slow variables $S$ and $\varphi$ that are introduced by
 \begin{equation}
S=\epsilon^{(i)}s\;,\ \ \varphi=\epsilon^{(i)}\phi\;.
 \end{equation}

 Then, we deform the additional parameters in the wave function (e.g. the time frequency $\omega$ and the wave numbers $\kappa$ in the wave function)\begin{equation}a\to A(S,\varphi)\;,\end{equation} by the slow variables $S$ and $\varphi$ as a power series in $\epsilon^{(i)}$:\cite{WKB}\begin{equation}
R\Psi=\sum_{\ell\in{\boldsymbol{Z}}_{\ge 0}}(\epsilon^{(i)})^{\ell}\Psi^{(\ell)}[{{{g}}},{s(S)},\phi(\varphi)|A(S,\varphi)]\;,\label{eq:WKB}
\end{equation} satisfying the equation \begin{equation}({{Q}}R)\Psi=0\;,\label{eq:Whitham}\end{equation}
with the BRST charge ${Q}$. 
In Eq.(\ref{eq:WKB}), the phase parameters in $R\Psi$, denoted by $\theta$, are shifted by
\begin{equation}
\theta\to \xi(S,\varphi)+\theta\;,
\end{equation}
where we introduce the fast variables $\xi(S,\varphi)$ of the 
multi-phase functions $I(S,\varphi)$:
 \begin{equation}
 \xi(S,\varphi)=(\epsilon^{(i)})^{-1}I(S,\varphi)\;,
 \end{equation}
 which satisfy Eq.(\ref{eq:Whitham}).
Here, we impose the symmetries, whose order parameters $\Lambda$ are still zero at the critical scale $\epsilon^{(i)}$, on $R\Psi$.

In the Whitham method, we assume the following two periodicity conditions.

First, the periods $\tau^{(0)}$ of $\Psi^{(0)}$ in the fast variables $\xi(S,\varphi)$ are constants $c$ in the slow variables $S$ and $\varphi$:
\begin{equation}
\tau^{(0)}=c\;,
\end{equation}where both sides are vectors.

Second, to exclude the secular terms in $\Psi^{(\ell_+)}$ for $\ell_+=1,2,\ldots$, we set the periods $\tau^{(\ell_+)}$ of any wave function $\Psi^{(\ell_+)}$ in the fast variables $\xi(S,\varphi)$ to be the same as the periods of $\Psi^{(0)}$ in them, that is,
\begin{equation}
\tau^{(\ell_+)}=c\;,\ \ \ell_+=1,2,\ldots\;.
\end{equation}
We set all of the periods $c$ to $2\pi$.

With these deformed solutions $\Psi^{(\ell)}$ for $\ell=0,1,2,\ldots$, we average them by the fast variables $s$ and $\phi$ and regard $R\Psi$ as functions of the coupling constant ${{{g}}}$ and the slow variables $S$ and $\varphi$. (By Eq.(\ref{eq:Whitham}), the dependence of the effective vacuum $R\Psi$ on the coupling constant ${{{g}}}$ is specified.)

The consistency conditions on the $\Psi^{(0)}$ and $\Psi^{(\ell_+)}$ parts of the Whitham deformation, found by the comparisons of both sides of the Kugo-Ojima physical state condition at each power of $(\epsilon^{(i)})^{\ell}$ for $\ell=0,1,2,\ldots$, are the integrability conditions
\begin{equation}\omega=\frac{\partial I}{\partial S}\;,\ \ \kappa=\frac{\partial I}{\partial \varphi}\;,
\end{equation}
 and the solvability conditions as  the 
 local conservation laws under the averages of the contributions from the fast variables $s$ and $\phi$ respectively.\cite{Wh1,Wh2,WKB} The local conservation laws are just Eq.(\ref{eq:Whitham}). This is why we adopt the Whitham method as the renormalization.

\end{appendix}

\end{document}